\documentclass[11pt,twoside]{./atmp}

\usepackage{amsmath,amssymb}
\usepackage{pstricks}
\usepackage{epic, xypic}
\usepackage{amsfonts}
\usepackage{amscd}
\usepackage{txfonts}
\usepackage[english]{babel}
\usepackage[all]{xy}
\usepackage{color}

\newgray{vlgray}{.90}

\renewcommand{\AA}{\mb{A}}

\newcommand{\CC}{\mb{C}}

\newcommand{\FF}{\mb{F}}

\newcommand{\PP}{\mb{P}}
\newcommand{\QQ}{\mb{Q}}
\newcommand{\RR}{\mb{R}}

\newcommand{\ZZ}{\mb{Z}}

\newcommand{\cM}{{\mathcal {M}}}
\newcommand{\cO}{{\mathcal {O}}}

\newcommand{\cRR}{{\mathcal R}}
\newcommand{\cS}{{\mathcal S}}

\newcommand{\Pic}{{\rm Pic}}
\newcommand{\GC}{{\rm G-}\CC^3_6}
\newcommand{\RRC}{{\rm R_2-}\CC^3_6}
\newcommand{\RSC}{{\rm R_3-}\CC^3_6}
\newcommand{\RTC}{{\rm R_4-}\CC^3_6}
\newcommand{\RUC}{{\rm R_5-}\CC^3_6}
\newcommand{\ch}{{\rm ch}}
\newcommand{\td}{{\rm td}}

\ifx\mb\undefined
\newcommand{\mb}[1]{{\mathbb #1}}
\fi

\begin{document}

\title[D-Branes on $\mathbb {C}^3_6$ Part I: prepotential and $GW$--invariants]
{D-Branes on $\mathbb {C}^3_6$ Part I: prepotential and $GW$--invariants}

\arxurl{0806.2372v1}

\author{Sergio Luigi Cacciatori and Marco Compagnoni}
\address{Dipartimento di Fisica, Universit\`a degli Studi dell'Insubria,\\
Via Valleggio 11, 22100 Como, Italy\\
Dipartimento di Matematica, Politecnico di Milano,\\
Via Bonardi 9, 20133 Milano, Italy}
\addressemail{sergio.cacciatori@uninsubria.it\\ marco.compagnoni@polimi.it}

\begin{abstract}
This is the first of a set of papers having the aim to provide a
detailed description of brane configurations on a family of noncompact
threedimensional Calabi-Yau manifolds. The starting point is the
singular manifold defined by a given quotient $\mathbb {C}^3/\mathbb
{Z}_6$, which we called simply $\mathbb {C}^3_6$ and which admits five
distinct crepant resolutions. Here we apply local mirror symmetry to
partially determine the prepotential encoding the $GW$--invariants of
the resolved varieties. It results that such prepotential provides all
numbers but the ones corresponding to curves having null intersection
with the compact divisor.  This is realized by means of a conjecture,
due to S. Hosono, so that our results provide a check confirming at
least in part the conjecture.
\end{abstract}

\maketitle

\cutpage

\setcounter{page}{2}

\noindent

\section{Introduction}
We use (local) mirror symmetry to compute the Gromov-Witten
invariants\footnote{more precisely the lowest genus Gopakumar--Vafa
invariants} for a family of noncompact Calabi-Yau manifolds obtained
resolving an orbifold $X=\mathbb{C}^3/\mathbb{Z}_{6}$. The main
interest for this example is that it admit five distinct crepant
resolutions all birational to $X$, differing by flops. Going from a
resolution to another passing through the singular orbifold realizes a
geometrical transition.  Geometrically, the transition is obtained by
moving the K\"ahler moduli $t$ through an orbifold point, where the
manifold becomes singular with a curve which shrinks down and
reemerges as a flopped curve.  As it is well known, in string theory
such transition can correspond to smooth physical processes.  This can
be understood for example by means of a (physically equivalent) dual
description using mirror symmetry. Because the orbifold was obtained
quotienting by an abelian group, the resulting smooth manifold are
indeed toric varieties, so that the powerful toric methods can be
employed to work out all details. Mirror symmetry for noncompact CY
varieties was developed quite recently in \cite{vafaecompagnia}. For
toric varieties the mirror manifold result to be defined as the zero
locus
$$ 
Y_x =\{ (\vec u, \vec v)\in \mathbb
{C}^2 \times \mathbb {C}_*^2 | F_x (\vec u, \vec v)=0 \}, 
$$
where $x$ determines a point in the complex structures moduli space of
the mirror, corresponding to the point $t$ specifying the K\"ahler
moduli of the starting manifold $X_t$, and $F_x (\vec u, \vec
v)=u_1^2+u_2^2+f_x(\vec v)$ is a certain polynomial fully determined
by the toric data describing the starting orbifold. Thus varying the
moduli $t$ corresponds to varying the moduli $x$ of the mirror
manifold. However, whereas $X_t$ undergoes a flop transition, $Y_t$
simply changes smoothly its complex moduli.\\ {F}rom the mathematical
point of view, the noncompactness of the variety and the particular
structure of its cohomology ring introduce some ambiguities in
defining the GW invariants and in their interpretation, thus requiring
a deeper understanding of the geometrical structures living on a
noncompact manifold.  {F}rom this point of view, a consistent step
forward was made by T.-M. Chiang, A. Klemm, S.-T. Yau and E. Zaslow
\cite{Chiang}, who gave an interpretation to the GW invariants from an
enumerative point of view. \\ {F}rom the physical point of view,
mirror symmetry looks like a generalization of T-duality equivalence
between different perturbative limits of the, supposed to exist,
unique M-theory. However, some non perturbative enhancements are
provided by adding D-brane configurations. At a semiclassical level,
D-branes are described by closed cycles (with bundles) which the
branes are supposed to wrap on. One may wonder if any possible cycle
is a good candidate as a wrapping locus. Indeed, this is actually a
still open question, even if many overcomes have been made in the last
decade. A consistence check must be stability, at first place
perturbative stability. A first step in favour of perturbative
stability is supersymmetry. This gives some strong constraint which
can depend on the kind of strings to work with.  For IIB strings on a
CY, supersymmetric configurations are represented by holomorphic (then
evendimensional) cycles, whereas for type IIA branes one finds
Lagrangian submanifolds (with respect to the K\"ahler form) as brane
representatives, that are halfdimensional subvarieties.  In the case
in our interest the lasts are three dimensional surfaces. Thus, mirror
symmetry must be extended in order to take account of nonperturbative
brane configurations. An astounding advance in this direction has been
proposed by M. Kontsevich \cite{kontsevich} who introduced the concept
of homological mirror symmetry. In this case type B branes are
described in terms of bounded derived categories of coherent sheaves
whereas A branes are substituted by derived Fukaya categories
\cite{Fukaya}. Such a description reconciles some apparent asymmetry
between A an B branes. Indeed, whereas A branes result to be
halfdimensional, the dimensions of B branes are heterogeneous so that
to any A cycle it can correspond a B cycle of different dimension; on
the other hand, lagrangian cycles can be linearly combined to compose
new lagrangian cycles (monodromies), which, by mirror symmetry, must
match with combinations of holomorphic cycles having different
dimensions.  In the B side, to monodromies correspond autoequivalences
of derived categories.  In some sense homological mirror symmetry
introduce some democracy since all branes are described in terms of
higher dimensional branes in an homogeneous way.  {F}rom a more
topological point of view, brane charges (central charges or masses)
are thus described in terms of K-theory groups (even if there are many
other indications for this beyond and independently from mirror
symmetry, see \cite{moore,witten}).  {F}rom the physical point of
view, some new insight in this direction for the case of noncompact CY
manifolds was done by X. de la Ossa, B. Florea and H. Skarke
\cite{ossa} who were able to select a distinguished K-theory basis for
B-branes configurations adapted to support monodromy correspondence,
generalizing (at least at a conjectural level) the corresponding
results quite well established in the compact case. \\
On this side, further progress is due to S. Hosono
\cite{Hosono:2004jp} who found an elegant way to describe local mirror
symmetry in terms of cohomology valued hypergeometric series. Mirror
symmetry identifies the K\"ahler moduli of a CY variety with the
periods of its mirror, which as functions of the complex structure
moduli must satisfy a set of Picard-Fuchs equations, the
Gel'fand-Kapranov-Zelevinski system \cite{GKZ}.  It results that a
particular cohomology valued hypergeometric series $w$ arises
naturally providing a basis of solutions for the GKZ system
\cite{HKTY1}, \cite{HKTY2}, \cite{HLY1}, \cite{HLY2}.
Hosono was able to recognize such series as a formula identifying the
BPS states of the associated physical theory, and proposed an
intriguing conjecture, which we dub ``{\it the Hosono conjecture}'',
see conjecture 6.3 in \cite{Hosono:2004jp}, which, beyond identifying
the central charge of a brane configuration $F\in K^c(X_t)$ in terms
of $w$, interprets the monodromy of the periods via a naturally
associated symplectic form on $K^c(X)$. Hosono checked very carefully
his conjecture for the toric quotients $\mathbb{C}^2/G$, and for the
examples $\mathbb{C}^3/\mathbb{Z}_3$, $\mathbb{C}^3/\mathbb{Z}_5$ in
three complex dimensions. Among others, a consequence of the Hosono
conjecture is to provide a closed formulation of a prepotential for
noncompact quotients also. Indeed, at cohomological level, mirror
symmetry provides a map
$$ {\rm mir}: K^c
(X_t)\stackrel{\sim}{\longrightarrow} H_3 (Y_x,\mathbb{Z}), 
$$
transferring the symplectic form on $K^c(X)$ to a symplectic structure
on $H_3 (Y,\mathbb{Z})$. This is the noncompact analog of the
symplectic structure that, combined with Griffiths transversality,
ensures the existence of a prepotential in the compact cases. However,
due to non compactness, the symplectic structure is generically
degenerate. On the $X$ side, it defines a correspondence between
$H^2(X,\mathbb{Q})$ and $H^4(X,\mathbb{Q})$ which permits a complete
determination of the prepotential (and correspondingly of all GW
invariants) only when it arises as a vector space isomorphism. \\ At
homological level, mirror symmetry is conjectured to define a map $$
{\rm Mir}: D^{\flat} Coh (X)\longrightarrow D Fuk^o(Y,\omega) $$ where
the symplectic form $\omega$ is the K\"ahler form corresponding to the
fixed complex structure in $X$. In this way, monodromies of Lagrangian
on $Y$ correspond to autoequivalences of derived categories on $X$
described by opportune Mukay transforms which are expected to realize
a (quiver) representation of the quotient group by the Mckay
correspondence. This has been analyzed for example by Karp
\cite{Karp,Canonaco}. The Hosono conjecture indeed works to this
higher level too, and gives some hints to get information on the
mirror map Mir.

In this paper we will work at the lower level, that is at
K-theoretical level, postponing the study of the higher (categorical)
level mirror map to a future paper. We apply the Hosono conjecture to
compute the GW invariants for a family of noncompact toric CY
varieties obtained as crepant resolutions of an orbifold quotient
$\mathbb{C}^3/\mathbb{Z}_6$. We chosen this model because it has quite
general properties which make it very interesting to test the
conjecture. To begin with, the second and fourth Betti numbers are
$b_2=4$ and $b_4=1$, so that the symplectic structure result to be
highly degenerate. Thus it defines a quite poor correspondence between
$H^2(X,\mathbb{Q})$ and $H^4(X,\mathbb{Q})$. Nevertheless, we will see
that the Hosono procedure permits to define a partial prepotential
containing a lot of information about local geometry. Indeed, from it
we are able to read out almost all GW invariants, leaving out only a
three dimensional subcone of the fourdimensional Mori
cone.\footnote{We are grateful to Professor S. Hosono to give us
explanations on this point.}  Indeed, it was proposed by B. Forbes and
M. Jinzenji \cite{FoJi1}, \cite{FoJi2}, a possible way to extend the
GKZ system obtaining a complete determination of all GW invariants. To
such extension we will devote a future paper. Here, we will only
discuss the possible origin for the ambiguity in defining the lacking
GW invariants.\\ A second interesting peculiarity of our model, yet
anticipated at the beginning, is that it admits five distinct crepant
resolutions, which differ by flops.  Thus one expects monodromy to
relate different resolution by means of different Fourier Mukay
transforms. This is indeed one of the main targets of this starting
studies, but as announced we will not tackle it here. We will limit
ourselves to compute the prepotentials and the computable GW
invariants for all resolutions, comparing with themselves.\\ Thus in
some sense this first paper can be thought as a preparatory one. In
this spirit we will try to be as much explicit as possible. In section
\ref{sec:introductive} we include a short overview of the main steps
which lead to the introduction of local mirror symmetry to arrive to
the Hosono conjecture.\\ In section \ref{sec:GHilb-resolution} we
present a detailed analysis of the first resolution, that is the
G-Hilbert resolution.  We will use the Hosono's conjecture to
construct the cohomological hypergeometric series generating the
periods of the mirror manifold. Due to noncompactness, the structure
of the cohomology ring does not consent a full definition of the
GW--invariants. However, as we will see, the procedure proposed by
Hosono permits to equally define a prepotential which generates all
GW--invariants associated to the curves in the Mori cone, excluding a
codimension one subcone.\\ In section \ref{flops} we repeat the
previous analysis for all the other resolutions, deriving a partial
determination of the GW--invariants for all of them.  \\ The results
will be commented in section \ref{sec:conclusion}.


\section{Local mirror symmetry and the Hosono conjecture}\label{sec:introductive}
Here we will recall some main step leading to the conjectures we are
testing in this and following papers. The literature on the subject is
quite huge, so that we will mainly refer to \cite{librone} and
references therein.

\subsection{Dualities and mirror symmetry}
Let us consider a string theory having a toric Calabi-Yau variety $X$
as target space. Thus, there is a nice interpretation of mirror
symmetry as a T-duality transformation. Indeed, string theory on $X$
can be described in terms of a two dimensional $U(1)^m$ supersymmetric
gauge theory, the so called ``gauged linear sigma model'' (see
\cite{librone}, sections 7.3, 7.4). It contains a certain number $n>m$
of complex scalar fields $Z=\{Z_\alpha\}_{\alpha=1}^n$ having charges
$Q_{\alpha,r}$ $r=1,2,\ldots, m$ with respect to the gauge group
$U(1)^m$, and with potential energy
$$
U(Z)=\frac 12 \sum_{r=1}^m g_r^2 \left( \sum_{\alpha=1}^n Q_{\alpha,r} Z_\alpha \bar Z_\alpha -{\rm r}_r\right)^2.
$$
Here $g_r$ and ${\rm r}_r$ are the gauge couplings and the
Fayet-Iliopoulos terms respectively. Supersymmetric ground states
require the vanishing of the potential energy:
$$
\sum_{\alpha=1}^n Q_{\alpha,r} Z_\alpha \bar Z_\alpha ={\rm r}_r.
$$

For a fixed choice of the $F-I$ parameters, these equations define a
toric variety $X$ associated to a fan, in an $(n-m)$--dimensional
lattice $N$, generated by an opportune set $\Sigma(1)={v_1,\ldots,
v_n}$ of vectors in $N$. From this it is possible to conclude that the
supersymmetric vacua are identified with the points of a toric variety
$X$.  Each vector $v_\alpha$ determines an invariant
divisor\footnote{t.i. invariant under the toric action},
$D_{v_\alpha}$. It is not hard to show (see \cite{librone}, sec. 7.4)
that one can chose a basis $\{ C_r \}_{r=1}^m$ of irreducible curves
of $H_2(X,\ZZ)$ (which indeed result to be $m$--dimensional) such that
the charges are given by the intersection numbers $Q_{\alpha,
r}=D_{v_\alpha}\cdot C_r$. Also note that the F--I parameters rescale
as $|Z|^2$ so that, if chosen to be positive, they indeed parameterize
the points of the K\"ahler cone. This means that the supersymmetric
configurations are completely characterized in geometrical terms.\\ At
this point mirror symmetry can be realized as a T--duality
transformation (\cite{librone}, sec. 20). Indeed, recall that roughly
speaking T--duality on a circle transforms a type A string theory on a
circle of radius $R$ in a type B string theory on a circle of radius
$\alpha'/R$.  If $Z_\alpha$ are taking value on a complex variety
(indeed the toric variety in the vacuum configuration) then we can
T--dualize their phases which define circles in the target
manifold. The result (\cite{librone}, sec. 13) is a Landau-Ginzburg
theory with superpotential
$$
W(Y,t)=\sum_{\alpha=1}^n e^{-Y_\alpha},
$$
for a set of chiral superfields related by the set of constraints
$$
\sum_{\alpha=1}^n Q_{\alpha,r} Y_\alpha =t_r
$$
where $t_r$ are the complexified K\"ahler parameters (${\rm Re}
(t_r)={\rm r}_r$).
In this way, the mirror transformation applied to the twodimensional
sigma model gives rise to a Landau--Ginzburg model with superpotential
$W(Y,t)$. To take contact with the Batyrev's geometric construction of
mirror manifolds for toric varieties, let us proceed as follows (see
\cite{vafaecompagnia}) for the cases when the starting linear sigma
model describes strings on a crepant resolution of some abelian
quotient $\CC_3/G$. Being crepant, it will be described by a set of
vectors $v_1,\ldots, v_n$ in a three-dimensional lattice such that for
some isomorphism $\phi : N\longrightarrow \ZZ^3$ one has
$\phi(v_\alpha)=(n_{\alpha,1},n_{\alpha,2},1)$. The solutions of the
constraints can thus be written in terms of three independent fields
$y_0$, $y_1$, $y_2$ as $Y_\alpha =y_0 +n_{\alpha,1} y_1+n_{\alpha,2}
y_2+c_\alpha$ where $c_\alpha$ are some constant satisfying
$\sum_{\alpha=1}^n Q_{\alpha,r} Y_\alpha =t_r$. These linear
redefinitions do not affect the functional measure, and setting
$w_a=\exp (-y_a)$, $a=0,1,2$ and $a_\alpha=\exp (-c_\alpha)$ we get
for the superpotential
$$
W(w,a)=w_0 \sum_{\alpha=1}^n a_\alpha w_1^{n_{\alpha,1}}
w_2^{n_{\alpha,2}}, \qquad w_a\in \mathbb {C_*} .
$$
As discussed in \cite{vafaecompagnia}, we can note that, for what
concerns the BPS configurations, this LG model is equivalent to
another one, where $w_0\in \mathbb {C}$ and with two extra chiral
fields $U,V\in \mathbb {C}$, whose superpotential is
$$
\tilde W(U,V;w;a)=W(w,a)-w_0 UV.
$$
Integrating the field $w_0$ thus gives a delta function $\delta
(\sum_{\alpha=1}^n a_\alpha w_1^{n_{\alpha,1}} w_2^{n_{\alpha,2}}-UV)$
so that the mirror LG model is equivalent to a geometrical theory on a
Calabi-Yau manifold
$$
Y_a =\{ (\vec u, \vec w)\in \mathbb {C}^2 \times \mathbb {C}_*^2 | F_a
(\vec u, \vec w)=0 \},
$$
where
$$
F_a (\vec u, \vec w)=u_1^2+u_2^2 +f_a (\vec u, \vec w)=u_1^2+u_2^2
+\sum_{\alpha=1}^n a_\alpha w_1^{n_{\alpha,1}} w_2^{n_{\alpha,2}}.
$$
The K\"ahler parameters $t$ now parameterize the complex moduli of
$Y$. This is indeed local mirror symmetry as discovered for the first
time at physical level in \cite{vafa1}, \cite{vafa2}.


\subsection{Branes and homological mirror symmetry}
The intuitive picture described above does not takes into account the
presence of brane configurations. Because we are looking for
supersymmetric vacua, we need to know what kind of brane
configurations are admitted on a Calabi-Yau manifold $X$. In other
words, one must search for boundary condition compatible with
supersymmetry. This is described for example in \cite{vafaecompagnia},
sec. 3.  The answer depends on the type of string theory one
considers. For type A strings, supersymmetric branes are represented
(at classical level) by halfdimensional subvarieties $S$,
$\iota:S\hookrightarrow X$, where the K\"ahler form $\omega$ of the
C-Y manifold vanishes, $\iota^* \omega=0$, and supporting flat vector
bundles.  Thus A--branes are Lagrangian submanifolds with respect to
the symplectic structure $\omega$. For type B strings one finds that
supersymmetric branes must wrap holomorphic cycles of $X$ supporting
holomorphic vector bundles.  In our models it means that type B brane
configurations will be described classically by compact divisors,
curves of the Mori cone and points.\\ Thus mirror symmetry should map
BPS states of a model into the BPS states of the mirror model,
converting A--branes to B--branes and viceversa. However, there is an
odd asymmetry between A and B configurations: indeed all A--branes
have the same dimensions, whereas this does not happen for
B--branes. Now, the point is that in the LG model description branes
configurations can change when moduli vary. In this picture, BPS
states will correspond to critical points of the superpotential.
Essentially they determine the points of $Y_t$ around which the
supersymmetric three--cycles are defined. Varying $t$, the critical
points move on the $W$--plane; when some of these points moves around
a branch point, a monodromy transformation can give rise to a new
brane configuration (\cite{vafaecompagnia}). The boundary states
corresponding to the branes are described by the periods of the
holomorphic three-form $\Omega$ of $Y_t$ (in the geometric picture).
The monodromy thus acts on a basis of cycles recasting them in some
linear recombination or equivalently on the periods in the same linear
recombination. On the mirror $X$ it should correspond to a
recombination of the holomorphic cycles, hard to understand in the
na\"ive geometrical picture where they have different dimensions.\\ To
solve this point a first aid comes from a K-theoretical description,
where lower dimensional branes can be described in terms of the top
dimensional branes and a tachyon field \cite{witten}. K-theory mainly
captures topological aspects of the problem, carrying important
information on the admissible brane configurations, but it is quite
poor from the geometrical point of view. In \cite{douglas} it was
argued that a deeper geometrical understanding of (stable) brane
configuration in (topological) type B superstring can be understood in
terms of triangulated categories, in particular the derived category
of coherent sheaves on the manifold (see also \cite{aspinwall}, or
\cite{bridgeland} for a more mathematical point of view). This
provided a deep contact between physics and the ``homological mirror
symmetry'' conjectured by Kontsevich \cite{kontsevich} who proposed
that the usual geometrical mirror symmetry should enhance to
homological level as an equivalence between triangulated categories:
the derived category of coherent sheaves on a CY manifold $X$ with a
fixed complex structure on one side\footnote{for clarity we confine
ourselves to the case of Calabi-Yau varieties} and the derived
${\mathcal A}^\infty$ Fukaya's category over the mirror manifold $Y$
on the other side, essentially generated by the Lagrangian
submanifolds of $\{Y, \omega\}$, where the symplectic structure
$\omega$ is given by the fixed K\"ahler form on $Y$, dual to the
complex form on $X$:
$$
{\rm Mir}: D^{\flat} Coh (X)\stackrel{\simeq}{\longrightarrow} D
Fuk^o(Y,\omega).
$$


\subsection{The Hosono conjecture}
As we said, BPS states in the mirror type A string model are described
by periods that are integrals of the holomorphic three form $\Omega$
on $Y$ over the Lagrangian cycles. For the noncompact quotients we are
describing, the holomorphic three form on the mirror $Y$ is
$$
\Omega=\frac 1{4\pi^3} {\rm Res}_{F=0} \left[ \frac {du^1\wedge du^2
\wedge dw^1 \wedge dw^2}{w^1 w^2 F(\vec u; \vec w; a)} \right].
$$
Here we have fixed the K\"ahler form, however $\Omega$ depends
explicitly on the complex moduli of $Y$ (as shown by the explicit
dependence on $a$ of the polynomial $F$) so that the periods
$$
\Pi_{C_i}(a) =\int_{C_i} \Omega,
$$
of any set of Lagrangian cycles $C_i$, will be locally holomorphic
functions of the moduli. Indeed, they are forced to satisfy a set of
hypergeometric differential equations known as the GKZ hypergeometric
system, largely studied in \cite{GKZ}.\\ For compact varieties the
knowledge of a complete set of solutions for the GKZ system correspond
to an exhaustive description of the set of BPS brane configurations on
the $A$ side. Furthermore, the special K\"ahler geometry of the
complex structure moduli space of a C-Y manifold can be described in
terms of periods \cite{strominger}.  If $x$ parameterizes the
structure complex moduli of $Y$ then the K\"ahler potential of the
moduli space can be written as
$$
K(x,\bar x)=-\log \left[ i \sum_{I=0}^{h^{2,1} (Y)} \left( X^I \frac
{\bar {\partial G}}{\partial \bar X^I}-\bar X^I
\frac {{\partial G}}{\partial X^I} \right)\right],
$$
where
$$
X^I (x) =\int_A^I \Omega(x)
$$
are the periods with respect to a canonical symplectic basis
$\{A^I,B_I\}$ of $H_3 (Y,\mathbb {Z})$. Finally $G(x)$ is the
prepotential
$$
G(x)=\frac 12 \sum_{I=0}^{h^{1,2}(Y)} \int_{A^I} \Omega \int_{B_I} \Omega.
$$
Mirror symmetry gives a correspondence between K\"ahler moduli $t_i$
of $X$ and complex moduli of $Y$ so that
$$
t_i=\frac {X^i}{X^0}, \qquad\ i=1,\ldots, h^{1,2}(Y)=h^{1,1}(X).
$$
On the other side, also the K\"ahler moduli space of $X$ is a special
K\"ahler manifold which can thus be described in terms of a
prepotential function $F(t)$. At classical level such geometry is
described by the prepotential
$$
F^{c}(t)=\frac 16 d_{ijk} t^i t^j t^k
$$
where $d_{ijk}=J_i \cdot J_j \cdot J_k$ are the intersection numbers
of the K\"ahler cone generators. Physically they determine the Yukawa
couplings of the chiral fields \cite{candelas}. However these
couplings receive quantum corrections which come from worldsheet
instantons. At lowest order they corresponds to wrapping of the
worldsheet on rational curves in $X$. The energy of such a wrapping is
given by the volume of the wrapped cycle as measured by the K\"ahler
metric. Any given (class of) rational curve of degree $\vec d$ results
to contribute to the prepotential with a term
$$
n_{\vec d}\ Li_3(e^{2\pi i \vec d \cdot t}),
$$
$n_{\vec d}$ being the number of classes of curves with the given
degree, so that it can be shown that the quantum corrected
prepotential takes the form \cite{Chiang}
$$
F(t)=\frac 16 d_{ijk} t^i t^j t^k-\frac 1{24} c_2(X) \cdot J_i t^i -i
\frac {\zeta (3)}{16\pi^3} c_3(X)
+\sum_{d\in \mathbb {Z}_>^{h^{1,1}}} n_{\vec d}\ Li_3(e^{2\pi i \vec d
\cdot t}).
$$
More precisely $n_{\vec d}$ are the Gromov-Witten invariant (in the
Gopakumar-Vafa interpretation \cite{gopak-vafa-I},
\cite{gopak-vafa-II}).  See \cite{Chiang} for a mathematical
enumerative interpretation. By means of the identification (making use
of the Griffith transversality, \cite{Chiang}, \cite{strominger})
$$
\left\{\int_{A^I} \Omega; \int_{B_I} \Omega \right\}_{I=0}^{h^{2,1}(Y)}=
\left\{1,\ t^i;\ \partial_{t^i} F,\
2F-\!\!\!\!\!\sum_{j=1}^{h^{1,1}(X)}\!\!\! t^j \partial_{t^i} F
\right\}_{i=1}^{h^{1,1}(X)}
$$
mirror symmetry thus gives a simple way to compute the
$GW$--invariants of $X$.

In a series of papers (see for example \cite{HKTY1}, \cite{HKTY2},
\cite{HLY1}, \cite{HLY2}) it was provided an efficient strategy to
characterize a complete set of the GKZ system for a C-Y hypersurface,
which is summarized in \cite{Hosono-birk}. In particular there was
introduced a cohomological valued power series whose expansion in the
Chow ring
$$
A^*(X)\otimes \mathbb {C} [[x]] [\log x]
$$
gives a basis for the period integrals of the mirror manifold $Y$ in
the large complex structure limit (LCSL), (see \cite{Hosono-birk},
Claim 5.11).  Thus the cohomological series encodes many geometrical
information on both the manifolds $X$ and $Y$ so summarizing several
fundamental aspects of mirror symmetry.\\ In \cite{Hosono:2000eb},
\cite{Hosono:2004jp} Hosono extended this picture to local mirror
symmetry for noncompact C-Y manifolds, in particular for resolutions
of abelian quotients $\mathbb {C}^k/G$, with $k=2,3$. For convenience
we will state the conjecture in section \ref{conjecture}. In
\cite{Hosono:2004jp}, Hosono verified his conjecture carefully for the
case $k=2$ and reported the analysis for the cases $\mathbb
{C}^3/\mathbb {Z}_3$ and $\mathbb {C}^3/\mathbb {Z}_5$, where it was
shown the existence of a prepotential for the noncompact cases also.

Here we use Hosono conjecture to analyze the geometry of a quotient
$X=\mathbb {C}^3/\mathbb {Z}_6$, which we call for simplicity $\mathbb
{C}^3_6$.  As stated in the introduction, this singular orbifold
admits five distinct crepant resolutions. All these resolutions are
related by flop transformations. To noncompactness of the manifold $X$
it corresponds the ambiguity in defining the $GW$--invariants. In our
model this reflects in the fact that the symplectic structure on the
half dimensional homology of the mirror $Y$ is degenerate. On the
mirror, such structure should determine a pairing between two
dimentional and four dimensional cohomology, permitting the
reconstruction of the prepotential, but which now becomes
degenerate. We determine the LCSL cohomological series for all the
resolutions.  {F}rom each of them, using Hosono's prescriptions we
will able to partially determine a prepotential which codifies all the
$GW$--invariants of the (four dimensional) Mori cone excluding a three
dimensional subcone.


\section{The tridimensional orbifold $\CC^3_6$ and the $G$-Hilb
resolution}\label{sec:GHilb-resolution}
\subsection{Definition of $\CC^3_6$}
We briefly review the homogeneous coordinates construction of toric
varieties \cite{Cox}. The data of a $d$-dimensional toric variety
$X(\Delta)$ can always be specified in terms of a fan $\Delta$ in a
lattice $N$ isomorphic to $\ZZ^d$. Let $\rho_1,\ldots,\rho_r$ be the
$1$-dimensional cones of $\Delta$ and let $v_i\in \ZZ^n$ denote the
primitive element of $\rho_i$, i.e. the generator of $\rho_i\cap
\ZZ^n$. Then introduce variables $x_i$ for $i=1,\ldots,r$ in the
affine complex space $\CC^r$.
The homogeneous coordinates construction represents $X(\Delta)$ as the
quotient
\begin{eqnarray*}
X(\Delta)=(\CC^r\backslash Z)/G
\end{eqnarray*}
for a certain variety $Z$ and some abelian group $G\subset (C^*)^r$.

$Z$ is determined as follows. We say that a set of edge generators
$I=\{v_{i_1},\ldots,v_{i_s}\}$ is primitive if they don't lie in any
cone of $\Delta$ but every proper subset does. Then
\begin{eqnarray*}
Z=\bigcup_{I\ \text{primitive}}\{x_{i_1}=0,\ldots,x_{i_s}=0\}\ .
\end{eqnarray*}
If $\{e_1,\ldots,e_d\}$ is the standard basis of the dual lattice $M$
and $<,>:M\times N \rightarrow \mathbb {Z}$ is the natural pairing,
the group $G$ is 
defined as the kernel of the following homomorphism
\begin{eqnarray*}
& \Phi :(\CC^*)^r \rightarrow (\CC^*)^d,\quad (\lambda_1,\ldots,
\lambda_r)\mapsto
\left(\prod_{i=1}^r\lambda_i^{\langle e_1,v_i\rangle},
\ldots,\prod_{i=1}^r\lambda_i^{\langle e_n,v_i\rangle}\right)
\end{eqnarray*}
and its actions on $\CC^r\backslash Z$ is by multiplication
\begin{eqnarray*}
(\lambda_1,\ldots, \lambda_r)\cdot(x_1,\ldots,x_r):=
(\lambda_1 x_1,\ldots, \lambda_r x_r)\ .
\end{eqnarray*}

In this paper we study the threedimensional orbifold $\CC^3_6$ defined
as the toric variety associated to the fan generated by the vectors
\begin{eqnarray}
v_1=\left(\begin{array}{c} -1\\-1\\1 \end{array} \right) \ , \qquad
v_2=\left(\begin{array}{c} 2\\-1\\1 \end{array} \right) \ , \qquad
v_3=\left(\begin{array}{c} -1\\1\\1 \end{array} \right) \ ,
\end{eqnarray}
in $N\simeq\ZZ^3$.
In this case $Z=$\O\ and the associated homomorphism is
\begin{eqnarray}
&\Phi :(\CC^*)^3 \rightarrow (\CC^*)^3,\quad
(\lambda_1, \lambda_2, \lambda_3)\mapsto
(\lambda_1^{-1}\lambda_2^2 \lambda_3^{-1}, \lambda_1^{-1}
\lambda_2^{-1}\lambda_3, \lambda_1 \lambda_2 \lambda_3)\ ,
\end{eqnarray}
which has kernel
\begin{eqnarray}
G:=\ker \Phi =<(\epsilon, \epsilon^2, \epsilon^3)>\subset (\CC^*)^3\ ,
\ {\rm with} \ \epsilon=e^{\frac{2\pi i}{6}}\ .
\end{eqnarray}
Thus $G\simeq \ZZ_6$ and $\CC^3_6=\CC^3 /\ZZ_6$ where the action on
the coordinates is
\begin{eqnarray}
\epsilon\cdot(x_1,x_2,x_3)=
(\epsilon\, x_1,\epsilon^2\, x_2, \epsilon^3\, x_3)\ .
\end{eqnarray}
$\CC^3_6$ is a non compact Calabi-Yau ($K_{\CC^3_6}$ is trivial)
threefold with an isolated quotient singularity at the origin, because
all vectors $v_i$ lie in the plane $z=1$ (if $(x,y,z)$ are the
coordinates on the lattice).\footnote{
We refer to section \ref{Chow ring} for an explanation about this CY
condition.}
In this way, all relevant information is included in the two
dimensional intersection of the fan $\Delta$ with the plane $z=1$. In
the figure \ref{fig:1} we have drawn this section for the fan of
$\CC_6^3$.
\begin{figure}
\centering
\begin{picture}(300, 190)(-110,-90)
\dashline{3}(0,-80)(0,95) 
\put(-3,95){\makebox(0, 0)[t]{\scriptsize{\boldmath{$y$}}}}
\dashline{3}(-100,0)(180,0) 
\put(180,-4){\makebox(0,0)[t]{\scriptsize{\boldmath{$x$}}}}
\put(-80,80){\circle*{4}}
\put(160,-80){\circle*{4}}
\put(-80,-80){\circle*{4}}
\thicklines
\drawline(-80,80)(-80,-80)
\drawline(-80,-80)(160,-80)
\drawline(-80,80)(160,-80)
\put(-83,-86){\makebox(0, 0)[r]{\scriptsize{\boldmath{$v_1$}}}}
\put(-83,86){\makebox(0, 0)[r]{\scriptsize{\boldmath{$v_3$}}}}
\put(164,-86){\makebox(0, 0)[r]{\scriptsize{\boldmath{$v_2$}}}}
\end{picture}
\caption{$\CC_6^3$ fan}
\label{fig:1}
\end{figure}

\subsection{Crepant resolutions of $\CC^3_6$}
\label{toric-resolution}
A crepant resolution of a variety $X$ is a smooth variety $Y$
together with a a proper birational morphism $\tau: Y \rightarrow X$
such that $K_Y=\tau^* K_X$. If $X$ is a Calabi-Yau variety this means
that $K_Y$ has to be trivial. Any crepant resolution of a toric
Calabi-Yau orbifold $X(\Delta)=\CC^3/G$ can be obtained in two simple
steps (see \cite{FultonT,Oda}). First, add to $\Delta$ all possible
edges $\rho_i$ that are generated by the integer vectors $v_i\in N$
intersecting the fan and
lying on the plane determined by $v_1,v_2,v_3$. Next, let one completely
triangulate $\Delta$, to obtain the regular fan $\Delta'$ of the toric
resolution $X(\Delta')$. If there exist several complete
triangulations this means that the orbifold admits multiple crepant
resolutions, all related by flops of curves.

Therefore, to obtain the resolutions of the $\CC^3_6$ singular
variety we add to $\Delta$ the four vectors
\begin{eqnarray}
& v_4=\left(\begin{array}{c} 0\\-1\\1 \end{array} \right) \ , \qquad
v_5=\left(\begin{array}{c} 1\\-1\\1 \end{array} \right) \ , \qquad
v_6=\left(\begin{array}{c} -1\\0\\1 \end{array} \right) \ , \qquad
v_7=\left(\begin{array}{c} 0\\0\\1 \end{array} \right) \ .
\end{eqnarray}
It is easy to show that we have five admissible complete triangulations.

\subsubsection{Toric $G$-Hilbert resolution}
We start considering the $G$-Hilbert resolution, which we call
$\GC$. Its general toric construction is given in \cite{ReidCraw99}
and we refer to it for a detailed explanation. We can think to
$G$-Hilb fan as the ``more symmetric'' triangulation. We try to
illustrate this concept in our case.
\begin{figure}
\centering
\begin{picture}(300, 190)(-110,-90)
\dashline{3}(0,-80)(0,95) \put(-3,95){\makebox(0, 0)[t]{\scriptsize{\boldmath{$y$}}}}
\dashline{3}(-100,0)(180,0) \put(180,-4){\makebox(0,0)[t]{\scriptsize{\boldmath{$x$}}}}
\put(0,0){\circle*{4}}
\put(0,-80){\circle*{4}}
\put(-80,0){\circle*{4}}
\put(-80,80){\circle*{4}}
\put(80,-80){\circle*{4}}
\put(160,-80){\circle*{4}}
\put(-80,-80){\circle*{4}}
\thicklines
\drawline(0,0)(0,-80)
\drawline(-80,80)(80,-80)
\drawline(-80,80)(-80,-80)
\drawline(-80,-80)(160,-80)
\drawline(-80,0)(0,0)
\drawline(0,0)(160,-80)
\drawline(-80,80)(160,-80)
\drawline(-80,0)(0,-80)
\put(-83,6){\makebox(0, 0)[r]{\scriptsize{\boldmath{$v_6$}}}}
\put(-83,-86){\makebox(0, 0)[r]{\scriptsize{\boldmath{$v_1$}}}}
\put(-83,86){\makebox(0, 0)[r]{\scriptsize{\boldmath{$v_3$}}}}
\put(4,6){\makebox(0, 0)[r]{\scriptsize{\boldmath{$v_7$}}}}
\put(4,-86){\makebox(0, 0)[r]{\scriptsize{\boldmath{$v_4$}}}}
\put(84,-86){\makebox(0, 0)[r]{\scriptsize{\boldmath{$v_5$}}}}
\put(164,-86){\makebox(0, 0)[r]{\scriptsize{\boldmath{$v_2$}}}}
\end{picture}
\caption{Fan of the G-Hilbert resolution of $\CC^3_6$}
\label{fig:2}
\end{figure}
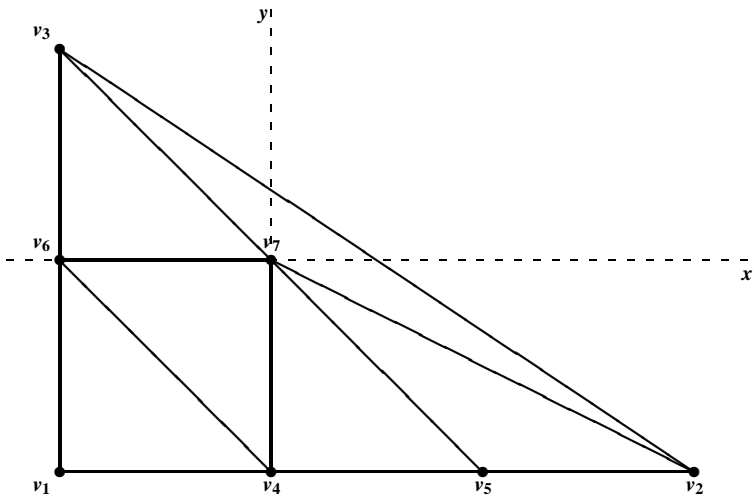
First we add to $\Delta$ the two dimensional cones generated by
$(v_2,v_7)$ and $(v_3,v_7)$, that are necessary to obtain any complete
triangulation. Then we extend the line $(v_3,v_7)$ to $v_5$ so
obtaining a subdivision of the fan into regular triangles, three of
them with edges of length one and the bigger one with edges of length
two. Finally we complete the triangulation subdividing this last
triangle with a regular tessellation, obtained by drawing all possible
internal lines parallel to its edges.

\subsubsection{$G$-Hilbert resolution as the moduli space of
$G$-clusters of $\mathbb {C}^3$}
\label{G-Hilb}
Given an algebraic variety $M$ and a finite group $G$ with an action
on $M$, the $G$-Hilb($M$) is defined as the
moduli space of $G$-clusters $Z\subset M$. A $G$-cluster is a
$G$-invariant zero dimensional subscheme $Z$, with defining ideal
$\mathcal{I}_{Z}\subset \mathcal{O}_M$ and structure sheaf
$\mathcal{O}_Z=\mathcal{O}_M/\mathcal{I}_Z$ isomorphic to the regular
representation of $G$, i.e. $H^0(Z,\mathcal{O}_Z)\simeq R(G)$ with
dim~$H^0(Z,\mathcal{O}_Z)=|G|$. The simplest example of $G$-cluster
is a general orbit of $G$ consisting of $N$ distinct point.

We will study the simple example of $\ZZ_2$-Hilb($\CC^2$).
Let us consider $\CC^2=\text{Spec}\;\CC[X,Y]$ and the action of $\ZZ_2$,
with generator $\epsilon=-1$, defined on the coordinates as
\begin{eqnarray}
\epsilon\cdot(X,Y)=
(\epsilon\, X,\epsilon\, Y)\ .
\end{eqnarray}
The orbits of $\ZZ_2$ are the sets of couple of points
\begin{eqnarray}
\{(p_1,p_2)\in\CC^2\times\CC^2\,|\,X(p_1)=-X(p_2),Y(p_1)=-Y(p_2)\}\ .
\end{eqnarray}
If $p_1$ has coordinates $(a,b)$, on the open set $X\neq 0$ the
$\ZZ_2$-cluster $Z$ with support over $(p_1,p_2)$ is defined by the
equations
\begin{eqnarray}
\label{g-clusters}
&X^2=a,\quad Y=\frac{b}{a}X\qquad\Longrightarrow\qquad
\mathcal{O}_Z=\frac{\CC[X,Y]}{(X^2-a,Y-\frac{b}{a}X)}
\simeq\CC\oplus\CC\cdot X \ ,
\end{eqnarray}
and on the open set $Y\neq 0$ by
\begin{eqnarray}
\label{g-clusters2}
&Y^2=b,\quad X=\frac{a}{b}Y\qquad\Longrightarrow\qquad
\mathcal{O}_Z=\frac{\CC[X,Y]}{(Y^2-b,X-\frac{a}{b}Y)}
\simeq\CC\oplus\CC\cdot Y \ .
\end{eqnarray}
It is easy to verify all the properties of the $\ZZ_2$-clusters. Thus
we have a bijective relation between generic orbits and
$\ZZ_2$-cluster having support on them. On the set $\{X=0,Y=0\}$ we have
$\ZZ_2$-clusters $Z$ of type
\begin{eqnarray}
\label{g-clusters3}
&X^2=0,\quad X=\frac{\beta}{\alpha}Y\qquad\Longrightarrow\qquad
\mathcal{O}_Z=\frac{\CC[X,Y]}{(X^2,X-\frac{\beta}{\alpha}Y)}
\simeq\CC\oplus\CC\cdot X \ ,
\end{eqnarray}
for any $(\alpha,\beta)$ with $\alpha\neq 0$, or, in alternative, of type
\begin{eqnarray}
\label{g-clusters4}
&Y^2=0,\quad Y=\frac{\alpha}{\beta}X\qquad\Longrightarrow\qquad
\mathcal{O}_Z=\frac{\CC[X,Y]}{(Y^2,Y-\frac{\alpha}{\beta}X)}
\simeq\CC\oplus\CC\cdot Y \ ,
\end{eqnarray}
for any $(\alpha,\beta)$ with $\beta\neq 0$.
It is evident that the $\ZZ_2$-Hilb($\CC^2$) has the structure of the
blow-up of $\CC^2/\ZZ_2$ at the origin and, with the map
\begin{eqnarray}
\tau :\ZZ_2\text{-Hilb}(\CC^2) \longrightarrow \CC^2/G \qquad
Z\,(p_1,p_2)\longmapsto(p_1,p_2)\ ,
\end{eqnarray}
it becomes the (crepant) resolution of the orbifold.
Let us prove this fact explicitly using toric geometry.

We will follow the construction of toric orbifold given in \cite{FultonT}. Let
$L=\ZZ^2+\frac{1}{2}(1,1)$ be the lattice over $\ZZ^2$; in $L$ the
fan of $\CC^2/\ZZ_2$ is the junior simplex $\Delta_{\text{junior}}$
generated by the standard base $(e_1,e_2)$ of $\ZZ_2$.
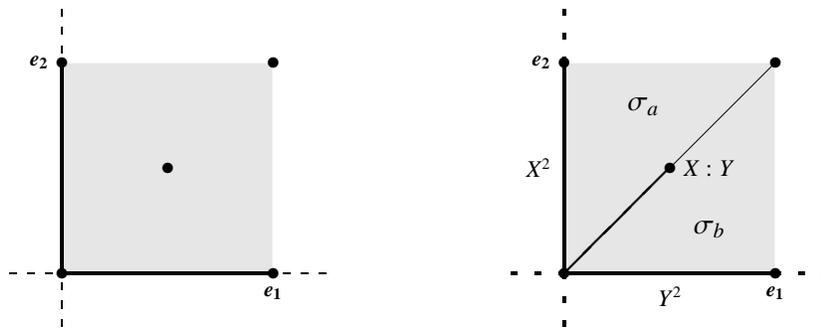
\begin{figure}[hbtp]
\centering
\setlength{\unitlength}{1pt}
\pssetlength{\unitlength}{1pt}
\begin{picture}(370,135)(-40,-30)
\pspolygon[linecolor=white,fillstyle=solid,fillcolor=vlgray](0,0)(0pt,80pt)(80pt,80pt)(80pt,0)
\dashline{3}(0,-20)(0,100)
\dashline{3}(-20,0)(100,0)
\put(0,0){\circle*{4}}
\put(80,0){\circle*{4}}
\put(40,40){\circle*{4}}
\put(80,80){\circle*{4}}
\put(0,80){\circle*{4}}
\thicklines
\drawline(0,0)(0,80)
\put(-5,80){\makebox(0,0)[r]{\scriptsize{\boldmath{$e_2$}}}}
\drawline(0,0)(80,0)
\put(80,-5){\makebox(0,0)[t]{\scriptsize{\boldmath{$e_1$}}}}
\pspolygon[linecolor=white,fillstyle=solid,fillcolor=vlgray]
(190pt,0)(190pt,80pt)(270pt,80pt)(270pt,0)
\dashline{3}(190,-20)(190,100)
\dashline{3}(170,0)(290,0)
\put(190,0){\circle*{4}}
\put(270,0){\circle*{4}}
\put(230,40){\circle*{4}}
\put(270,80){\circle*{4}}
\put(190,80){\circle*{4}}
\psline[linewidth=.1mm](230pt,40pt)(270pt,80pt)
\thicklines
\drawline(190,0)(190,80)
\put(185,80){\makebox(0,0)[r]{\scriptsize{\boldmath{$e_2$}}}}
\drawline(190,0)(270,0)
\put(270,-5){\makebox(0,0)[t]{\scriptsize{\boldmath{$e_1$}}}}
\drawline(190,0)(230,40)
\put(230,-5){\makebox(0,0)[t]{\footnotesize$Y^2$}}
\put(185,40){\makebox(0,0)[r]{\footnotesize$X^2$}}
\put(235,40){\makebox(0,0)[l]{\footnotesize$X:Y$}}
\put(220,60){\makebox(0,0)[b]{$\sigma_a$}}
\put(245,20){\makebox(0,0)[t]{$\sigma_b$}}
\end{picture}
\caption{{Fan for $C^2/\ZZ_2$ and $\ZZ_2$-Hilb($C^2$) in
$L=\ZZ^2+\frac{1}{2}(1,1)$}}
\label{fig:3}
\end{figure}\\
Using the ``old construction'' of toric variety \cite{FultonT}, we have
\begin{eqnarray}
X_{\Delta_{\text{junior}}}=\text{Spec}\; \CC[X^2,XY,Y^2]=
\text{Spec}\; \frac{\CC[U,V,W]}{(UW-V^2)}\ .
\end{eqnarray}
The toric resolution of $X_{\Delta_{\text{junior}}}$ is obtained
adding to $\Delta_{\text{junior}}$ the edge generated by
$\frac{1}{2}(e_1+e_2)$. In the right side of the figure \ref{fig:3} we
have drawn the toric fan of the resolution marked with the coordinates
related to the toric curves, expressed as $\ZZ_2$-invariant ratios of
monomials in the orbifold coordinates. Geometrically this is the blow
up of $X_{\Delta_{\text{junior}}}=\CC^2/\ZZ_2$ in the origin. The two
affine open sets are
\begin{eqnarray}
U_{\sigma_a}=\text{Spec}\; \CC[X^2,Y/X]
\ , \qquad\qquad
U_{\sigma_b}=\text{Spec}\; \CC[Y^2,X/Y]
\ .
\end{eqnarray}
Thus $U_{\sigma_a}$, for example, parameterizes equations of the form
\begin{eqnarray}
X^2=\xi_a\qquad\ Y=\eta_a X\ ,
\end{eqnarray}
which define the $\ZZ_2$-clusters (\ref{g-clusters}).
Similar $U_{\sigma_b}$ parameterizes clusters (\ref{g-clusters2}) and
their intersection $U_{\sigma_a\cap\sigma_b}$ the clusters
(\ref{g-clusters3},\ref{g-clusters4}).
Therefore the crepant toric resolution of $\CC^2/\ZZ_2$ is exactly
$\ZZ_2$-Hilb($\CC^2$).

\begin{figure}[hbtp]
\centering
\setlength{\unitlength}{1pt}
\pssetlength{\unitlength}{1pt}
\begin{picture}(230,200)(-10,-10)
\pspolygon
(0pt,0)(210pt,0pt)(105pt,180pt)
\put(0,0){\circle*{4}}
\put(210,0){\circle*{4}}
\put(105,180){\circle*{4}}
\put(87.5,30){\circle*{4}}
\put(175,60){\circle*{4}}
\put(52.5,90){\circle*{4}}
\put(140,120){\circle*{4}}
\thicklines
\drawline(0,0)(175,60)
\drawline(52.5,90)(140,120)
\drawline(87.5,30)(52.5,90)
\drawline(87.5,30)(140,120)
\drawline(87.5,30)(210,0)
\put(105,185){\makebox(0,0)[b]{\scriptsize{\boldmath{$e_1$}}}}
\put(215,-5){\makebox(0,0)[l]{\scriptsize{\boldmath{$e_2$}}}}
\put(-5,-5){\makebox(0,0)[r]{\scriptsize{\boldmath{$e_3$}}}}
\put(105,-5){\makebox(0,0)[t]{\footnotesize$X^6$}}
\put(21.75,45){\makebox(0,0)[r]{\footnotesize$Y^3$}}
\put(74.25,135){\makebox(0,0)[r]{\footnotesize$Y^3$}}
\put(197.5,30){\makebox(0,0)[l]{\footnotesize$Z^2$}}
\put(162.5,90){\makebox(0,0)[l]{\footnotesize$Z^2$}}
\put(127.5,150){\makebox(0,0)[l]{\footnotesize$Z^2$}}
\put(148.75,20){\makebox(0,0)[l]{\footnotesize$X^3:Z$}}
\put(50,14){\makebox(0,0)[l]{\footnotesize$X^2:Y$}}
\put(135.75,42){\makebox(0,0)[l]{\footnotesize$X^2:Y$}}
\put(50,50){\makebox(0,0)[l]{\footnotesize$Z:XY$}}
\put(114,70){\makebox(0,0)[l]{\footnotesize$Y^2:XZ$}}
\put(105,115){\makebox(0,0)[b]{\footnotesize$X:Y^2Z$}}
\put(148.75,20){\makebox(0,0)[l]{\footnotesize$X^3:Z$}}
\end{picture}
\caption{{Fan for $\ZZ_6$-Hilb($C^3$) in lattice
$L=\ZZ^3+\frac{1}{6}(1,2,3)$}}
\label{fig:4}
\end{figure}
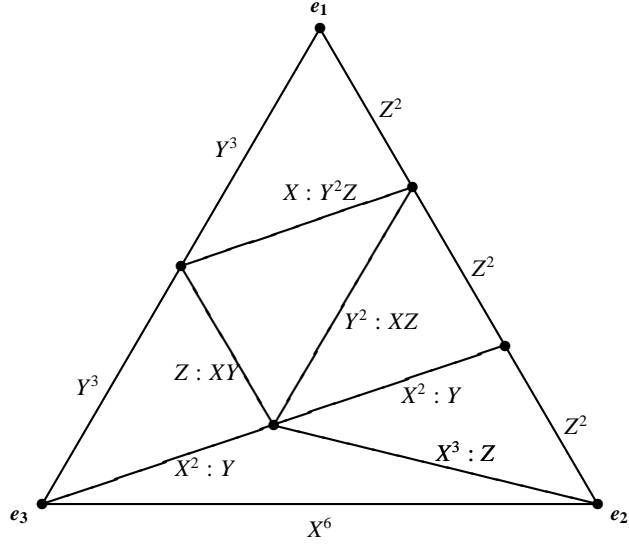

In a similar way it has been proved in \cite{Ito-Nakajima,ReidCraw99}
that the toric resolutions of $\CC^3/G$ defined in the previous
section (for $G\subset SL(3,\CC)$ abelian) are exactly the
$G$-Hilb($\CC^3$). In figure \ref{fig:4} we report the
$\ZZ_6$-Hilb($\CC^3$) fan  marked with the $\ZZ_6$-invariant ratios
associated to the curves of the resolution.

\subsection{Intersection theory for $\GC$}
We are interested in finding the Chow ring
$A^*(\GC)$, the module $A^c_*(\GC)$ and the intersection pairing
$A^*(\GC)\otimes A_*^c(\GC)\rightarrow A_*^c(\GC)$ (for this section
we refer to \cite{Fulton,FultonT,Oda}).

\subsubsection{Chow ring $A^*(\GC)$}
\label{Chow ring}
On any variety $X$ the Chow group $A_k(X)$ is defined to be the free
abelian group on the $k$-dimensional irreducible closed subvarieties of
$X$, modulo the subgroup generated by the cycles of the form
$(f)$, where $f$ is a nonzero rational function on a
$k+1$-dimensional subvariety of $X$.
\footnote{Recall that $(f)$ is the cycle obtained as the
sum of the zeros of $f$ minus the poles of $f$, each counted with
its multiplicity.}
For a toric variety $X=X(\Delta)$,
the Chow group $A_k(X)$ is generated by the classes of the
closures $V(\sigma)=\overline{O(\sigma)}$ of orbits of the $n-k$
dimensional cones $\sigma\in\Delta$ under the action of the
torus $\mathbb {C}_*^n$.
If $\tau$ is a cone of $\Delta$, we
define $N_\tau:=\ZZ
\cdot \tau$ and $N(\tau):=N/N_\tau$. The relations in $A_k(X)$ are
generated by the cycles of the form $(\chi^u):=\sum_i\langle
u,v_i\rangle\, V(\rho_i)$, where $u$ is an element in the dual lattice
$M(\tau)=N(\tau)^*$, $\rho_i$ are the one dimensional subcones of the
projection of $\tau$ in $N(\tau)$ with primitive vectors $v_i$,
$\langle\ ,\ \rangle $ is the natural pairing between $M(\tau)$ and
$N(\tau)$, for any cone $\tau\in\Delta$ of dimension $n-k-1$.

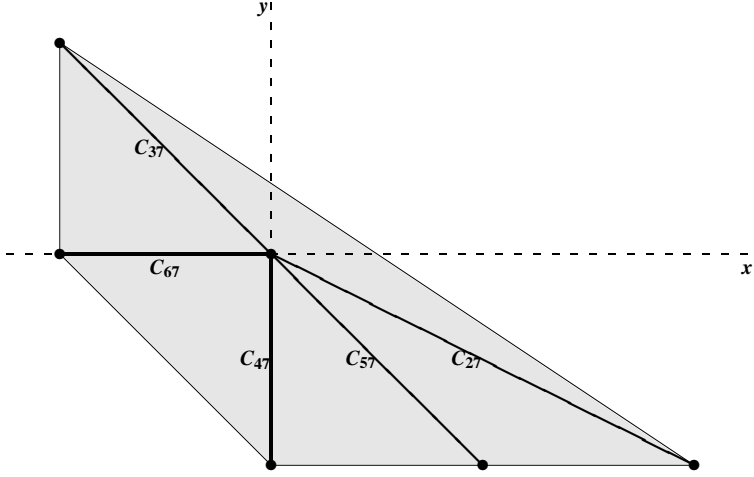
\begin{figure}[hbtp]
\centering
\setlength{\unitlength}{1pt}
\pssetlength{\unitlength}{1pt}
\begin{picture}(280,190)(-100,-90)
\pspolygon[linecolor=white,fillstyle=solid,fillcolor=vlgray](-80pt,80pt)(-80pt,0)(0,-80pt)(80pt,-80pt)(160pt,-80pt)
\dashline{3}(0,-80)(0,95) \put(-3,95){\makebox(0, 0)[t]{\scriptsize{\boldmath{$y$}}}}
\dashline{3}(-100,0)(180,0) \put(180,-4){\makebox(0,0)[t]{\scriptsize{\boldmath{$x$}}}}
\put(0,0){\circle*{4}}
\put(0,-80){\circle*{4}}
\put(-80,0){\circle*{4}}
\put(-80,80){\circle*{4}}
\put(80,-80){\circle*{4}}
\put(160,-80){\circle*{4}}
\thicklines
\drawline(0,0)(0,-80) \put(0,-40){\makebox(0,0)[r]{\scriptsize{\boldmath{$C_{47}$}}}}
\drawline(-80,80)(0,0) \put(-40,40){\makebox(0,0)[r]{\scriptsize{\boldmath{$C_{37}$}}}}
\drawline(80,-80)(0,0) \put(40,-40){\makebox(0,0)[r]{\scriptsize{\boldmath{$C_{57}$}}}}
\drawline(-80,0)(0,0) \put(-40,0){\makebox(0,-2)[t]{\scriptsize{\boldmath{$C_{67}$}}}}
\drawline(0,0)(160,-80) \put(80,-40){\makebox(0,0)[r]{\scriptsize{\boldmath{$C_{27}$}}}}
\psline[linewidth=.1mm](-80pt,80pt)(-80pt,0pt)
\psline[linewidth=.1mm](-80pt,0pt)(0pt,-80pt)
\psline[linewidth=.1mm](0pt,-80pt)(160pt,-80pt)
\psline[linewidth=.1mm](160pt,-80pt)(-80pt,80pt)
\end{picture}
\caption{Fan for $D_7\ $ ($Star(\rho_7)$)}
\label{fig:6}
\end{figure}

We will study explicitly this construction for $X=\GC$.\\
Let us first decorate the fan in figure \ref{fig:5} with labels for
the toric invariant subvarieties related to the cones of $\Delta$:
\begin{figure}
\centering
\begin{picture}(300, 190)(-110,-90)
\put(0,0){\circle*{4}}
\put(0,-80){\circle*{4}}
\put(-80,0){\circle*{4}}
\put(-80,80){\circle*{4}}
\put(80,-80){\circle*{4}}
\put(160,-80){\circle*{4}}
\put(-80,-80){\circle*{4}}
\thicklines
\drawline(0,0)(0,-80) \put(-0,-40){\makebox(0,0)[r]{\scriptsize{\boldmath{$C_{47}$}}}}
\drawline(-80,80)(0,0) \put(-40,40){\makebox(0,0)[r]{\scriptsize{\boldmath{$C_{37}$}}}}
\drawline(80,-80)(0,0) \put(40,-40){\makebox(0,0)[r]{\scriptsize{\boldmath{$C_{57}$}}}}
\drawline(-80,80)(-80,0) \put(-80,40){\makebox(0,0)[r]{\scriptsize{\boldmath{$C_{36}$}}}}
\drawline(-80,0)(-80,-80) \put(-80,-40){\makebox(0,0)[r]{\scriptsize{\boldmath{$C_{16}$}}}}
\drawline(-80,-80)(0,-80) \put(-40,-80){\makebox(0,-2)[t]{\scriptsize{\boldmath{$C_{14}$}}}}
\drawline(0,-80)(80,-80) \put(40,-80){\makebox(0,-2)[t]{\scriptsize{\boldmath{$C_{45}$}}}}
\drawline(80,-80)(160,-80) \put(120,-80){\makebox(0,-2)[t]{\scriptsize{\boldmath{$C_{25}$}}}}
\drawline(-80,0)(0,0) \put(-40,0){\makebox(0,-2)[t]{\scriptsize{\boldmath{$C_{67}$}}}}
\drawline(-80,80)(160,-80) \put(60,0){\makebox(5,-8)[r]{\scriptsize{\boldmath{$C_{23}$}}}}
\drawline(0,0)(160,-80) \put(80,-40){\makebox(0,0)[r]{\scriptsize{\boldmath{$C_{27}$}}}}
\drawline(-80,0)(0,-80) \put(-40,-40){\makebox(0,0)[r]{\scriptsize{\boldmath{$C_{46}$}}}}
\put(-83,6){\makebox(0, 0)[r]{\scriptsize{\boldmath{$D_6$}}}}
\put(-83,-86){\makebox(0, 0)[r]{\scriptsize{\boldmath{$D_1$}}}}
\put(-83,86){\makebox(0, 0)[r]{\scriptsize{\boldmath{$D_3$}}}}
\put(-12,-6){\makebox(0, 0)[l]{\scriptsize{\boldmath{$D_7$}}}}
\put(4,-86){\makebox(0, 0)[r]{\scriptsize{\boldmath{$D_4$}}}}
\put(84,-86){\makebox(0, 0)[r]{\scriptsize{\boldmath{$D_5$}}}}
\put(164,-86){\makebox(0, 0)[r]{\scriptsize{\boldmath{$D_2$}}}}
\put(90,-61){\makebox(0, 0)[c]{\scriptsize{\boldmath{$P_6$}}}}
\put(-55,-61){\makebox(0, 0)[c]{\scriptsize{\boldmath{$P_1$}}}}
\put(-55,20){\makebox(0, 0)[c]{\scriptsize{\boldmath{$P_2$}}}}
\put(-25,-25){\makebox(0, 0)[c]{\scriptsize{\boldmath{$P_3$}}}}
\put(25,-61){\makebox(0, 0)[c]{\scriptsize{\boldmath{$P_4$}}}}
\put(18,6){\makebox(0, 0)[r]{\scriptsize{\boldmath{$P_5$}}}}
\end{picture}
\caption{}
\label{fig:5}
\end{figure}\\
$A_3(X)$ has only one generator, corresponding to the unique
zero dimensional cone of $\Delta$, and obviously without relations.
\begin{eqnarray}
A_3(X)=\ZZ \cdot X\qquad X =V(0)
\end{eqnarray}
$A_2(X)$ has seven generators, related to the seven one dimensional
cones of the fan. The relations are generated by the cycles
$(\chi^u)$ for $u$ in $M(0)=M$:
\begin{eqnarray}
A_2(X)=\frac{\bigoplus_{i=1}^7\ZZ \cdot D_i}{<(\chi^u)>},
\qquad D_i =V(\rho_i),\qquad
\rho_i=\RR_{\geq 0}\cdot v_i
\end{eqnarray}
We choose as $u$ the standard basis of the lattice $M$,
$e_1^*,e_2^*,e_3^*$, so we obtain these three independent relations
\begin{eqnarray}
\label{Linear_eq_div}
&& -D_1+2D_2 -D_3 +D_5-D_6 =0\ , \\
&& -D_1-D_2+D_3-D_4-D_5=0 \ , \\
\label{Linear_eq_div3}
&& D_1+D_2+D_3+D_4+D_5+D_6+D_7=0 \ .
\end{eqnarray}
In $A_2(X)$ the divisors $D_5,\ D_6,\ D_7$ can be expressed in terms
of the others
\begin{eqnarray}
&& D_5=-D_1-D_2+D_3-D_4\ , \\
&& D_6=-2 D_1+D_2-D_4\ , \\
&& D_7=2 D_1-D_2-2 D_3+D_4\ .
\end{eqnarray}
It follows that
\begin{eqnarray}
A_2(X)=\bigoplus_{i=1}^4\ZZ \cdot D_i\simeq \ZZ^4.
\end{eqnarray}
Since the variety $\GC$ is non singular, we have
\begin{eqnarray}
\Pic (\GC) \simeq A_2(\GC)\simeq \ZZ^4 \ .
\end{eqnarray}
Let us make a remark about the canonical divisor. It is a standard
fact in toric geometry that the canonical divisor of a variety $X$ is
given by $K_X=-\sum_i D_i$ where the sum is over all toric invariant
divisors. By relation (\ref{Linear_eq_div3}) it then follows that $K_{\GC}=0$
in $\Pic (\GC)$ so that $\GC$ is a Calabi-Yau variety. This is true
for any toric variety with all integer vectors generating the fan lying
on the same (hyper)plane.\\
$A_1(X)$ has twelve generators, the toric invariant curves. The
relations are generated by the cycles $(\chi^u)$ for $u$ in
$M(\rho_i)$:
\begin{eqnarray}
A_1(X)=\frac{\bigoplus\ZZ \cdot C_{ij}}{<(\chi^u)>}.
\end{eqnarray}
Therefore, to find the relations we have to study the geometry of any
toric invariant divisor. Recall that $D_i=V(\rho_i)$ is a toric variety
for any $i$; its fan is called $Star(\rho_i)$ and is obtained by the
projection of the cones containing $\rho_i$ into the quotient lattice
$N(\rho_i)$. As an example we plot $Star(\rho_7)$ in figure \ref{fig:6}.
\\
$D_7$ has five toric invariant divisors and two relations between them
\begin{eqnarray}
&& -C_{27}+C_{37}-C_{47}-C_{57}=0\ ,\\
&& 2C_{27}-C_{37}+C_{57}-C_{67}=0\ .
\end{eqnarray}
Doing the same for any divisor $D_i$ we obtain all relations
between curves. At the end we find that any two given curves are
equivalent:
\begin{eqnarray}
A_1(X)=\ZZ \cdot C, \qquad C=[C_{46}].
\end{eqnarray}
Any other invariant curve is related to $C$ by the relations
expressed in the decorated fan of figure \ref{fig:7}.
\begin{figure}[hbtp]
\centering
\begin{picture}(300, 190)(-110,-90)
\put(0,0){\circle*{4}}
\put(0,-80){\circle*{4}}
\put(-80,0){\circle*{4}}
\put(-80,80){\circle*{4}}
\put(80,-80){\circle*{4}}
\put(160,-80){\circle*{4}}
\put(-80,-80){\circle*{4}}
\thicklines
\drawline(0,0)(0,-80)
\put(-2,-42){\makebox(0,0)[r]{\scriptsize{\boldmath{$-C$}}}}
\drawline(-80,80)(0,0)
\put(-42,38){\makebox(0,0)[r]{\scriptsize{\boldmath{$-3C$}}}}
\drawline(80,-80)(0,0)
\put(38,-42){\makebox(0,0)[r]{\scriptsize{\boldmath{$0$}}}}
\drawline(-80,80)(-80,0)
\put(-82,38){\makebox(0,0)[r]{\scriptsize{\boldmath{$C$}}}}
\drawline(-80,0)(-80,-80)
\put(-82,-42){\makebox(0,0)[r]{\scriptsize{\boldmath{$0$}}}}
\drawline(-80,-80)(0,-80)
\put(-42,-82){\makebox(0,-2)[t]{\scriptsize{\boldmath{$0$}}}}
\drawline(0,-80)(80,-80)
\put(38,-82){\makebox(0,-2)[t]{\scriptsize{\boldmath{$C$}}}}
\drawline(80,-80)(160,-80)
\put(120,-82){\makebox(0,-2)[t]{\scriptsize{\boldmath{$C$}}}}
\drawline(-80,0)(0,0)
\put(-42,-2){\makebox(0,0)[t]{\scriptsize{\boldmath{$-C$}}}}
\drawline(-80,80)(160,-80)
\put(36,0){\makebox(5,-8)[r]{\scriptsize{\boldmath{$C$}}}}
\drawline(0,0)(160,-80)
\put(76,-42){\makebox(0,0)[r]{\scriptsize{\boldmath{$-2C$}}}}
\drawline(-80,0)(0,-80)
\put(-41,-41){\makebox(0,0)[r]{\scriptsize{\boldmath{$C$}}}}
\end{picture}
\caption{}
\label{fig:7}
\end{figure}\\
$A_0(X)$ is generated by the six toric invariant points of $X$. Every
toric variety contains only two kind of toric curves:
compact curves isomorphic to $\PP_{\CC}^1$ and noncompact curves
isomorphic to $\AA_{\CC}^1$. Any compact curve gives a rational relation
between two invariant points, and in such way we find that any two
given points are rationally equivalent. Finally linear equivalence on
affine curves says us that points are rational equivalent to
zero. Therefore $A_0(X)$ is the trivial group (this is true for any
noncompact toric variety).

On a nonsingular $n$-dimensional variety $X$, one sets
$A^p(X):=A_{n-p}(X)$. There is an intersection product $A^p(X)\times
A^q(X)\rightarrow A^{p+q}(X)$, making $A^*(X):=\bigoplus A^p(X)$ into a
commutative graded ring. For a general toric variety $X(\Delta)$, if $\sigma$
and $\tau$ are cones in $\Delta$, then
\begin{equation*}
V(\sigma) \cap V(\tau) =
\begin{cases}
V(\gamma)   & \text{if $\sigma$ and $\tau$ span the cone $\gamma$},\\
\varnothing & \text{if $\sigma$ and $\tau$ do not span a cone in $\Delta$}.
\end{cases}
\end{equation*}
If $X(\Delta)$ is nonsingular and the intersection is proper,
i.e. each component of the intersection has codimension equal to the
sum of the codimension of the two subvarieties, or empty, then
$V(\sigma)$ and $V(\tau)$ meet transversally in $V(\gamma)$ (or
$\varnothing$). In this case we define $[V(\sigma)]\cdot
[V(\tau)]=[V(\gamma)]$ (or $0$). Otherwise if $V(\sigma)$ and
$V(\tau)$ do not meet properly, we can always use rational equivalence
to replace in $A^*(X)$ a subvariety (i.e. $V(\sigma)$) with
another one in the same class and such that it meets $V(\tau)$ in a proper
way.\\
Again, let us apply these considerations to our example $X=\GC$.\\
First, note that the intersection between $X$ and any
subvarieties $V(\sigma)$ is obviously equal to $V(\sigma)$. Therefore
$X$ is the multiplicative identity in $A^*(X)$.\\
Any divisor $D_i$ meets each other properly (or not at all), and so
their products give the curves $D_i\cdot D_j=[C_{ij}]$ (or $0$). We have
to use the linear equivalences (\ref{Linear_eq_div}) only to find the
autointersections $D_i\cdot D_i$.\\
Finally, when we intersect divisors and curves we obtain a point (or
$\varnothing$), but, as we have seen, they are rational equivalent to
zero. Any other intersection is always equivalent to the empty set.\\
The intersection products in $A^*(X)$ are summarized in table
\ref{tab:divdivCR1}.
\begin{table}[hbtp]
\begin{center}
\begin{eqnarray*}
\begin{array}{c|cccccc}
\phantom{x} & X & D_1 & D_2 & D_3 & D_4 & C\\
\hline
X   & X   & D_1 & D_2 & D_3 & D_4 & C\\
D_1 & D_1 & 0   & 0   & 0   & 0   & 0\\
D_2 & D_2 & 0   & 0   & C   & 0   & 0\\
D_3 & D_3 & 0   & C   & C   & 0   & 0\\
D_4 & D_4 & 0   & 0   & 0   &-C   & 0\\
C   & C   & 0   & 0   & 0   & 0   & 0\\
\end{array}
\end{eqnarray*}
\end{center}
\caption{Intersection product in $A^*(X)$}
\label{tab:divdivCR1}
\end{table}\\
We can see that the product is symmetric and respects the grading. If
we call $R$ the set of relations given by the intersection product we
find
\begin{eqnarray}
\label{Chow_ring}
A^*(X) =\ZZ[X,D_1,D_2,D_3,D_4,C]/R\ .
\end{eqnarray}

\subsubsection{Group $A_*^c(\GC)$ of compactly supported subvarieties}
On a noncompact variety $X$ the group $A_*^c(X)$ is defined to be the
direct limit of the groups $A^*(Z)$, where $Z$ are the closed and
compact subvarieties of $X$ ordered by inclusion.
This means $A_*^c(X)=\bigoplus A^*(Z)/R$, where the direct sum is
over all compact subvarieties of $X$ and the relations $R$ say that
two elements $[Z_1]$ and $[Z_2]$ of $\bigoplus A^*(Z)$ must be
identified if exists a compact subvariety $Z_3$ that contains them and
such that in $A^*(Z_3)$ they represent the same cycle class. As usual
in toric geometry we can restrict our analysis to compact toric invariant
subvarieties $V(\sigma)$; recall that $X(\Delta)$ is compact in the
classical topology if and only if its support $|\Delta|$ is the whole
space $N_{\RR}$.

Our example has one compact invariant divisor $D_7$, six compact curves
($C_{27}$, $C_{37}$,$C_{47}$, $C_{57}$,$C_{67}$, $C_{46}$) and six
points $P_i$. It's easy to see that (as a group)
\begin{eqnarray}
A^*(P_i)=\ZZ \cdot P_i^c\ ,\qquad
A^*(C_{ij})=\ZZ \cdot C_{ij}^c \oplus \ZZ \cdot P_{ij}^c\ ,
\end{eqnarray}
where $P_{ij}^c$ represents the point class in the curve $C_{ij}^c$.\\
The divisor $D_7$ is the toric variety associated to the fan of figure
\ref{fig:6}, therefore
\begin{eqnarray}
\label{Chow_ringD7}
A^*(D_7) =\ZZ \cdot D_7^c \oplus \ZZ \cdot C_{47}^c \oplus
\ZZ \cdot C_{57}^c \oplus \ZZ \cdot C_{67}^c \oplus \ZZ \cdot P_7^c
\end{eqnarray}
and the relations with other curves are
\begin{eqnarray}
\label{relcaurveH1R1}
&& C_{27}^c=C_{47}^c+C_{67}^c\ ,\\
\label{relcaurveH2R1}
&& C_{37}^c=2C_{47}^c+C_{57}^c+C_{67}^c\ .
\end{eqnarray}
Now we have to sum all these groups and find relations between
different generators. It results that all point classes have to be
identified, exactly as the classes of the same curve. The
group of compact subvarieties of $\GC$ is then isomorphic to $\ZZ^6$:
\begin{eqnarray}
\label{CompactChow_groupR1}
A_*^c(X) =\ZZ \cdot D_7^c \oplus \ZZ \cdot C_{46}^c \oplus
\ZZ \cdot C_{57}^c \oplus \ZZ \cdot C_{67}^c \oplus \ZZ \cdot C_{47}^c
\oplus \ZZ \cdot P^c\ .
\end{eqnarray}

\subsubsection{Intersection pairing}
There is a well defined intersection pairing $A^*(X)\otimes A_*^c(X)
\rightarrow A_*^c(X)$. For any two generators $[Z_1]\in A^*(X)$ and
$[Z_2]\in A_*^c(X)$ it is possible to find two representatives which
meet properly. Their intersection is a compact subvariety and defines
the above pairing $[Z_1].[Z_2]:=[Z_1\cap Z_2]\in A_*^c(X)$, which is
extendable by linearity to all elements in $A^*(X)\otimes A_*^c(X)$.
This product gives the group $A_*^c(X)$ the structure of an
$A^*(X)-$module.\\
For $X=\GC$ we obtain the intersection pairing of table
\ref{tab:divdivIPR1}:
\footnote{We can quickly obtain the product between divisors and
curves which do not intersect properly in this way: suppose $v_1,v_2$ are
the minimal lattice points on the edges of $\sigma_C$ and let $v',v''$
be the minimal lattice points of the threedimensional cones containing
$\sigma_C$, then $v'+v''=a_1 v_1 + a_2 v_2$ and $D_k\cdot C= -a_k P^c$.}
\begin{table}[hbtp]
\begin{center}
\begin{eqnarray*}
\begin{array}{c|cccccc}
\phantom{x} & D_7^c    & C_{46}^c & C_{57}^c & C_{67}^c & C_{47}^c & P^c\\
\hline
    X       & D_7^c    & C_{46}^c & C_{57}^c & C_{67}^c & C_{47}^c & P^c\\
    D_1     & 0        & P^c      & 0        & 0        & 0        & 0\\
    D_2     & C_{27}^c & 0        & P^c      & 0        & 0        & 0\\
    D_3     & C_{37}^c & 0        & 0        & P^c      & 0        & 0\\
    D_4     & C_{47}^c & -P^c     & P^c      & P^c      & -P^c     & 0\\
    C       & P^c      & 0        & 0        & 0        & 0        & 0\\
\end{array}
\end{eqnarray*}
\end{center}
\caption{Intersection pairing $A^*(X)\otimes A_*^c(X) \rightarrow A_*^c(X)$}
\label{tab:divdivIPR1}
\end{table}\\

\subsection{Homology, cohomology, Mori and K\"ahler cones}
For any compact smooth variety $X$ we have two natural homomorphisms
\begin{eqnarray*}
cl_X:A_*(X)\rightarrow H_*(X,\ZZ)\ , \qquad\quad
cl^X:A^*(X)\rightarrow H^*(X,\ZZ) \ .
\end{eqnarray*}
The map $cl_X$ sends the representant $V$ of an algebraic cycle to the
homological cycle $[V]$; it's well defined because algebraic
equivalence implies homological equivalence.
The map $cl^X$ is defined by composition of $cl_X$ with Poincar\'e
duality, which associate to an homological $k$-cycle $V$ the
$(n-k)$-form $\eta_V$ such that
\begin{eqnarray*}
\int_V \theta=\int_X \theta\wedge\eta_V\ .
\end{eqnarray*}
In the case of crepant resolutions of toric orbifolds it is possible to
prove \cite{Craw} that it exists the following module isomorphism
\begin{eqnarray*}
A^c_*(X)\simeq H^c_*(X,\ZZ)\ , \qquad A^*(X)\simeq H^*(X,\ZZ)
\end{eqnarray*}
which respects the intersection product
\footnote{The restriction to compact homology is necessary because of
the problem in defining integration over non compact cycles.}
\begin{eqnarray*}
H^*(X,\ZZ)\otimes H_*^c(X,\ZZ) \rightarrow H_*^c(X,\ZZ)\ .
\end{eqnarray*}

We are interested in determining the K\"ahler cone of $X$, which is the
set of all forms $J$ in $H^2(X,\QQ)$ such that
\begin{eqnarray*}
\int_C J\geq 0
\end{eqnarray*}
for all effective cycles in $H^c_2(X,\QQ)$.
We describe the K\"ahler cone using the module isomorphism of the
previous paragraph. We begin defining the Mori cone, i.e. the polyhedral
cone in $A_2^c(X)\otimes \QQ$ generated by effective toric invariant
compact curves of $X$, which are the compact algebraic cycles
$\sum_{i=1}^l a_{ij} C_{ij}^c$ where all the $a_{ij}$ are non-negative.
Now we can think at the K\"ahler cone of $X$ as the dual polyhedral
cone in $A^2(X)\otimes \QQ$ of the Mori cone with respect to the
intersection pairing.

For $X=\GC$, in view of the relations
(\ref{relcaurveH1R1},\ref{relcaurveH2R1}), the Mori cone is generated by:
\begin{eqnarray}
C_1:=C_{46}^c\ ,\qquad C_2:=C_{57}^c\ ,\qquad
C_3:=C_{67}^c\ ,\qquad C_4:=C_{47}^c\ .
\end{eqnarray}
Then the K\"ahler cone has the following dual generators, that
satisfied $T_a \cdot C_b= \delta_{ab}\, P^c\,$:
\begin{eqnarray}
&T_1:=D_1\, ,\quad T_2:=D_2\, ,
\quad T_3:=D_3\, ,\quad T_4:=-D_1+D_2+D_3-D_4\, .
\end{eqnarray}
For completeness we report in table \ref{tab:morikahler} the products
between the $T_i$ in the Chow ring $A^*(X)$:
\begin{table}[hbtp]
\begin{center}
\begin{eqnarray*}
\begin{array}{c|cccc}
\phantom{x} & T_1 & T_2 & T_3 & T_4  \\
\hline
     T_1    & 0 & 0 & 0  & 0  \\
     T_2    & 0 & 0 & C  & C  \\
     T_3    & 0 & C & C  & 2C \\
     T_4    & 0 & C & 2C & 2C \\
\end{array}
\end{eqnarray*}
\end{center}
\caption{Products between the K\"ahler generators in $A^*(X)$.}
\label{tab:morikahler}
\end{table}\\
If we call $J_i$ the K\"ahler generators in $H^2(X,\QQ)$
corresponding to the $T_i$, then we find the cohomology ring
\begin{eqnarray}
&H^*(X,\QQ)=\frac{\QQ[J_1,J_2,J_3,J_4]}{(J_1^2,J_1J_2,
J_1J_3,J_1J_4,J_2^2,J_2J_4-J_2J_3,J_3^2-J_2J_3,J_3J_4-2J_2J_3,
J_4^2-2J_2J_3)}.
\end{eqnarray}

\subsection{K-theory}
\subsubsection{Preliminaries}
The K-theory ring of a variety $X$ is related to the cohomology via
the Chern character map. More precisely this is an injective homomorphism of
rings from $K(X)$ to the Chow ring with rational coefficients
$A^*(X)_{\QQ}$. Then composition with $cl^X$ gives the homomorphism
with $H^*(X,\QQ)$:
\begin{eqnarray*}
\text{ch}\ :\ K(X) \rightarrow A^*(X)\otimes \QQ \simeq H^*(X,\QQ)\ .
\end{eqnarray*}
Here we summarize some general properties of Chern map that will be
useful in the next sections. The Chern class $c_i$ is a map from
$K(X)$ to $A^i(X)\otimes{\QQ}$; the total Chern class is defined as the sum of
all Chern class
\begin{eqnarray*}
\text{c}(\mathcal{F}) := \text{c}_0(\mathcal{F}) +
\text{c}_1(\mathcal{F}) + \ldots + \text{c}_n(\mathcal{F})\ ,
\end{eqnarray*}
where $n$ is the dimension of $X$.
A divisorial $\mathcal{O}_X(D)$ sheaf has a very simple total Chern class
\begin{eqnarray*}
\text{c}(\mathcal{O}_X(D)) = X + D \
\end{eqnarray*}
and, using multiplicative properties of the Chern classes, this implies
\begin{eqnarray*}
\mathcal{F}=\bigoplus_{i=1}^r \mathcal{O}_X(D_i)
\qquad \Longrightarrow \qquad
\text{c}(\mathcal{F}) = \prod_{i=1}^r (X + D_i) \ .
\end{eqnarray*}
The Chern character is defined for such sheaf as
\begin{eqnarray*}
\text{ch}(\mathcal{F}):=\sum_{i=1}^r e^{D_i}=
r\, X+\text{c}_1(\mathcal{F}) + \frac{1}{2} (\text{c}_1(\mathcal{F})^2-2
\text{c}_2(\mathcal{F}))\ ,
\end{eqnarray*}
where the expansion is stopped to second order in view of the cohomology
ring structure of our non compact threefold varieties. In particular for
a divisorial sheaf we have
\begin{eqnarray*}
\text{ch}(\mathcal{O}_X(D)) = X + D + \frac{1}{2} D^2\ .
\end{eqnarray*}
We recall also the definition of the Todd class:
\begin{eqnarray*}
\text{td}(\mathcal{F}):=\prod_{i=1}^r \frac{D_i}{1-e^{-D_i}}=
X+\frac{1}{2}\text{c}_1(\mathcal{F}) + \frac{1}{12}
(\text{c}_1(\mathcal{F})^2+ \text{c}_2(\mathcal{F}))\ .
\end{eqnarray*}
In particular, we need the Todd class of the tangent bundle $T_X$; for a
toric variety its total Chern class is
\begin{eqnarray*}
\text{c}(T_X) = \sum_{\sigma\in\Delta}[V(\sigma)] \
\end{eqnarray*}
and so if $X$ is a Calabi-Yau non compact toric threefold we have
\begin{eqnarray}
\label{cherntangent}
\text{c}(T_X) = X + \sum[C_{ij}]
\quad \Rightarrow \quad
\text{td}(X)=\text{td}(T_X)= X + \frac{1}{12} \sum[C_{ij}]\ ,
\end{eqnarray}
where the sum is over all (compact and non compact) toric curves in
$X$.

In the context of non compact varieties we also have to work with the
compactly supported K-theory group $K^c(X)$. This group is
related to the compactly supported Chow group with rational
coefficients $A^c_*(X)_{\QQ}$, and therefore to $H_*^c(X,\QQ)$, via
the local Chern character map \cite{Iversen}:
\begin{eqnarray*}
\text{ch}^c\ :\ K^c(X) \rightarrow A^c_*(X)\otimes \QQ\ \simeq
H^c_*(X,\QQ)\ .
\end{eqnarray*}
Let us briefly review its definition and properties.
Any element $S$ of $K^c(X)$ can be represented by
coherent sheaves $S_V$ on a compact subvariety $V$ of $X$.
If $i: V \hookrightarrow X$ is the embedding of $V$ in $X$, we
can define the local Chern character of $S$ by
\begin{eqnarray*}
\text{ch}^c(S)=\text{ch}(i_*S_V)\ .
\end{eqnarray*}
Actually we can compute the local Chern characters with the help of
the Grothendieck-Riemann-Roch theorem:
\begin{eqnarray*}
i_*(\text{ch}(S_V)\text{td}(V))=\text{ch}(i_*S_V)\text{td}(X)\
\end{eqnarray*}
for any compact subvarieties $V$ of $X$, which implies
\begin{eqnarray*}
\text{ch}^c(S)=\text{td}(X)^{-1}\, i_*(\text{ch}(S_V)\text{td}(V))\ .
\end{eqnarray*}
The local Chern classes of the divisorial sheaves over the compact
subvarieties in a Calabi-Yau non compact toric threefold $X$ are:
\begin{eqnarray}
\label{LocalChernclasses}
&&\text{ch}^c(\mathcal{O}_{p^c})=p^c   \qquad
\text{ch}^c(\mathcal{O}_{C^c}(n))=C^c+(n+1)p^c\nonumber\\[1mm]
&&\text{ch}^c(\mathcal{O}_{D^c}(C))=i_*\biggl(
D^c+\Bigl(C+\frac{1}{2}\text{c}_1(D^c)\Bigr)+\nonumber\\
&&\qquad\qquad\qquad\ +\frac12\Bigl({C}^2+\text{c}_1(D^c)\,C 
+\frac{1}{6}\bigl(\text{c}_1(D^c)^2+\text{c}_2(D^c)\bigr)\Bigr)\biggr)
-\frac{1}{12}\text{c}_2(X)\,D^c.
\end{eqnarray}
In the last character, $C$ is a divisor in $D^c$ and the
$\text{c}_i(D^c)$ are the Chern classes $\text{c}_i(T_{D^c})$, which
naturally live in $A^*(D^c)$ and that can be calculated using formula
(\ref{cherntangent}). Morover all the products excepted the last are
in $A^*(D^c)$.

\subsubsection{K-theory generators}
\label{K-theory generators}
Let $G$ be an abelian subgroup of $SL(3,\CC)$ which acts on the affine
space $\CC^3$. We write $\pi:\CC^3\rightarrow Y=\CC^3/G$ for the
quotient, $X=G$-Hilb($\CC^3$) for the Hilbert scheme with crepant
resolution $\tau:X\rightarrow Y$ and the universal scheme
$\mathcal{Z}=\{(Z(x),x)\in X\times\CC^3\}$ where $Z(x)$ is the
$G$-cluster over $x$ (see section \ref{G-Hilb}). Thus we have the
commutative diagram
\begin{eqnarray*}
\begin{CD}
\mathcal{Z}    @>q>>                \CC^3\\
@VV{p}V                          @VV{\pi}V\\
X              @>{\tau}>>             Y
\end{CD}
\end{eqnarray*}\\
Let us consider the sheaf $\mathcal R:=p_*\mathcal{O_Z}$ on the
resolution $X$. Over any point $x\in X$ the fibre of $\mathcal{R}$ is
$H^0(Z(x),\mathcal{O}_{Z(x)})$ which supports the regular
representation of $G$. In particular the rank of $\mathcal{R}$ is
equal to the order of the group $G$. The decomposition of the
regular representation into irreducible submodules induces the
decomposition
\begin{eqnarray*}
\mathcal{R}=\bigoplus_k\mathcal{R}_k\otimes\rho_k\qquad \text{for}\
\mathcal{R}_k=\text{Hom}_G(\rho_k,\mathcal{R})
\end{eqnarray*}
into locally free sheaves of rank$\,\mathcal{R}_i=\text{dim}\,
\rho_i=1$. We called $\mathcal{R}_k$ the tautological line bundle on $X$
associated to the irreducible representation $\rho_k$ of $G$.
At the level of $K$-theory the McKay correspondence states the
equivalence of the $G$-equivariant $K$-theory of $\CC^n$ and the
$K$-theory of the crepant resolutions. In \cite{Ito-Nakajima} it has
been determined the ring isomorphism
\begin{eqnarray*}
\varphi: K^G(\CC^3)\xrightarrow{\sim} K(X)
\end{eqnarray*}
showing that $\varphi(\rho_i\otimes\cO_{\CC^3})=\mathcal{R}_i$ and
therefore, that the tautological line bundles form a $\ZZ$-basis of
$K(X)$.

In section \ref{G-Hilb} we studied the orbifold $\CC^2/\ZZ_2$ and its
crepant resolution $\ZZ^2$-Hilb($\CC^2$). We have given a description of
$\mathcal{R}$ and its decomposition into line bundles on
the two open sets $U_{\sigma_a}$ and $U_{\sigma_b}$. In figure
\ref{fig:8} we report the monomial generators of $\mathcal{R}_i$ on
the affine pieces.
\begin{figure}[hbtp]
\centering
\setlength{\unitlength}{1pt}
\pssetlength{\unitlength}{1pt}
\begin{picture}(370,135)(-40,-30)
\pspolygon
[linecolor=white,fillstyle=solid,fillcolor=vlgray]
(0,0)(0pt,80pt)(80pt,80pt)(80pt,0)
\dashline{3}(0,-20)(0,100)
\dashline{3}(-20,0)(100,0)
\put(0,0){\circle*{4}}
\put(80,0){\circle*{4}}
\put(40,40){\circle*{4}}
\put(80,80){\circle*{4}}
\put(0,80){\circle*{4}}
\psline[linewidth=.1mm](40pt,40pt)(80pt,80pt)
\thicklines
\drawline(0,0)(0,80)
\put(-5,80){\makebox(0,0)[r]{\scriptsize{\boldmath{$e_2$}}}}
\drawline(0,0)(80,0)
\put(80,-5){\makebox(0,0)[t]{\scriptsize{\boldmath{$e_1$}}}}
\drawline(0,0)(40,40)
\put(30,60){\makebox(0,0)[b]{\footnotesize$1$}}
\put(55,20){\makebox(0,0)[t]{\footnotesize$1$}}
\pspolygon[linecolor=white,fillstyle=solid,fillcolor=vlgray]
(190pt,0)(190pt,80pt)(270pt,80pt)(270pt,0)
\dashline{3}(190,-20)(190,100)
\dashline{3}(170,0)(290,0)
\put(190,0){\circle*{4}}
\put(270,0){\circle*{4}}
\put(230,40){\circle*{4}}
\put(270,80){\circle*{4}}
\put(190,80){\circle*{4}}
\psline[linewidth=.1mm](230pt,40pt)(270pt,80pt)
\thicklines
\drawline(190,0)(190,80)
\put(185,80){\makebox(0,0)[r]{\scriptsize{\boldmath{$e_2$}}}}
\drawline(190,0)(270,0)
\put(270,-5){\makebox(0,0)[t]{\scriptsize{\boldmath{$e_1$}}}}
\drawline(190,0)(230,40)
\put(220,60){\makebox(0,0)[b]{\footnotesize$X$}}
\put(245,20){\makebox(0,0)[t]{\footnotesize$Y$}}
\end{picture}
\caption{Monomial generators of $\mathcal{R}_0$ and $\mathcal{R}_1$
for $\ZZ_2$-Hilb($\CC^2$)}
\label{fig:8}
\end{figure}
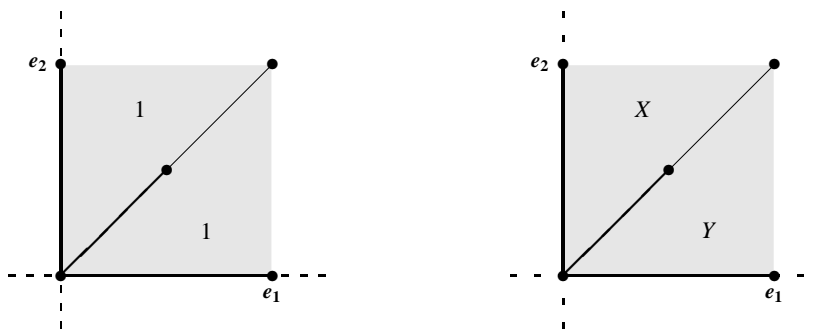\\
With the same procedure we can give the generators of $\mathcal{R}_i$
on $G$-Hilb($\CC^3$) for any abelian $G\subset SL(3,\CC)$. We report in
the figure \ref{fig:9} the monomial generators for $\ZZ_6$-Hilb($\CC^3$).
\begin{figure}[hbtp]
\centering
\setlength{\unitlength}{1pt}
\pssetlength{\unitlength}{1pt}
\begin{picture}(240,165)(-5,-100)

\put(35,-5){\makebox(0, 0)[t]{$\mathcal{R}_0$}}
\pspolygon (0pt,0)(70pt,0pt)(35pt,60pt)

\put(115,-5){\makebox(0, 0)[t]{$\mathcal{R}_1$}}
\pspolygon (80pt,0)(150pt,0pt)(115pt,60pt)

\put(195,-5){\makebox(0, 0)[t]{$\mathcal{R}_2$}}
\pspolygon (160pt,0)(230pt,0pt)(195pt,60pt)

\put(35,-95){\makebox(0, 0)[t]{$\mathcal{R}_3$}}
\pspolygon (0pt,-90pt)(70pt,-90pt)(35pt,-30pt)

\put(115,-95){\makebox(0, 0)[t]{$\mathcal{R}_4$}}
\pspolygon (80pt,-90pt)(150pt,-90pt)(115pt,-30pt)

\put(195,-95){\makebox(0, 0)[t]{$\mathcal{R}_5$}}
\pspolygon (160pt,-90pt)(230pt,-90pt)(195pt,-30pt)

\thicklines

\drawline(0,0)(58.33,20)
\drawline(46.66,40)(17.42,30)
\drawline(29.16,10)(17.42,30)
\drawline(29.16,10)(46.66,40)
\drawline(29.16,10)(70,0)
\put(30,2){\makebox(0,0)[b]{\footnotesize$1$}}
\put(14,15){\makebox(0,0)[l]{\footnotesize$1$}}
\put(30,23){\makebox(0,0)[b]{\footnotesize$1$}}
\put(45,26){\makebox(0,0)[l]{\footnotesize$1$}}
\put(32,41){\makebox(0,0)[b]{\footnotesize$1$}}
\put(52,8){\makebox(0,0)[b]{\footnotesize$1$}}

\drawline(80,0)(138.33,20)
\drawline(126.66,40)(97.42,30)
\drawline(109.16,10)(97.42,30)
\drawline(109.16,10)(126.66,40)
\drawline(109.16,10)(150,0)
\put(110,2){\makebox(0,0)[b]{\footnotesize$X$}}
\put(94,15){\makebox(0,0)[l]{\footnotesize$X$}}
\put(110,23){\makebox(0,0)[b]{\footnotesize$X$}}
\put(125,26){\makebox(0,0)[l]{\footnotesize$X$}}
\put(114.5,41){\makebox(0,0)[b]{\footnotesize$Y^2Z$}}
\put(132,8){\makebox(0,0)[b]{\footnotesize$X$}}

\drawline(160,0)(218.33,20)
\drawline(206.66,40)(177.42,30)
\drawline(189.16,10)(177.42,30)
\drawline(189.16,10)(206.66,40)
\drawline(189.16,10)(230,0)
\put(190,2){\makebox(0,0)[b]{\footnotesize$X^2$}}
\put(174,15){\makebox(0,0)[l]{\footnotesize$Y$}}
\put(190,23){\makebox(0,0)[b]{\footnotesize$Y$}}
\put(205,26){\makebox(0,0)[l]{\footnotesize$Y$}}
\put(192,41){\makebox(0,0)[b]{\footnotesize$Y$}}
\put(212,8){\makebox(0,0)[b]{\footnotesize$X^2$}}

\drawline(0,-90)(58.33,-70)
\drawline(46.66,-50)(17.42,-60)
\drawline(29.16,-80)(17.42,-60)
\drawline(29.16,-80)(46.66,-50)
\drawline(29.16,-80)(70,-90)
\put(30,-89){\makebox(0,0)[b]{\footnotesize$X^3$}}
\put(12,-75){\makebox(0,0)[l]{\footnotesize$XY$}}
\put(30,-67){\makebox(0,0)[b]{\footnotesize$Z$}}
\put(45,-64){\makebox(0,0)[l]{\footnotesize$Z$}}
\put(32,-49){\makebox(0,0)[b]{\footnotesize$Z$}}
\put(52,-82){\makebox(0,0)[b]{\footnotesize$Z$}}

\drawline(80,-90)(138.33,-70)
\drawline(126.66,-50)(97.42,-60)
\drawline(109.16,-80)(97.42,-60)
\drawline(109.16,-80)(126.66,-50)
\drawline(109.16,-80)(150,-90)
\put(110,-89){\makebox(0,0)[b]{\footnotesize$X^4$}}
\put(93,-75){\makebox(0,0)[l]{\footnotesize$Y^2$}}
\put(110,-67){\makebox(0,0)[b]{\footnotesize$Y^2$}}
\put(122,-64){\makebox(0,0)[l]{\footnotesize$XZ$}}
\put(114,-49){\makebox(0,0)[b]{\footnotesize$Y^2$}}
\put(132,-82){\makebox(0,0)[b]{\footnotesize$XZ$}}

\drawline(160,-90)(218.33,-70)
\drawline(206.66,-50)(177.42,-60)
\drawline(189.16,-80)(177.42,-60)
\drawline(189.16,-80)(206.66,-50)
\drawline(189.16,-80)(230,-90)
\put(190,-89){\makebox(0,0)[b]{\footnotesize$X^5$}}
\put(170,-75){\makebox(0,0)[l]{\footnotesize$XY^2$}}
\put(190,-68){\makebox(0,0)[b]{\footnotesize$YZ$}}
\put(202,-64){\makebox(0,0)[l]{\footnotesize$YZ$}}
\put(194,-49){\makebox(0,0)[b]{\footnotesize$YZ$}}
\put(212,-82){\makebox(0,0)[b]{\footnotesize$X^2Z$}}

\end{picture}
\caption{Monomial generators of
$\mathcal{R}_i$ for $\ZZ_6$-Hilb($\CC^3$)}
\label{fig:9}
\end{figure}
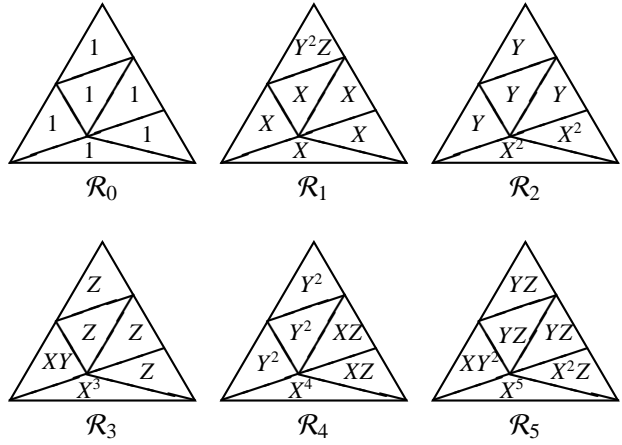\\
The action of $\ZZ_6$ on the coordinate ring of $\CC^3$ is
\begin{eqnarray}
\epsilon\cdot(X,Y,Z)=(\epsilon X,\epsilon^2 Y,\epsilon^3 Z),
\qquad\qquad \epsilon=e^{\frac{2\pi i}{6}}.
\end{eqnarray}
Therefore, it is simple to verify that each $\mathcal{R}_i$ supports the
irreducible representation $\rho_i$ of $\ZZ_6$.

Any line bundle on a smooth algebraic variety is a divisorial
bundle. We briefly sketch the standard procedure to find the divisor
related to the $\mathcal{R}_i$ defined by Reid and proved by
Craw, and refer to \cite{Craw} for a detailed explanation.\\
The first step consists in decorating the $G$-Hilb fan with the
characters of the 
group. Any curve has to be marked with the character of the monomials
in its associated ratio. For any internal vertex $v$ there exists a
recipe to associate one or two characters of $G$, depending primarly
on the valency of $v$ (i.e the number of lines meeting at $v$). For a
$G$-Hilb fan this is always $3,4,5$ or $6$. There are the following cases:
\begin{itemize}
\item
A vertex $v$ of valency $3$ defines an exceptional $\PP^2$. A single
character $\chi_k$ marks all three lines meeting at $v$. Mark the
vertex $v$ with the character $\chi_m:=\chi_k\otimes\chi_k$.
\item
A vertex $v$ of valency $4$ defines an exceptional Hirzebruch surface
$\FF_r$. There are distinct characters $\chi_k$ and $\chi_l$
each one marking a pair of lines meeting at $v$. Mark the vertex $v$ with the
character $\chi_m:=\chi_k \otimes \chi_l$.
\item
A vertex $v$ of valency $5$ or $6$ (exluding three straight lines
meeting at a point) defines an Hirzebruch surface $\FF_r$ blown-up in
one or two points. There are uniquely determined characters $\chi_k$
and $\chi_l$ each one marking a pair of lines meeting at $v$. Mark the
vertex $v$ with $\chi_m:=\chi_k \otimes \chi_l$.
\item
A vertex $v$ at the intersection of three straight lines defines an
exceptional Del Pezzo surface of degree six, denoted dP$_6$. The
monomials defining the pair of morphisms dP$_6\rightarrow \PP^2$ lie
in uniquely determined character spaces $\chi_l$ and $\chi_m$
satisfying
\begin{eqnarray*}
\chi_l \otimes \chi_m=\chi_i \otimes \chi_j \otimes \chi_k\ ,
\end{eqnarray*}
where $\chi_i,\ \chi_j$ and $\chi_k$ mark the straight lines through
the vertex $v$. Mark the vertex $v$ with both $\chi_l$ and $\chi_m$.
\end{itemize}
Each character of $G$ appears once on the fan $\Delta$.\\
By analyzing of the monomial generators of the tautological
line bundles $\mathcal{R}_i$, in \cite{Craw} the author proved that:
\begin{itemize}
\item
If $\chi_k$ marks the line defining the compact curve $C_k\in
H_2^c(X,\ZZ)$ on the resolution $X$, the first Chern class
$c_1(\mathcal{R}_k)$ is the dual to $C_k$ in $H^2(X,\ZZ)$:
\begin{eqnarray*}
\int_{C_l} c_1(\mathcal{R}_k)=\delta_{lk}\ .
\end{eqnarray*}
This means that $R_k=\mathcal{O}_X(T_k)$, where $T_k$ is the generator
of the K\"ahler cone dual to $C_k$.
\item
In $\Pic(X)$ all relations between tautological line bundles are of the
following forms:
\begin{itemize}
\item
$\mathcal{R}_m=\mathcal{R}_k\otimes\mathcal{R}_k$ when
$\chi_m=\chi_k\otimes\chi_k$ marks a vertex $v$ of valency $3$;
\item
$\mathcal{R}_m=\mathcal{R}_k\otimes\mathcal{R}_l$ when
$\chi_m=\chi_k\otimes\chi_l$ marks a vertex $v$ of valency $4$;
\item
$\mathcal{R}_m=\mathcal{R}_k\otimes\mathcal{R}_l$ when
$\chi_m=\chi_k\otimes\chi_l$ marks a vertex $v$ of valency $5$ or $6$
(exluding three straight lines meeting at a point);
\item
$\mathcal{R}_l\otimes\mathcal{R}_m=
\mathcal{R}_i\otimes\mathcal{R}_j\otimes\mathcal{R}_k$ when
the pair of characters $\chi_l$ and $\chi_m$ satisfying  $\chi_l
\otimes \chi_m=\chi_i \otimes \chi_j \otimes \chi_k$ marks the
intersection point $v$ of three straight lines.
\end{itemize}
\end{itemize}

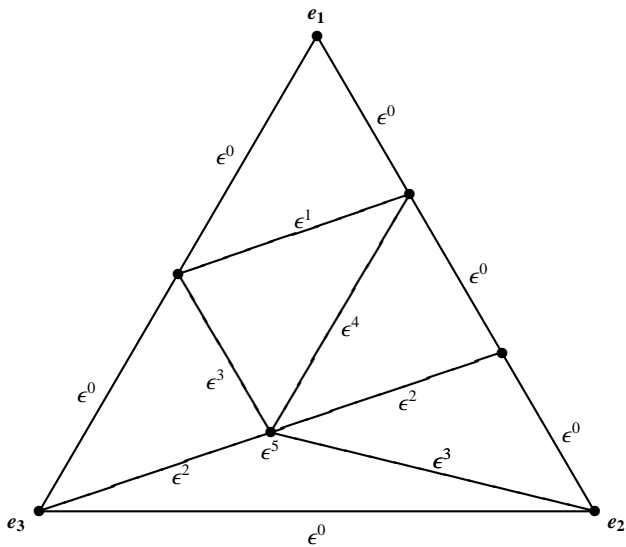
\begin{figure}[hbtp]
\centering
\setlength{\unitlength}{1pt}
\pssetlength{\unitlength}{1pt}
\begin{picture}(230,200)(-10,-10)
\pspolygon(0pt,0)(210pt,0pt)(105pt,180pt)
\put(0,0){\circle*{4}}
\put(210,0){\circle*{4}}
\put(105,180){\circle*{4}}
\put(87.5,30){\circle*{4}}
\put(175,60){\circle*{4}}
\put(52.5,90){\circle*{4}}
\put(140,120){\circle*{4}}
\thicklines
\drawline(0,0)(175,60)
\drawline(52.5,90)(140,120)
\drawline(87.5,30)(52.5,90)
\drawline(87.5,30)(140,120)
\drawline(87.5,30)(210,0)
\put(105,185){\makebox(0,0)[b]{\scriptsize{\boldmath{$e_1$}}}}
\put(215,-5){\makebox(0,0)[l]{\scriptsize{\boldmath{$e_2$}}}}
\put(-5,-5){\makebox(0,0)[r]{\scriptsize{\boldmath{$e_3$}}}}
\put(105,-5){\makebox(0,0)[t]{\footnotesize$\epsilon^0$}}
\put(21.75,45){\makebox(0,0)[r]{\footnotesize$\epsilon^0$}}
\put(74.25,135){\makebox(0,0)[r]{\footnotesize$\epsilon^0$}}
\put(197.5,30){\makebox(0,0)[l]{\footnotesize$\epsilon^0$}}
\put(162.5,90){\makebox(0,0)[l]{\footnotesize$\epsilon^0$}}
\put(127.5,150){\makebox(0,0)[l]{\footnotesize$\epsilon^0$}}
\put(148.75,20){\makebox(0,0)[l]{\footnotesize$\epsilon^3$}}
\put(50,14){\makebox(0,0)[l]{\footnotesize$\epsilon^2$}}
\put(135.75,42){\makebox(0,0)[l]{\footnotesize$\epsilon^2$}}
\put(63,50){\makebox(0,0)[l]{\footnotesize$\epsilon^3$}}
\put(114,70){\makebox(0,0)[l]{\footnotesize$\epsilon^4$}}
\put(100,107){\makebox(0,0)[b]{\footnotesize$\epsilon^1$}}
\put(148.75,20){\makebox(0,0)[l]{\footnotesize$\epsilon^3$}}
\put(87.5,27){\makebox(0,0)[t]{\footnotesize$\epsilon^5$}}
\end{picture}
\caption{Fan for $\ZZ_6$-Hilb($C^3$) decorated with Reid's recipe}
\label{fig:10}
\end{figure}

As usual we apply these considerations to our case
$X=\ZZ_6$-Hilb($\CC_3^6$) and we summarize them in the decorated fan of
figure \ref{fig:10}.
\\
The resulting tautological line bundles, that give a $\ZZ$-basis of
$K(X)$, are:
\begin{eqnarray}
&& \cRR_0=\cO_X,\quad \cRR_1=\cO_X(D_1),\quad \cRR_2=\cO_X(D_2),\quad
\cRR_3=\cO_X(D_3),\cr
&& \cRR_4=\cO_X(-D_1+D_2+D_3-D_4),\quad
\cRR_5=\cRR_2\otimes\cRR_3=\cO_X(D_2+D_3).
\end{eqnarray}

\subsubsection{$K(X)$ and $K^c (X)$}
Chosen a base of generators for $K(X)$ we can find the dual basis for
the compact K-theory $K^c(X)$ as in \cite{Ito-Nakajima} using the
perfect pairing
\begin{eqnarray*}
&(\ |\ ): K(X)\times K^c (X) \longrightarrow \ZZ ,\quad
(\cRR, \cS)\longmapsto(\cRR|\cS)=\int_X \ch(\cRR)\ch^c(\cS)\td(X)\ ,
\end{eqnarray*}
so that
\begin{eqnarray}
(\cRR_i| \cS_j)=\delta_{ij}. \label{dual}
\end{eqnarray}
As usual, the integral is by definition the coefficient of the point
class. Using this fact, the standard computations of Chern and Todd
characters and the intersection product table \ref{tab:divdivIPR1},
from condition (\ref{dual}) we find
\begin{eqnarray}
&& \ch^c(\cS_0)=D_7^c-\left(C_{46}^c+C_{57}^c+\frac 32 C_{67}^c+2 C_{47}^c\right)+\frac 76 P^c\ ,\cr
&& \ch^c(\cS_1)=C_{46}^c\ ,\cr
&& \ch^c(\cS_2)=-D_7^c+\left(C_{57}^c+\frac 12 C_{67}^c+C_{47}^c\right)-\frac 16 P^c\ ,\cr
&& \ch^c(\cS_3)=-D_7^c+\left(\frac 32 C_{67}^c+C_{47}^c\right) -\frac 16 P^c\ ,\cr
&& \ch^c(\cS_4)=C_{47}^c\ ,\cr
&& \ch^c(\cS_5)=D_7^c-\left(\frac 12 C_{67}^c+C_{47}^c\right)+\frac 16 P^c\ .
\end{eqnarray}

In the spirit of the paper \cite{Hosono:2004jp} we now express the
elements $\cS_i$ in terms of a symplectic D-brane basis of
$K^c(X)$. Such basis can be constructed starting from the
generators of the compact Chow ring. We choose
\begin{eqnarray}
B_0:= \cO_P^c\ ;\quad\ B_a:=\cO_{C_a^c} (-T_a)\ ;\quad\ B_5:=\cO_{D_7^c}(-T_2-T_3)\ ,
\end{eqnarray}
with $a=1,\ldots,4$ and $\cO_{C_a^c} (-T_a):= \cO_{C_a^c}\otimes \cO_X
(-T_a)$, $\cO_{D_7^c}(-T_2-T_3):= \cO_{D_7^c}\otimes \cO_X(-T_2-T_3)$.
To express the basis $\cS_i$ in terms of $B_j$ we can compare their
compact Chern characters.
Using (\ref{LocalChernclasses}) and the multiplicative property of Chern
character we find
\begin{eqnarray}
&\text{ch}^c(B_0)=P^c   \quad
\text{ch}^c(B_a)=C^c_a \quad
\text{ch}^c(B_5)=D_7^c - \left(\frac 12 C_{67}^c+C_{47}^c\right)+\frac 16 P^c
\end{eqnarray}
and then
\begin{eqnarray}
&& \cS_0=B_0-B_1-B_2-B_3-B_4+B_5\ ,\cr
&& \cS_1=B_1\ ,\cr
&& \cS_2=B_2-B_5\ ,\cr
&& \cS_3=B_3-B_5\ ,\cr
&& \cS_4=B_4\ ,\cr
&& \cS_5=B_5\ ,
\end{eqnarray}
Finally we write the $B_i$ basis of $K^c(X)$ in terms of the $\cS_i$
and its dual basis $\Phi_i$ of $K(X)$ in term of $\cRR_i$:
\begin{align}
& B_0=\cS_0+\cS_1+\cS_2+\cS_3+\cS_4+\cS_5\ ,
&& \Phi_0=\cRR_0 \ ,\cr
& B_1=\cS_1\ ,
&& \Phi_1=-\cRR_0+\cRR_1\ ,\cr
& B_2=\cS_2+\cS_5\ ,
&& \Phi_2=-\cRR_0+\cRR_2\ ,\cr
& B_3=\cS_3+\cS_5\ ,
&& \Phi_3=-\cRR_0+\cRR_3\ ,\cr
& B_4=\cS_4\ ,
&& \Phi_4=-\cRR_0+\cRR_4\ ,\cr
& B_5=\cS_5\ .
&& \Phi_5=\cRR_0-\cRR_2-\cRR_3+\cRR_5\ .
\end{align}

\subsection{The Hosono conjecture}\label{conjecture}
We will restate shortly here the Hosono conjecture \cite{Hosono:2004jp},
for convenience. The main point is that the periods for the mirror
manifold are solutions of a set of Picard-Fuchs equations, and the
general solution can be expressed in terms of an hypergeometric
function with value in the cohomology of $X$:
\begin{eqnarray*}
w=w\left(x_1,\ldots,x_4;\frac{J_1}{2\pi i},\ldots,\frac{J_4}{2\pi i}\right)
\end{eqnarray*}
Then the conjecture (adapted to our case) states as follows.
\subsubsection{Hosono conjecture}
Define the basis for $H^*(X,\QQ)$
\begin{eqnarray*}
Q_i:=\ch(\Phi_i)\ , \quad\ i=0,\ldots,5
\end{eqnarray*}
and expand the cohomology-valued hypergeometric series $w$ with
respect to this basis:
\begin{eqnarray*}
w\left(x_1,\ldots,x_4;\frac{J_1}{2\pi i},\ldots,\frac{J_4}{2\pi
i}\right)=\sum_{i=0}^5 w_i(x_1,\ldots,x_4) Q_i\ .
\end{eqnarray*}
Thus
\begin{enumerate}
\item\label{primo}
the coefficient hypergeometric series $w_i(x_1,\ldots,x_4)$ may be
identified with the period integrals over the cycles $mir(B_i)$,
\begin{eqnarray*}
w_i(x_1,\ldots,x_4)=\int_{mir(B_i)} \Omega(Y_x)\ ;
\end{eqnarray*}
\item\label{secondo}
the monodromy of the hypergeometric series is integral and symplectic
with respect to the symplectic form defined in $K^c(X)$
\begin{eqnarray*}
\chi(B_i,B_j)= \int_X \ch(B_i^\vee) \ch(B_j) \td(X)\ ;
\end{eqnarray*}
\item\label{terzo}
the central charge of an element $F\in K^c(X)$ is expressed in terms
of the cohomology valued hypergeometric $w$ as
\begin{eqnarray*}
Z(F)=\int_X \ch(F) w\left(x_1,\ldots,x_4;\frac{J_1}{2\pi
i},\ldots,\frac{J_4}{2\pi i}\right) \td(X)\ .
\end{eqnarray*}
\end{enumerate}

The symplectic form of point \ref{secondo} can be easily computed with
respect to the basis $\cS^-_i$ following the paper of $Ito-Nakajima$
\cite{Ito-Nakajima}. Let $Q$ be the $3$-dimensional representation
given by the inclusion $G\subset SL(3,\CC)$ and $\{\rho_i\}_{i=0}^{r}$
be the irreducible representations. The decomposition
\begin{eqnarray*}
Q\otimes\rho_j=\bigoplus_k a_{ij}\rho_i
\end{eqnarray*}
is related to the symplectic form by
\begin{eqnarray*}
\chi(\cS_i,\cS_j)=a_{ji}-a_{ij}\ .
\end{eqnarray*}
In our example
\begin{eqnarray}
\chi(\cS_i,\cS_j)=\left(
\begin{array}{cccccc}
0 & 1 & 1 & 0 & -1 & -1 \\
-1 & 0 & 1 & 1 & 0 & -1 \\
-1 & -1 & 0 & 1 & 1 & 0 \\
0 & -1 & -1 & 0 & 1 & 1 \\
1 & 0 & -1 & -1 & 0 & 1 \\
1 & 1 & 0 & -1 & -1 & 0
\end{array}
\right)\ ,
\end{eqnarray}
and then, for our chosen basis,
\begin{eqnarray}\label{simplecticform}
\chi(B_i,B_j)=\left(
\begin{array}{cccccc}
0 & 0 & 0 & 0 & 0 & 0 \\
0 & 0 & 0 & 0 & 0 & 1 \\
0 & 0 & 0 & 0 & 0 & 0 \\
0 & 0 & 0 & 0 & 0 & -1 \\
0 & 0 & 0 & 0 & 0 & -1 \\
0 & -1 & 0 & 1 & 1 & 0
\end{array}
\right)\ .
\end{eqnarray}
This matrix gives a symplectic correspondence between the space
$H^4(X,\QQ)$ and a one dimensional subspace of $H^2(X,\QQ)$. There is
an obvious ambiguity in such a correspondence, but we will turn back
to it later.

\subsubsection{The cohomological hypergeometric series}
The vectors $\ell_a$, $a=1,\ldots,4$ are given by the intersection numbers
between the Mori cone generators and the invariant divisors of $X$, so
that we find
\begin{eqnarray}
C_1 \quad\ &:& \quad\  \ell_1=(1,0,0,-1,0,-1,1)\ ,\cr
C_2 \quad\ &:& \quad\  \ell_2=(0,1,0,1,-2,0,0)\ ,\cr
C_3 \quad\ &:& \quad\  \ell_3=(0,0,1,1,0,-1,-1)\ ,\cr
C_4 \quad\ &:& \quad\  \ell_4=(0,0,0,-1,1,1,-1)\ .
\end{eqnarray}
The hypergeometric series is then
\begin{eqnarray}
\left.w=\sum_{\vec m\in \ZZ^4_{\geq 0}}
\frac {x_1^{m_1+\rho_1} x_2^{m_2+\rho_2}x_3^{m_3+\rho_3}x_4^{m_4+\rho_4}}{\prod_{i=1}^7 \Gamma_i (\vec m +\vec \rho)}\right|_{\vec \rho=\frac {\vec J}{2\pi i}}
\label{hypergeometric}
\end{eqnarray}
\begin{eqnarray}
&& \Gamma_1(\vec m)=\Gamma(1+m_1)\ ,\cr
&& \Gamma_2(\vec m)=\Gamma(1+m_2)\ ,\cr
&& \Gamma_3(\vec m)=\Gamma(1+m_3)\ ,\cr
&& \Gamma_4(\vec m)=\Gamma(1-m_1+m_2+m_3-m_4)\ ,\cr
&& \Gamma_5(\vec m)=\Gamma(1-2m_2+m_4)\ ,\cr
&& \Gamma_6(\vec m)=\Gamma(1-m_1-m_3+m_4)\ ,\cr
&& \Gamma_7(\vec m)=\Gamma(1+m_1-m_3-m_4)\ .
\end{eqnarray}
We need to expand this function in power series in $\vec J$. Because
of the ring relations for $H^*(X,\QQ)$, we see that the expansion
stops at order two. The coefficient functions, with respect to the
basis $\{1,\vec J, C\}$ of $H^*(X,\QQ)$, are computed in the
appendix.\\
However we chosen the basis $B_i$ in $K^c(X)$ so that we need to
rewrite the expansion in terms of the dual basis $Q_i=\ch (\Phi_i)$:
\begin{eqnarray}
&Q_0 =1,\ Q_1 =J_1,\ Q_2 =J_2,\ Q_3 =J_3 +\frac 12 C,\ Q_4 =J_4+C,\ Q_5 =C.
\end{eqnarray}
If we make this change of basis and use the mirror symmetry identification
\begin{eqnarray}
w\left(\vec x, \frac {\vec J}{2\pi i} \right)= Q_0 1 +\sum_{a=1}^4 Q_a
t_a +Q_5 g(t_1,\ldots,t_4)\ ,
\end{eqnarray}
then we find
\begin{eqnarray}
&& 2\pi i t_1= \log x_1 +\Psi(x_1 x_3) +\Phi(x_2,x_1 x_4)+\aleph (\vec x) \ ,\cr
&& 2\pi i t_2= \log x_2 -\Phi(x_2,x_1 x_4) +2\Phi(x_1 x_4,x_2) \ ,\cr
&& 2\pi i t_3= \log x_3 -\Phi(x_2, x_1 x_4) +\Psi(x_1 x_3) -\aleph (\vec x) \ ,\cr
&& 2\pi i t_4= \log x_4 -\Phi(x_1 x_4,x_2) +\Phi(x_2,x_1 x_4)-\Psi(x_1 x_3) -\aleph (\vec x)  \ ,\label{moduli}
\end{eqnarray}
and
\begin{eqnarray}
&& (2\pi i)^2g(\vec t)=-\frac {\pi^2}3-\pi i (\log x_3 -\Phi(x_2, x_1
x_4) +\Psi(x_1 x_3) -\aleph (\vec x))\cr
&&\qquad\ -2\pi i(\log x_4 -\Phi(x_1 x_4,x_2) +\Phi(x_2,x_1
x_4)-\Psi(x_1 x_3) -\aleph (\vec x))\cr
&&\qquad\ +7\aleph^{(1)}(\vec x)-3\aleph^{(2)}(\vec
x)-2\aleph^{(3)}(\vec x)-\aleph^{(4)}(\vec x)-\aleph^{(5)}(\vec x)\cr
&&\qquad\ -\Psi_1(x_2,x_1 x_4)+\Psi_2(x_2,x_1 x_4)-\Psi_1(x_1 x_4,x_2)
+\Psi_2(x_1 x_4,x_2)\cr
&&\qquad\ -\Psi_1(x_1 x_4,x_2)+\Psi_3(x_1 x_4,x_2)-\Psi_4 (x_1 x_3)
+\Psi_5 (x_1 x_3)+\Psi_6(x_2,x_4)\cr
&&\qquad\ +\Lambda_1(\vec x)-\Lambda_2(\vec x)-\Lambda_3(\vec x)\cr
&&\qquad\ +\frac 12 (\log x_3)^2 +\log x_3 [\Psi(x_1x_3) -\Phi(x_2,x_1
x_4)-\aleph(\vec x)]\cr
&&\qquad\ + (\log x_4)^2 +2\log x_4 [\Phi(x_2,x_1 x_4)-\Phi(x_1
x_4,x_2)-\Psi(x_1 x_3)-\aleph(\vec x)] \cr
&&\qquad\ + \log x_2 \log x_3 + \log x_2 [\Psi(x_1 x_3)-\Phi(x_2, x_1
x_4)-\aleph(\vec x)] \cr
&&\qquad\ + \log x_3 [2\Phi(x_1 x_4,x_2)-\Phi(x_2,x_1 x_4)] + \log x_2
\log x_4 \cr
&&\qquad\ +\log x_2 [\Phi(x_2,x_1 x_4)-\Phi(x_1 x_4,x_2)-\Psi(x_1
x_3)-\aleph(\vec x)]\cr
&&\qquad\ +\log x_4 [2\Phi(x_1 x_4,x_2)-\Phi(x_2,x_1 x_4)] +2\log x_3 \log x_4 \cr
&&\qquad\ +2\log x_3 [\Phi(x_2,x_1 x_4)-\Phi(x_1 x_4,x_2)-\Psi(x_1 x_3)-\aleph(\vec x)]\cr
&&\qquad\ +2\log x_4 [\Psi(x_1x_3) -\Phi(x_2,x_1 x_4)-\aleph(\vec x)]\ .
\end{eqnarray}
Using the above expressions we find
\begin{eqnarray}
&& g(\vec t ) =P_2(\vec t)+\frac 1{(2\pi i)^2} \phi(\vec t)\ ,
\end{eqnarray}
where $P_2$ is the degree two polynomial part
\begin{eqnarray}
P_2(\vec t )=\frac 1{12}-\frac 12 t_3 -t_4+\frac 12 t_3^2 +t_4^2 +t_2 t_3 +t_2 t_4 +2t_3 t_4\ ,
\end{eqnarray}
and
\begin{eqnarray}
&& \phi(\vec t)=7\aleph^{(1)}(\vec x)-3\aleph^{(2)}(\vec x)-2\aleph^{(3)}(\vec x)-\aleph^{(4)}(\vec x)-\aleph^{(5)}(\vec x)\cr
&&\qquad\ -\Psi_1(x_2,x_1 x_4)+\Psi_2(x_2,x_1 x_4) -\Psi_1(x_1
x_4,x_2)+\Psi_2(x_1 x_4,x_2)\cr
&&\qquad\ -\Psi_1(x_1 x_4,x_2)+\Psi_3(x_1 x_4,x_2)-\Psi_4 (x_1
x_3)+\Psi_5 (x_1 x_3)+\Psi_6(x_2,x_4)\cr
&&\qquad\ +\Lambda_1(\vec x)-\Lambda_2(\vec x)-\Lambda_3(\vec x)\cr
&&\qquad\ +\frac 12 \Psi^2 (x_1 x_3)+\frac 12 \Phi^2 (x_2, x_1 x_4)
+\Phi^2 (x_1 x_4, x_2) -\frac 72 \aleph^2 (\vec x)\cr
&&\qquad\  -\Psi(x_1 x_3)\aleph(\vec x) -\Phi(x_2, x_1 x_4)\aleph(\vec x)\cr
&&\qquad\ -\Phi(x_2, x_1 x_4) \Psi(x_1 x_3) -\Phi(x_2, x_1
x_4)\Phi(x_1 x_4, x_2) \ , \label{mistero}
\end{eqnarray}
with $\vec x$ expressed as a function of $\vec t$ by inverting system
(\ref{moduli}), is the part corresponding to instantonic
contributions. Following Hosono and using (\ref{simplecticform}) we
find
\begin{eqnarray}
(\partial_{t_1}- \partial_{t_3}-\partial_{t_4}) F(\vec t) =g(\vec t )\ ,\label{struttura}
\end{eqnarray}
where $F$ is the prepotential. To integrate this equation we must
expect for the prepotential to be as usual the sum of a classical
term, a cubic polynomial in $\vec t$ and a quantum instantonic
contribution. Setting
\begin{eqnarray}
q_k :=e^{2\pi i t_k}\ ,
\end{eqnarray}
we then find
\begin{eqnarray}
&& F(\vec t)=-\frac {t_4}{12}+\frac {t_4^2}{4} +\frac {t_3 t_4}2 -\frac {t_4^3}6 -\frac {t_3 t_4}{2} (t_3+t_4+2t_2)\cr
&&\qquad\quad +F_{{\rm inst}} (\vec q) +P_{{\rm class}} (t_2, t_1+t_3, t_1+t_4)
 +Q_{{\rm inst}} (q_2, q_1 q_3, q_1 q_4) \ , \label{prepotential}
\end{eqnarray}
where $F_{{\rm inst}}$ are the instantonic corrections, obtained
integrating $\phi$, $P_{{\rm class}}$ is an arbitrary cubic polynomial
of three variables and $Q_{{\rm inst}}$ an arbitrary function of three
variables which we assume analytic in $(0,0,0)$.
$P_{{\rm class}}$ and $Q_{{\rm inst}}$ represent the contributions
which are undetermined by the equation (\ref{struttura}).

As an example, let us compute the Gromov-Witten (GW) invariants up to
degree six. Here we use the Gopakumar-Vafa (G-V) reinterpretation, so
that by $GW$--invariants we mean the G-V integral invariants for
rational curves, in place of the original fractional
$GW$--invariants.\footnote{See \cite{Bouchard:2007nr}
for a nice introduction to $GW$ invariants and their interpretation
in physics and in mathematics.}
We will use $[d_1,d_2,d_3,d_4]$ to indicate the
degree of the curves in the Mori cone, corresponding to the generators
$J_1,\ldots,J_4$. Thus we consider the curves with degree
$d_1+d_2+d_3+d_4\leq 6$. The curves with degree in the integer cone
generated by $[0,1,0,0]$, $[1,0,1,0]$, $[1,0,0,1]$ must be excluded,
because corresponding to the undetermined part of the
prepotential. The only nonvanishing invariants in the considered range
are
\begin{eqnarray}
&& GW_{[0,0,0,1]}=GW_{[0,0,1,0]}=GW_{[1,0,0,0]}=\cr
&& GW_{[1,0,1,1]}=GW_{[0,1,0,1]}=GW_{[1,1,1,1]}=1;\cr
&& GW_{[0,0,1,1]}=GW_{[0,1,1,1]}=GW_{[1,1,1,2]}=-2;\cr
&& GW_{[0,1,1,2]}=GW_{[1,1,2,2]}=3;\cr
&& GW_{[0,1,2,2]}=-4; \qquad GW_{[0,1,2,3]}=5.
\end{eqnarray}


\section{The flopped resolutions}\label{flops}
In this section we study the remaining four crepant resolutions $X$ of
the orbifold $\CC^3_6$.
\subsection{Intersection theory}
For any resolution we have
\begin{eqnarray}
&& A^0(X)=\ZZ\cdot X\ , \cr
&& A^1(X)=\bigoplus_{i=1}^4\ZZ \cdot D_i\simeq \ZZ^4\ , \cr
&& A^2(X)=\ZZ \cdot C\ , \cr
&& A^3(X)=0\ .
\end{eqnarray}
In $A_2(X)$ the divisors $D_5,\ D_6,\ D_7$ can be expressed in terms
of the others
\begin{eqnarray}
&& D_5=-D_1-D_2+D_3-D_4\ , \\
&& D_6=-2 D_1+D_2-D_4\ , \\
&& D_7=2 D_1-D_2-2 D_3+D_4\ .
\end{eqnarray}
The curve $C$ depends on the resolution, as well as the intersection
product in $A^*(X)$. If we call $R$ the set of relations given by the
intersection product, we have
\begin{eqnarray}
\label{Chow_ring2}
A^*(X) =\ZZ[X,D_1,D_2,D_3,D_4,C]/R\ .
\end{eqnarray}
$A^c_*(X)$ is an $A^*(X)$-module, it is generated as group by the
compact divisor $D_7^c$, the four compact curves depending on the
resolution and the point class $P^c$. Finally the intersection
pairing $A^*(X)\otimes A_*^c(X) \rightarrow A_*^c(X)$ depends on the
resolution.

\newpage
\subsubsection*{Resolution $X=\RRC$}
\begin{figure}[hbtp]
\centering
\begin{picture}(300, 190)(-110,-90)
\put(0,0){\circle*{4}}
\put(0,-80){\circle*{4}}
\put(-80,0){\circle*{4}}
\put(-80,80){\circle*{4}}
\put(80,-80){\circle*{4}}
\put(160,-80){\circle*{4}}
\put(-80,-80){\circle*{4}}
\thicklines
\drawline(0,0)(0,-80)
\put(-0,-40){\makebox(0,0)[r]{\scriptsize{\boldmath{$C_{47}$}}}}
\drawline(-80,80)(0,0)
\put(-40,40){\makebox(0,0)[r]{\scriptsize{\boldmath{$C_{37}$}}}}
\drawline(80,-80)(0,0)
\put(40,-40){\makebox(0,0)[r]{\scriptsize{\boldmath{$C_{57}$}}}}
\drawline(-80,80)(-80,0)
\put(-80,40){\makebox(0,0)[r]{\scriptsize{\boldmath{$C_{36}$}}}}
\drawline(-80,0)(-80,-80)
\put(-80,-40){\makebox(0,0)[r]{\scriptsize{\boldmath{$C_{16}$}}}}
\drawline(-80,-80)(0,-80)
\put(-40,-80){\makebox(0,-2)[t]{\scriptsize{\boldmath{$C_{14}$}}}}
\drawline(0,-80)(80,-80)
\put(40,-80){\makebox(0,-2)[t]{\scriptsize{\boldmath{$C_{45}$}}}}
\drawline(80,-80)(160,-80)
\put(120,-80){\makebox(0,-2)[t]{\scriptsize{\boldmath{$C_{25}$}}}}
\drawline(-80,0)(0,0)
\put(-40,0){\makebox(0,-2)[t]{\scriptsize{\boldmath{$C_{67}$}}}}
\drawline(-80,80)(160,-80)
\put(60,0){\makebox(5,-8)[r]{\scriptsize{\boldmath{$C_{23}$}}}}
\drawline(0,0)(160,-80)
\put(80,-40){\makebox(0,0)[r]{\scriptsize{\boldmath{$C_{27}$}}}}
\drawline(-80,-80)(0,0)
\put(-43,-40){\makebox(0,0)[r]{\scriptsize{\boldmath{$C_{17}$}}}}
\put(-83,6){\makebox(0, 0)[r]{\scriptsize{\boldmath{$D_6$}}}}
\put(-83,-86){\makebox(0, 0)[r]{\scriptsize{\boldmath{$D_1$}}}}
\put(-83,86){\makebox(0, 0)[r]{\scriptsize{\boldmath{$D_3$}}}}
\put(-15,6){\makebox(0, 0)[l]{\scriptsize{\boldmath{$D_7$}}}}
\put(4,-86){\makebox(0, 0)[r]{\scriptsize{\boldmath{$D_4$}}}}
\put(84,-86){\makebox(0, 0)[r]{\scriptsize{\boldmath{$D_5$}}}}
\put(164,-86){\makebox(0, 0)[r]{\scriptsize{\boldmath{$D_2$}}}}
\put(90,-61){\makebox(0, 0)[c]{\scriptsize{\boldmath{$P_3$}}}}
\put(-25,-61){\makebox(0, 0)[c]{\scriptsize{\boldmath{$P_1$}}}}
\put(-55,20){\makebox(0, 0)[c]{\scriptsize{\boldmath{$P_5$}}}}
\put(-55,-25){\makebox(0, 0)[c]{\scriptsize{\boldmath{$P_6$}}}}
\put(25,-61){\makebox(0, 0)[c]{\scriptsize{\boldmath{$P_2$}}}}
\put(18,6){\makebox(0, 0)[r]{\scriptsize{\boldmath{$P_4$}}}}
\end{picture}
\caption{}
\label{fig:11}
\end{figure}
This resolution differs from the $G$-Hilb by the flop
\begin{eqnarray}
& C_{46} \longrightarrow C_{17}\ .
\end{eqnarray}
We define $C=[C_{14}]$ and we report the relations between any other
toric curve and $C$ in the decorated fan of figure \ref{fig:12}.
\begin{figure}[hbtp]
\centering
\begin{picture}(300,180)(-110,-90)
\put(0,0){\circle*{4}}
\put(0,-80){\circle*{4}}
\put(-80,0){\circle*{4}}
\put(-80,80){\circle*{4}}
\put(80,-80){\circle*{4}}
\put(160,-80){\circle*{4}}
\put(-80,-80){\circle*{4}}
\thicklines
\drawline(0,0)(0,-80)
\put(-2,-40){\makebox(0,0)[r]{\scriptsize{\boldmath{$0$}}}}
\drawline(-80,80)(0,0)
\put(-42,40){\makebox(0,0)[r]{\scriptsize{\boldmath{$-3C$}}}}
\drawline(80,-80)(0,0)
\put(38,-40){\makebox(0,0)[r]{\scriptsize{\boldmath{$0$}}}}
\drawline(-80,80)(-80,0)
\put(-80,40){\makebox(0,0)[r]{\scriptsize{\boldmath{$C$}}}}
\drawline(-80,0)(-80,-80)
\put(-80,-40){\makebox(0,0)[r]{\scriptsize{\boldmath{$C$}}}}
\drawline(-80,-80)(0,-80)
\put(-40,-80){\makebox(0,-2)[t]{\scriptsize{\boldmath{$C$}}}}
\drawline(0,-80)(80,-80)
\put(40,-80){\makebox(0,-2)[t]{\scriptsize{\boldmath{$C$}}}}
\drawline(80,-80)(160,-80)
\put(120,-80){\makebox(0,-2)[t]{\scriptsize{\boldmath{$C$}}}}
\drawline(-80,0)(0,0)
\put(-40,0){\makebox(0,-2)[t]{\scriptsize{\boldmath{$0$}}}}
\drawline(-80,80)(160,-80)
\put(54,0){\makebox(5,-8)[r]{\scriptsize{\boldmath{$C$}}}}
\drawline(0,0)(160,-80)
\put(76,-40){\makebox(0,0)[r]{\scriptsize{\boldmath{$-2C$}}}}
\drawline(-80,-80)(0,0)
\put(-43,-40){\makebox(0,0)[r]{\scriptsize{\boldmath{$-C$}}}}
\end{picture}
\caption{}
\label{fig:12}
\end{figure}\\
In table \ref{tab:divdivCR2} we summarize the intersection products,
which give the relations $R$ in the Chow ring
$A^*(X)=\ZZ[X,D_1,D_2,D_3,D_4,C]/R$.\\
\begin{table}[hbtp]
\begin{center}
\begin{eqnarray*}
\begin{array}{c|cccccc}
\phantom{x} & X & D_1 & D_2 & D_3 & D_4 & C\\
\hline
X   & X   & D_1 & D_2 & D_3 & D_4 & C\\
D_1 & D_1 & C   & 0   & 0   & -C  & 0\\
D_2 & D_2 & 0   & 0   & -C  & 0   & 0\\
D_3 & D_3 & 0   & -C  & -C  & 0   & 0\\
D_4 & D_4 & -C  & 0   & 0   &2C   & 0\\
C   & C   & 0   & 0   & 0   & 0   & 0\\
\end{array}
\end{eqnarray*}
\end{center}
\caption{Intersection product in $A^*(X)$}
\label{tab:divdivCR2}
\end{table}\\
The group of compact subvarieties of $X$ is
\begin{eqnarray}
\label{CompactChow_groupR2}
&A_*^c(X) =\ZZ \cdot D_7^c \oplus \ZZ \cdot C_{17}^c \oplus
\ZZ \cdot C_{57}^c \oplus \ZZ \cdot C_{67}^c \oplus \ZZ \cdot C_{47}^c
\oplus \ZZ \cdot P^c\ ,
\end{eqnarray}
with relations to other compact curves
\begin{eqnarray}
\label{relcaurveH1R2}
&& C_{27}^c=2C_{17}+C_{67}^c+C_{47}^c\ ,\\
\label{relcaurveH2R2}
&& C_{37}^c=3C_{17}+C_{57}^c+C_{67}^c+2C_{47}^c\ .
\end{eqnarray}
In table \ref{tab:divdivIPR2} we summarize the intersection pairing.\\
\begin{table}[hbtp]
\begin{center}
\begin{eqnarray*}
\begin{array}{c|cccccc}
\phantom{x} & D_7^c    & C_{17}^c & C_{57}^c & C_{67}^c & C_{47}^c & P^c\\
\hline
    X       & D_7^c    & C_{17}^c & C_{57}^c & C_{67}^c & C_{47}^c & P^c\\
    D_1     & C_{17}^c & -P^c     & 0        & P^c      & P^c      & 0\\
    D_2     & C_{27}^c & 0        & P^c      & 0        & 0        & 0\\
    D_3     & C_{37}^c & 0        & 0        & P^c      & 0        & 0\\
    D_4     & C_{47}^c & P^c      & P^c      & 0        & -2P^c    & 0\\
    C       & P^c      & 0        & 0        & 0        & 0        & 0\\
\end{array}
\end{eqnarray*}
\end{center}
\caption{Intersection pairing $A^*(X)\otimes A_*^c(X) \rightarrow A_*^c(X)$}
\label{tab:divdivIPR2}
\end{table}\\
The Mori cone generators are $C_a$, $a=1,\ldots,4$, with
\begin{eqnarray}
C_1=C_{17}\ ,\qquad C_2=C_{57}\ ,\qquad C_3=C_{67}\ ,\qquad C_4=C_{47}\ .
\end{eqnarray}
The K\"ahler cone is generated by the dual elements $T_a$,
$a=1,\ldots,4$ with
\begin{eqnarray}
&&T_1=-2D_1+D_2+2D_3-D_4\ ,\qquad T_2=D_2\ ,\cr
&&T_3=D_3\ ,\qquad T_4=-D_1+D_2+D_3-D_4\ .
\end{eqnarray}
\begin{table}[hbtp]
\begin{center}
\begin{eqnarray*}
\begin{array}{c|cccc}
\phantom\vline & T_1 & T_2 & T_3 & T_4  \\
\hline
T_1  &  6C  &  2C  &  3C  &  4C  \\
T_2  &  2C  &   0  &   C  &   C  \\
T_3  &  3C  &   C  &   C  &  2C  \\
T_4  &  4C  &   C  &  2C  &  2C  \\
\end{array}
\end{eqnarray*}
\end{center}
\caption{Intersection between K\"ahler generators}
\label{2tab:morikahler}
\end{table}
\\
If we call $J_a$ the K\"ahler generators in $H^2(X,\QQ)$
corresponding to the $T_a$ then the cohomology ring is
\begin{eqnarray}
H^*(X,\QQ)=\QQ[J_1,\ldots,J_4]/\sim \ ,
\end{eqnarray}
with $\sim$ given by table \ref{2tab:morikahler}.

\subsubsection*{Resolution $X=\RSC$}
\begin{figure}[hbtp]
\centering
\begin{picture}(300, 190)(-110,-90)
\put(0,0){\circle*{4}}
\put(0,-80){\circle*{4}}
\put(-80,0){\circle*{4}}
\put(-80,80){\circle*{4}}
\put(80,-80){\circle*{4}}
\put(160,-80){\circle*{4}}
\put(-80,-80){\circle*{4}}
\thicklines
\drawline(0,0)(0,-80)
\put(-0,-40){\makebox(0,0)[r]{\scriptsize{\boldmath{$C_{47}$}}}}
\drawline(-80,80)(0,0)
\put(-40,40){\makebox(0,0)[r]{\scriptsize{\boldmath{$C_{37}$}}}}
\drawline(80,-80)(0,0)
\put(40,-40){\makebox(0,0)[r]{\scriptsize{\boldmath{$C_{57}$}}}}
\drawline(-80,80)(-80,0)
\put(-80,40){\makebox(0,0)[r]{\scriptsize{\boldmath{$C_{36}$}}}}
\drawline(-80,0)(-80,-80)
\put(-80,-40){\makebox(0,0)[r]{\scriptsize{\boldmath{$C_{16}$}}}}
\drawline(-80,-80)(0,-80)
\put(-40,-80){\makebox(0,-2)[t]{\scriptsize{\boldmath{$C_{14}$}}}}
\drawline(0,-80)(80,-80)
\put(40,-80){\makebox(0,-2)[t]{\scriptsize{\boldmath{$C_{45}$}}}}
\drawline(80,-80)(160,-80)
\put(120,-80){\makebox(0,-2)[t]{\scriptsize{\boldmath{$C_{25}$}}}}
\drawline(0,-80)(-80,80)
\put(-42,0){\makebox(0,0)[r]{\scriptsize{\boldmath{$C_{34}$}}}}
\drawline(-80,80)(160,-80)
\put(60,0){\makebox(5,-8)[r]{\scriptsize{\boldmath{$C_{23}$}}}}
\drawline(0,0)(160,-80)
\put(80,-40){\makebox(0,0)[r]{\scriptsize{\boldmath{$C_{27}$}}}}
\drawline(-80,0)(0,-80)
\put(-40,-40){\makebox(0,0)[r]{\scriptsize{\boldmath{$C_{46}$}}}}
\put(-83,6){\makebox(0, 0)[r]{\scriptsize{\boldmath{$D_6$}}}}
\put(-83,-86){\makebox(0, 0)[r]{\scriptsize{\boldmath{$D_1$}}}}
\put(-83,86){\makebox(0, 0)[r]{\scriptsize{\boldmath{$D_3$}}}}
\put(-9,-3){\makebox(0, 0)[l]{\scriptsize{\boldmath{$D_7$}}}}
\put(4,-86){\makebox(0, 0)[r]{\scriptsize{\boldmath{$D_4$}}}}
\put(84,-86){\makebox(0, 0)[r]{\scriptsize{\boldmath{$D_5$}}}}
\put(164,-86){\makebox(0, 0)[r]{\scriptsize{\boldmath{$D_2$}}}}
\put(90,-61){\makebox(0, 0)[c]{\scriptsize{\boldmath{$P_5$}}}}
\put(-55,-61){\makebox(0, 0)[c]{\scriptsize{\boldmath{$P_1$}}}}
\put(-65,10){\makebox(0, 0)[c]{\scriptsize{\boldmath{$P_2$}}}}
\put(-25,0){\makebox(0, 0)[c]{\scriptsize{\boldmath{$P_3$}}}}
\put(25,-61){\makebox(0, 0)[c]{\scriptsize{\boldmath{$P_4$}}}}
\put(18,6){\makebox(0, 0)[r]{\scriptsize{\boldmath{$P_6$}}}}
\end{picture}
\caption{}
\label{fig:13}
\end{figure}
This resolution differs from the $G$-Hilb by the flop
\begin{eqnarray}
& C_{67} \longrightarrow C_{34}\ .
\end{eqnarray}
We set $C=[C_{34}]$ and we report the relations between any other
toric curve and $C$ in the decorated fan of figure \ref{fig:14}.\\
\begin{figure}[hbtp]
\centering
\begin{picture}(300,180)(-110,-90)
\put(0,0){\circle*{4}}
\put(0,-80){\circle*{4}}
\put(-80,0){\circle*{4}}
\put(-80,80){\circle*{4}}
\put(80,-80){\circle*{4}}
\put(160,-80){\circle*{4}}
\put(-80,-80){\circle*{4}}
\thicklines
\drawline(0,0)(0,-80)
\put(-0,-40){\makebox(0,0)[r]{\scriptsize{\boldmath{$-2C$}}}}
\drawline(-80,80)(0,0)
\put(-42,40){\makebox(0,0)[r]{\scriptsize{\boldmath{$-4C$}}}}
\drawline(80,-80)(0,0)
\put(38,-40){\makebox(0,0)[r]{\scriptsize{\boldmath{$0$}}}}
\drawline(-80,80)(-80,0)
\put(-82,40){\makebox(0,0)[r]{\scriptsize{\boldmath{$0$}}}}
\drawline(-80,0)(-80,-80)
\put(-82,-40){\makebox(0,0)[r]{\scriptsize{\boldmath{$0$}}}}
\drawline(-80,-80)(0,-80)
\put(-40,-80){\makebox(0,-2)[t]{\scriptsize{\boldmath{$0$}}}}
\drawline(0,-80)(80,-80)
\put(40,-80){\makebox(0,-2)[t]{\scriptsize{\boldmath{$C$}}}}
\drawline(80,-80)(160,-80)
\put(120,-80){\makebox(0,-2)[t]{\scriptsize{\boldmath{$C$}}}}
\drawline(0,-80)(-80,80)
\put(-42,0){\makebox(0,0)[r]{\scriptsize{\boldmath{$C$}}}}
\drawline(-80,80)(160,-80)
\put(54,0){\makebox(5,-8)[r]{\scriptsize{\boldmath{$C$}}}}
\drawline(0,0)(160,-80)
\put(76,-40){\makebox(0,0)[r]{\scriptsize{\boldmath{$-2C$}}}}
\drawline(-80,0)(0,-80)
\put(-42,-42){\makebox(0,0)[r]{\scriptsize{\boldmath{$0$}}}}
\end{picture}
\caption{}
\label{fig:14}
\end{figure}
\newpage\noindent
In table \ref{tab:divdivCR3} we summarize the intersection products,
which give the relations $R$ in the Chow ring
$A^*(X)=\ZZ[X,D_1,D_2,D_3,D_4,C]/R$.\\
\begin{table}[hbtp]
\begin{center}
\begin{eqnarray*}
\begin{array}{c|cccccc}
\phantom{x} & X & D_1 & D_2 & D_3 & D_4 & C\\
\hline
X   & X   & D_1 & D_2 & D_3 & D_4 & C\\
D_1 & D_1 & 0   & 0   & 0   & 0   & 0\\
D_2 & D_2 & 0   & 0   & C   & 0   & 0\\
D_3 & D_3 & 0   & C   & 2C  & C   & 0\\
D_4 & D_4 & 0   & 0   & C   & 0   & 0\\
C   & C   & 0   & 0   & 0   & 0   & 0\\
\end{array}
\end{eqnarray*}
\end{center}
\caption{Intersection product in $A^*(X)$}
\label{tab:divdivCR3}
\end{table}\\
The group of compact subvarieties of $X$ is
\begin{eqnarray}
\label{CompactChow_groupR3}
&A_*^c(X) =\ZZ \cdot D_7^c \oplus \ZZ \cdot C_{46}^c \oplus
\ZZ \cdot C_{57}^c \oplus \ZZ \cdot C_{67}^c \oplus \ZZ \cdot C_{47}^c
\oplus \ZZ \cdot P^c\ ,
\end{eqnarray}
with relations to other compact curves
\begin{eqnarray}
\label{relcaurveH1R3}
&& C_{27}^c=C_{47}^c\ ,\\
\label{relcaurveH2R3}
&& C_{37}^c=C_{57}+2C_{47}^c\ .
\end{eqnarray}
In table \ref{tab:divdivIPR3} we summarize the intersection pairing.\\
\begin{table}[hbtp]
\begin{center}
\begin{eqnarray*}
\begin{array}{c|cccccc}
\phantom{x} & D_7^c    & C_{46}^c & C_{57}^c & C_{34}^c & C_{47}^c & P^c\\
\hline
    X       & D_7^c    & C_{46}^c & C_{57}^c & C_{34}^c & C_{47}^c & P^c\\
    D_1     & 0        & P^c      & 0        & 0        & 0        & 0\\
    D_2     & C_{27}^c & 0        & P^c      & 0        & 0        & 0\\
    D_3     & C_{37}^c & P^c      & 0        & -P^c     & P^c      & 0\\
    D_4     & C_{47}^c & 0        & P^c      & -P^c     & 0        & 0\\
    C       & P^c      & 0        & 0        & 0        & 0        & 0\\
\end{array}
\end{eqnarray*}
\end{center}
\caption{Intersection pairing $A^*(X)\otimes A_*^c(X) \rightarrow A_*^c(X)$}
\label{tab:divdivIPR3}
\end{table}
\newpage\noindent
The Mori cone generators are $C_a$, $a=1,\ldots,4$, with
\begin{eqnarray}
C_1=C_{46}\ ,\qquad C_2=C_{57}\ ,\qquad C_3=C_{34}\ ,\qquad C_4=C_{47}\ .
\end{eqnarray}
The K\"ahler cone is generated by the dual elements $T_a$,
$a=1,\ldots,4$ with
\begin{eqnarray}
&&T_1=D_1\ ,\qquad T_2=D_2\ , \cr
&&T_3=D_2-D_4\ ,\qquad T_4=-D_1+D_2+D_3-D_4\ .
\end{eqnarray}
\begin{table}[hbtp]
\begin{center}
\begin{eqnarray*}
\begin{array}{c|cccc}
\phantom\vline & T_1 & T_2 & T_3 & T_4  \\
\hline
T_1  &   0  &   0  &   0  &   0  \\
T_2  &   0  &   0  &   0  &   C  \\
T_3  &   0  &   0  &   0  &   0  \\
T_4  &   0  &   C  &   0  &  2C  \\
\end{array}
\end{eqnarray*}
\end{center}
\caption{Intersection between K\"ahler generators}
 \label{3tab:morikahler}
\end{table}
\\
If $J_a$ are the K\"ahler generators in $H^2(X,\QQ)$
corresponding to the $T_a$ then the cohomology ring is
\begin{eqnarray}
H^*(X,\QQ)=\QQ[J_1,\ldots,J_4]/\sim \ ,
\end{eqnarray}
with $\sim$ given by table \ref{3tab:morikahler}.

\newpage
\subsubsection*{Resolution $X=\RTC$}
\begin{figure}[hbtp]
\centering
\begin{picture}(300, 190)(-110,-90)
\put(0,0){\circle*{4}}
\put(0,-80){\circle*{4}}
\put(-80,0){\circle*{4}}
\put(-80,80){\circle*{4}}
\put(80,-80){\circle*{4}}
\put(160,-80){\circle*{4}}
\put(-80,-80){\circle*{4}}
\thicklines
\drawline(80,-80)(-80,0)
\put(-2,-40){\makebox(0,0)[r]{\scriptsize{\boldmath{$C_{56}$}}}}
\drawline(-80,80)(0,0)
\put(-40,40){\makebox(0,0)[r]{\scriptsize{\boldmath{$C_{37}$}}}}
\drawline(80,-80)(0,0)
\put(40,-40){\makebox(0,0)[r]{\scriptsize{\boldmath{$C_{57}$}}}}
\drawline(-80,80)(-80,0)
\put(-80,40){\makebox(0,0)[r]{\scriptsize{\boldmath{$C_{36}$}}}}
\drawline(-80,0)(-80,-80)
\put(-80,-40){\makebox(0,0)[r]{\scriptsize{\boldmath{$C_{16}$}}}}
\drawline(-80,-80)(0,-80)
\put(-40,-80){\makebox(0,-2)[t]{\scriptsize{\boldmath{$C_{14}$}}}}
\drawline(0,-80)(80,-80)
\put(40,-80){\makebox(0,-2)[t]{\scriptsize{\boldmath{$C_{45}$}}}}
\drawline(80,-80)(160,-80)
\put(120,-80){\makebox(0,-2)[t]{\scriptsize{\boldmath{$C_{25}$}}}}
\drawline(-80,0)(0,0)
\put(-40,0){\makebox(0,-2)[t]{\scriptsize{\boldmath{$C_{67}$}}}}
\drawline(-80,80)(160,-80)
\put(60,0){\makebox(5,-8)[r]{\scriptsize{\boldmath{$C_{23}$}}}}
\drawline(0,0)(160,-80)
\put(76,-40){\makebox(0,0)[r]{\scriptsize{\boldmath{$C_{27}$}}}}
\drawline(-80,0)(0,-80)
\put(-40,-40){\makebox(0,0)[r]{\scriptsize{\boldmath{$C_{46}$}}}}
\put(-83,6){\makebox(0, 0)[r]{\scriptsize{\boldmath{$D_6$}}}}
\put(-83,-86){\makebox(0, 0)[r]{\scriptsize{\boldmath{$D_1$}}}}
\put(-83,86){\makebox(0, 0)[r]{\scriptsize{\boldmath{$D_3$}}}}
\put(-2,-2){\makebox(0, 0)[t]{\scriptsize{\boldmath{$D_7$}}}}
\put(4,-86){\makebox(0, 0)[r]{\scriptsize{\boldmath{$D_4$}}}}
\put(84,-86){\makebox(0, 0)[r]{\scriptsize{\boldmath{$D_5$}}}}
\put(164,-86){\makebox(0, 0)[r]{\scriptsize{\boldmath{$D_2$}}}}
\put(90,-61){\makebox(0, 0)[c]{\scriptsize{\boldmath{$P_4$}}}}
\put(-55,-61){\makebox(0, 0)[c]{\scriptsize{\boldmath{$P_1$}}}}
\put(-55,20){\makebox(0, 0)[c]{\scriptsize{\boldmath{$P_6$}}}}
\put(0,-25){\makebox(0, 0)[c]{\scriptsize{\boldmath{$P_3$}}}}
\put(0,-61){\makebox(0, 0)[c]{\scriptsize{\boldmath{$P_2$}}}}
\put(18,6){\makebox(0, 0)[r]{\scriptsize{\boldmath{$P_5$}}}}
\end{picture}
\caption{}
\label{fig:15}
\end{figure}
This resolution differs from the $G$-Hilb by the flop
\begin{eqnarray}
& C_{47} \longrightarrow C_{56}\ .
\end{eqnarray}
We set $C=[C_{56}]$  and we report the relations between any other
toric curve and $C$ in the decorated fan of figure \ref{fig:16}.\\
\begin{figure}[hbtp]
\centering
\begin{picture}(300,180)(-110,-90)
\put(0,0){\circle*{4}}
\put(0,-80){\circle*{4}}
\put(-80,0){\circle*{4}}
\put(-80,80){\circle*{4}}
\put(80,-80){\circle*{4}}
\put(160,-80){\circle*{4}}
\put(-80,-80){\circle*{4}}
\thicklines
\drawline(80,-80)(-80,0)
\put(-2,-42){\makebox(0,0)[r]{\scriptsize{\boldmath{$C$}}}}
\drawline(-80,80)(0,0)
\put(-42,38){\makebox(0,0)[r]{\scriptsize{\boldmath{$-3C$}}}}
\drawline(80,-80)(0,0)
\put(38,-42){\makebox(0,0)[r]{\scriptsize{\boldmath{$-C$}}}}
\drawline(-80,80)(-80,0)
\put(-82,40){\makebox(0,0)[r]{\scriptsize{\boldmath{$C$}}}}
\drawline(-80,0)(-80,-80)
\put(-82,-40){\makebox(0,0)[r]{\scriptsize{\boldmath{$0$}}}}
\drawline(-80,-80)(0,-80)
\put(-40,-80){\makebox(0,-2)[t]{\scriptsize{\boldmath{$0$}}}}
\drawline(0,-80)(80,-80)
\put(40,-80){\makebox(0,-2)[t]{\scriptsize{\boldmath{$0$}}}}
\drawline(80,-80)(160,-80)
\put(120,-80){\makebox(0,-2)[t]{\scriptsize{\boldmath{$C$}}}}
\drawline(-80,0)(0,0)
\put(-40,0){\makebox(0,-2)[t]{\scriptsize{\boldmath{$-2C$}}}}
\drawline(-80,80)(160,-80)
\put(54,0){\makebox(5,-8)[r]{\scriptsize{\boldmath{$C$}}}}
\drawline(0,0)(160,-80)
\put(76,-40){\makebox(0,0)[r]{\scriptsize{\boldmath{$-2C$}}}}
\drawline(-80,0)(0,-80)
\put(-42,-42){\makebox(0,0)[r]{\scriptsize{\boldmath{$0$}}}}
\end{picture}
\caption{}
\label{fig:16}
\end{figure}\\
In table \ref{tab:divdivCR3} we summarize the intersection products,
which give the relations $R$ in the Chow ring
$A^*(X)=\ZZ[X,D_1,D_2,D_3,D_4,C]/R$.\\
\begin{table}[hbtp]
\begin{center}
\begin{eqnarray*}
\begin{array}{c|cccccc}
\phantom{x} & X & D_1 & D_2 & D_3 & D_4 & C\\
\hline
X   & X   & D_1 & D_2 & D_3 & D_4 & C\\
D_1 & D_1 & 0   & 0   & 0   & 0   & 0\\
D_2 & D_2 & 0   & 0   & C   & 0   & 0\\
D_3 & D_3 & 0   & C   & C   & 0   & 0\\
D_4 & D_4 & 0   & 0   & 0   & 0   & 0\\
C   & C   & 0   & 0   & 0   & 0   & 0\\
\end{array}
\end{eqnarray*}
\end{center}
\caption{Intersection product in $A^*(X)$}
\label{tab:divdivCR4}
\end{table}\\
The group of compact subvarieties of $X$ is
\begin{eqnarray}
\label{CompactChow_groupR4}
&A_*^c(X) =\ZZ \cdot D_7^c \oplus \ZZ \cdot C_{46}^c \oplus
\ZZ \cdot C_{57}^c \oplus \ZZ \cdot C_{67}^c \oplus \ZZ \cdot C_{56}^c
\oplus \ZZ \cdot P^c\ ,
\end{eqnarray}
with relations to other compact curves
\begin{eqnarray}
\label{relcaurveH1R4}
&& C_{27}^c=C_{67}^c\ ,\\
\label{relcaurveH2R4}
&& C_{37}^c=C_{57}^c+C_{67}^c\ .
\end{eqnarray}
In table \ref{tab:divdivIPR4} we summarize the intersection pairing.\\
\begin{table}[hbtp]
\begin{center}
\begin{eqnarray}
\begin{array}{c|cccccc}
\phantom{x} & D_7^c    & C_{46}^c & C_{57}^c & C_{67}^c & C_{56}^c & P^c\\
\hline
    X       & D_7^c    & C_{46}^c & C_{57}^c & C_{67}^c & C_{56}^c & P^c\\
    D_1     & 0        & P^c      & 0        & 0        & 0        & 0\\
    D_2     & C_{27}^c & 0        & P^c      & 0        & 0        & 0\\
    D_3     & C_{37}^c & 0        & 0        & P^c      & 0        & 0\\
    D_4     & 0        & -2P^c    & 0        & 0        & P^c      & 0\\
    C       & P^c      & 0        & 0        & 0        & 0        & 0\\
\end{array}
\end{eqnarray}
\end{center}
\caption{Intersection pairing $A^*(X)\otimes A_*^c(X) \rightarrow A_*^c(X)$}
\label{tab:divdivIPR4}
\end{table}\\
The Mori cone generators are $C_a$, $a=1,\ldots,4$, with
\begin{eqnarray}
C_1=C_{46}\ ,\qquad C_2=C_{57}\ ,\qquad C_3=C_{67}\ ,\qquad C_4=C_{56}\ .
\end{eqnarray}
The K\"ahler cone is generated by the dual elements $T_a$,
$a=1,\ldots,4$ with
\begin{eqnarray}
&T_1=D_1\ ,\qquad T_2=D_2\ , \qquad
T_3=D_3\ ,\qquad T_4=2D_1+D_4\ .
\end{eqnarray}
\begin{table}[hbtp]
\begin{center}
\begin{eqnarray*}
\begin{array}{c|cccc}
\phantom\vline & T_1 & T_2 & T_3 & T_4  \\
\hline
T_1  &   0  &   0  &   0  &   0  \\
T_2  &   0  &   0  &   C  &   0  \\
T_3  &   0  &   C  &   C  &   0  \\
T_4  &   0  &   0  &   0  &   0  \\
\end{array}
\end{eqnarray*}
\end{center}
\caption{Intersection between K\"ahler generators}
 \label{4tab:morikahler}
\end{table}
\\
If $J_a$ are the K\"ahler generators in $H^2(X,\QQ)$
corresponding to the $T_a$ then the cohomology ring is
\begin{eqnarray}
H^*(X,\QQ)=\QQ[J_1,\ldots,J_4]/\sim \ ,
\end{eqnarray}
with $\sim$ given by table \ref{4tab:morikahler}.

\subsubsection*{Resolution $X=\RUC$}
\begin{figure}[hbtp]
\centering
\begin{picture}(300, 190)(-110,-90)
\put(0,0){\circle*{4}}
\put(0,-80){\circle*{4}}
\put(-80,0){\circle*{4}}
\put(-80,80){\circle*{4}}
\put(80,-80){\circle*{4}}
\put(160,-80){\circle*{4}}
\put(-80,-80){\circle*{4}}
\thicklines
\drawline(80,-80)(-80,0)
\put(-2,-40){\makebox(0,0)[r]{\scriptsize{\boldmath{$C_{56}$}}}}
\drawline(-80,80)(0,0)
\put(-40,40){\makebox(0,0)[r]{\scriptsize{\boldmath{$C_{37}$}}}}
\drawline(160,-80)(-80,0)
\put(32,-40){\makebox(0,0)[r]{\scriptsize{\boldmath{$C_{26}$}}}}
\drawline(-80,80)(-80,0)
\put(-80,40){\makebox(0,0)[r]{\scriptsize{\boldmath{$C_{36}$}}}}
\drawline(-80,0)(-80,-80)
\put(-80,-40){\makebox(0,0)[r]{\scriptsize{\boldmath{$C_{16}$}}}}
\drawline(-80,-80)(0,-80)
\put(-40,-80){\makebox(0,-2)[t]{\scriptsize{\boldmath{$C_{14}$}}}}
\drawline(0,-80)(80,-80)
\put(40,-80){\makebox(0,-2)[t]{\scriptsize{\boldmath{$C_{45}$}}}}
\drawline(80,-80)(160,-80)
\put(120,-80){\makebox(0,-2)[t]{\scriptsize{\boldmath{$C_{25}$}}}}
\drawline(-80,0)(0,0)
\put(-40,0){\makebox(0,-2)[t]{\scriptsize{\boldmath{$C_{67}$}}}}
\drawline(-80,80)(160,-80)
\put(60,0){\makebox(5,-8)[r]{\scriptsize{\boldmath{$C_{23}$}}}}
\drawline(0,0)(160,-80)
\put(76,-40){\makebox(0,0)[r]{\scriptsize{\boldmath{$C_{27}$}}}}
\drawline(-80,0)(0,-80)
\put(-40,-40){\makebox(0,0)[r]{\scriptsize{\boldmath{$C_{46}$}}}}
\put(-83,6){\makebox(0, 0)[r]{\scriptsize{\boldmath{$D_6$}}}}
\put(-83,-86){\makebox(0, 0)[r]{\scriptsize{\boldmath{$D_1$}}}}
\put(-83,86){\makebox(0, 0)[r]{\scriptsize{\boldmath{$D_3$}}}}
\put(-2,-2){\makebox(0, 0)[t]{\scriptsize{\boldmath{$D_7$}}}}
\put(4,-86){\makebox(0, 0)[r]{\scriptsize{\boldmath{$D_4$}}}}
\put(84,-86){\makebox(0, 0)[r]{\scriptsize{\boldmath{$D_5$}}}}
\put(164,-86){\makebox(0, 0)[r]{\scriptsize{\boldmath{$D_2$}}}}
\put(70,-61){\makebox(0, 0)[c]{\scriptsize{\boldmath{$P_4$}}}}
\put(-55,-61){\makebox(0, 0)[c]{\scriptsize{\boldmath{$P_1$}}}}
\put(-55,20){\makebox(0, 0)[c]{\scriptsize{\boldmath{$P_6$}}}}
\put(25,-25){\makebox(0, 0)[c]{\scriptsize{\boldmath{$P_3$}}}}
\put(0,-61){\makebox(0, 0)[c]{\scriptsize{\boldmath{$P_2$}}}}
\put(18,6){\makebox(0, 0)[r]{\scriptsize{\boldmath{$P_5$}}}}
\end{picture}
\caption{}
\label{fig:17}
\end{figure}
This resolution differs from the $G$-Hilb by the flops
\begin{eqnarray}
& C_{47} \longrightarrow C_{56}\ ,\qquad \qquad
C_{57} \longrightarrow C_{26}\ .
\end{eqnarray}
We set $C=[C_{56}]$ and we report the relations between any other
toric curve and $C$ in the decorated fan of figure \ref{fig:14}.\\
\begin{figure}[hbtp]
\centering
\begin{picture}(300,180)(-110,-90)
\put(0,0){\circle*{4}}
\put(0,-80){\circle*{4}}
\put(-80,0){\circle*{4}}
\put(-80,80){\circle*{4}}
\put(80,-80){\circle*{4}}
\put(160,-80){\circle*{4}}
\put(-80,-80){\circle*{4}}
\thicklines
\drawline(80,-80)(-80,0)
\put(-2,-42){\makebox(0,0)[r]{\scriptsize{\boldmath{$0$}}}}
\drawline(-80,80)(0,0)
\put(-42,38){\makebox(0,0)[r]{\scriptsize{\boldmath{$-3C$}}}}
\drawline(160,-80)(-80,0)
\put(32,-42){\makebox(0,0)[r]{\scriptsize{\boldmath{$C$}}}}
\drawline(-80,80)(-80,0)
\put(-82,40){\makebox(0,0)[r]{\scriptsize{\boldmath{$C$}}}}
\drawline(-80,0)(-80,-80)
\put(-82,-40){\makebox(0,0)[r]{\scriptsize{\boldmath{$0$}}}}
\drawline(-80,-80)(0,-80)
\put(-40,-80){\makebox(0,-2)[t]{\scriptsize{\boldmath{$0$}}}}
\drawline(0,-80)(80,-80)
\put(40,-80){\makebox(0,-2)[t]{\scriptsize{\boldmath{$0$}}}}
\drawline(80,-80)(160,-80)
\put(120,-80){\makebox(0,-2)[t]{\scriptsize{\boldmath{$0$}}}}
\drawline(-80,0)(0,0)
\put(-40,0){\makebox(0,-2)[t]{\scriptsize{\boldmath{$-3C$}}}}
\drawline(-80,80)(160,-80)
\put(54,0){\makebox(5,-8)[r]{\scriptsize{\boldmath{$C$}}}}
\drawline(0,0)(160,-80)
\put(76,-40){\makebox(0,0)[r]{\scriptsize{\boldmath{$-3C$}}}}
\drawline(-80,0)(0,-80)
\put(-42,-42){\makebox(0,0)[r]{\scriptsize{\boldmath{$0$}}}}
\end{picture}
\caption{}
\label{fig:18}
\end{figure}
\newpage\noindent
In table \ref{tab:divdivCR5} we summarize the intersection products,
which give the relations $R$ in the Chow ring
$A^*(X)=\ZZ[X,D_1,D_2,D_3,D_4,C]/R$.\\
\begin{table}[hbtp]
\begin{center}
\begin{eqnarray}
\begin{array}{c|cccccc}
\phantom{x} & X & D_1 & D_2 & D_3 & D_4 & C\\
\hline
X   & X   & D_1 & D_2 & D_3 & D_4 & C\\
D_1 & D_1 & 0   & 0   & 0   & 0   & 0\\
D_2 & D_2 & 0   & C   & C   & 0   & 0\\
D_3 & D_3 & 0   & C   & C   & 0   & 0\\
D_4 & D_4 & 0   & 0   & 0   & 0   & 0\\
C   & C   & 0   & 0   & 0   & 0   & 0\\
\end{array}
\end{eqnarray}
\end{center}
\caption{Intersection product in $A^*(X)$}
\label{tab:divdivCR5}
\end{table}\\
The group of compact subvarieties of $X$ is
\begin{eqnarray}
\label{CompactChow_groupR5}
&A_*^c(X) =\ZZ \cdot D_7^c \oplus \ZZ \cdot C_{46}^c \oplus
\ZZ \cdot C_{26}^c \oplus \ZZ \cdot C_{67}^c \oplus \ZZ \cdot C_{56}^c
\oplus \ZZ \cdot P^c\ ,
\end{eqnarray}
with relations to other compact curves
\begin{eqnarray}
\label{relcaurveH1R5}
&& C_{27}^c=C_{67}^c\ ,\\
\label{relcaurveH2R5}
&& C_{37}^c=C_{67}^c\ .
\end{eqnarray}
In table \ref{tab:divdivIPR4} we summarize the intersection pairing.\\
\begin{table}[hbtp]
\begin{center}
\begin{eqnarray}
\begin{array}{c|cccccc}
\phantom{x} & D_7^c    & C_{46}^c & C_{26}^c & C_{67}^c & C_{56}^c & P^c\\
\hline
    X       & D_7^c    & C_{46}^c & C_{26}^c & C_{67}^c & C_{56}^c & P^c\\
    D_1     & 0        & P^c      & 0        & 0        & 0        & 0\\
    D_2     & C_{27}^c & 0        & -P^c     & P^c      & P^c      & 0\\
    D_3     & C_{37}^c & 0        & 0        & P^c      & 0        & 0\\
    D_4     & 0        & -2P^c    & 0        & 0        & P^c      & 0\\
    C       & P^c      & 0        & 0        & 0        & 0        & 0\\
\end{array}
\end{eqnarray}
\end{center}
\caption{Intersection pairing $A^*(X)\otimes A_*^c(X) \rightarrow A_*^c(X)$}
\label{tab:divdivIPR5}
\end{table}
\newpage\noindent
The Mori cone generators are $C_a$, $a=1,\ldots,4$, with
\begin{eqnarray}
C_1=C_{46}\ ,\qquad C_2=C_{26}\ ,\qquad C_3=C_{67}\ ,\qquad C_4=C_{56}\ .
\end{eqnarray}
The K\"ahler cone is generated by the dual elements $T_a$,
$a=1,\ldots,4$ with
\begin{eqnarray}
&&T_1=D_1\ ,\qquad T_2=2D_1-D_2+D_3+D_4\ , \cr
&&T_3=D_3\ ,\qquad T_4=2D_1+D_4\ .
\end{eqnarray}
\begin{table}[hbtp]
\begin{center}
\begin{eqnarray}
\begin{array}{c|cccc}
\phantom\vline & T_1 & T_2 & T_3 & T_4  \\
\hline
T_1  &   0  &   0  &   0  &   0  \\
T_2  &   0  &   0  &   0  &   0  \\
T_3  &   0  &   0  &   C  &   0  \\
T_4  &   0  &   0  &   0  &   0  \\
\end{array}
\end{eqnarray}
\end{center}
\caption{Intersection between K\"ahler generators}
 \label{5tab:morikahler}
\end{table}
\\
If we call $J_a$ the K\"ahler generators in $H^2(X,\QQ)$
corresponding to the $T_a$ then the cohomology ring is
\begin{eqnarray}
H^*(X,\QQ)=\QQ[J_1,\ldots,J_4]/\sim \ ,
\end{eqnarray}
with $\sim$ given by table \ref{5tab:morikahler}.

\newpage
\subsection{K-theory generators}
\label{K-theory generators flop}
We have seen that $G$-Hilb is the moduli space of $G$-cluster in
$\CC^3$. The natural generalization of $G$-cluster is
$G$-constellation. For a finite group $G\subset GL(n,\CC)$, a
$G$-constellation is a $G$-equivariant coherent sheaf $F$ on $C^n$
with global sections $H^0(F)$ isomorphic as a $C[G]$-module to the
regular representation $R$ of $G$. Set
\begin{equation*}
\Theta:=\{\theta\in \text{Hom}_{\ZZ}(R(G),\QQ)|\theta(R)=0\}\ ,
\end{equation*}
where $R(G)$ is the representation ring of $G$. This is an hyperplane
in $\QQ^r$, where $r$ is the order of $G$. For $\theta\in\Theta$,
a $G$-constellation $F$ is said to be $\theta$-stable (or
$\theta$-semistable) if every proper $G$-equivariant coherent subsheaf
\ $0\subset E\subset F$ satisfies $\theta(E)>0$ (or $\theta(E)\geq 0$).
The moduli space $\cM_\theta$ of $\theta$-stable constellation is
constructed using GIT (cf. \cite{Sardo-Infirri}). The space $\Theta$ is
subdivided into polyhedral convex cones $C$ called GIT chamber. Given
$\theta$ and $\theta'$ in the same chamber $C$ the moduli spaces
$\cM_{\theta}$ and $\cM_{\theta'}$ are isomorphic, so we write $\cM_C$
in place of $\cM_{\theta}$ for any $\theta\in C$.
Ito and Nakajima \cite{Ito-Nakajima} observed that
$G$-Hilb$=\cM_{C_0}$ for some chamber $C_0\subset\Theta$ and more
generally the method of \cite{Bridgeland-King-Reid} shows that for any
chamber $C\subset\Theta$ there is a crepant resolution
$\tau:\cM_C\rightarrow \CC^3/G$ and an equivalence of
$\Phi_C:D(\cM_C)\rightarrow D^G(\CC^3)$ between derived categories of
coherent sheaves on $\cM_C$ and derived categories of $G$-equivariant
sheaves on $\CC^3$. Craw and Ishii \cite{Craw-Ishii} proved that in
the Abelian case every crepant resolution may be realized as a moduli
space $\cM_C$ for some chamber. Moreover they uncovered explicit
equivalence between the derived categories of moduli $\cM_\theta$ for
parameters lying in adjacent GIT chambers.
Therefore starting from $G$-Hilb and by analyzing
the chamber structure of $\Theta$, we can define the tautological
bundles $\cRR_{\rho}$ that generate $K(X)$ on every flopped resolution
$X$ and that are the Fourier-Mukay transforms of the original
tautological bundles on $G$-Hilb.

Here we summarize how calculate the chamber structure of $\Theta$ and
the transformation induced by crossing the walls $W$ of the chambers
$C$. We refer to \cite{Craw-Ishii,Degeratu} for detailed explanations.
The derived equivalence $\Phi_C$ induces a $\ZZ$-linear isomorphism
\begin{equation}
\varphi_C:K^c(\cM_C)\rightarrow R(G)\qquad \qquad
\sum a_i\cS_i \longmapsto \bigoplus a_i\rho_i
\label{isomKR}
\end{equation}
where as usual $\cS_i$ is the element of the basis of $K^c(\cM_C)$
dual of the tautological bundle $\cRR_i$ and $\rho_i$ is the
irreducible representation of character $i$.
Let $C\subset\Theta$ be a chamber. Then $\theta\in C$ if and only if
\begin{itemize}
\item
for every exceptional curve $\ell$, we have
$\theta(\varphi_C(\cO_\ell))\!=\!\sum_i \theta(\rho_i)\,
\text{deg}(\cRR_{\rho_i}|_\ell)>0$;
\footnote{Recall that if $\cRR_\rho=\cO_X(D')$ then
$\text{deg}(\cRR_{\rho}|_\ell)=D'\cdot\ell$}
\item
for every compact reduced divisor $D$
\footnote{I.e. $D=\sum a_i D_i$ where $D_i$ are compact invariant
divisors and the coefficient $a_i\in\{0,1,-1\}$.}
and irreducible representation
$\rho$, we have
\begin{equation*}
\theta(\varphi_C(\cRR^{-1}_{\rho}\otimes\omega_D))<0
\qquad\text{and}\qquad
\theta(\varphi_C(\cRR^{-1}_{\rho}|_D))>0\ ,
\end{equation*}
where $\omega_D$ is the canonical bundle of $D$.
\footnote{If $\cRR_\rho=\cO_X(D')$ and $\omega_D=\cO_D(K_D)$
then $\cRR^{-1}_{\rho}\otimes\omega_D=\cO_D(-D'\cdot D+K_D)$ and
$\cRR^{-1}_{\rho}|_D=\cO_D(-D'\cdot D)$. Then we calculate the
inequalities with the help of (\ref{LocalChernclasses}) and (\ref{isomKR}).}
\end{itemize}
These inequalities determine the walls of the chamber $C$, but we
have to pay attention that some of them may be redundant.\\
Let $\theta\in\Theta$ be a generic parameter, $C$ the chamber
containing it and $\theta_0$ a parameter on its wall $W$. The wall is
said to be of type 0, I, II or III as follows:
\begin{itemize}
\item
type 0 if $\cM_{\theta_0}$ isomorphic to $\cM_{\theta}$,
\item
type I if $\cM_{\theta_0}$ is obtained from $\cM_{\theta}$ by the
contraction of a curve to a point,
\item
type II if $\cM_{\theta_0}$ is obtained from $\cM_{\theta}$ by the
contraction of a divisor to a point,
\item
type III if $\cM_{\theta_0}$ is obtained from $\cM_{\theta}$ by the
contraction of a divisor to a curve.
\end{itemize}
The inequalities coming from curves determine walls of type I or III,
while the others determine walls of type 0. There are no walls of type
II.\\
If $C'$ is the chamber behind the wall $W$, the relatione between
$\cM_{C'}$ and $\cM_C$ and their tautological bundles depends on the
type of the wall.
\begin{itemize}
\item
$W$ of type 0: $\cM_{\theta'}$ is isomorphic to $\cM_{\theta}$; the
wall $W\subset\Theta$ is the zero locus of an equation of the form\\
$R(\theta_0(\rho_1),\ldots,\theta_0(\rho_r))=
a_1\theta_0(\rho_1)+\ldots+a_r\theta_0(\rho_r)=0$ and, if $D$ is the
divisor defining the wall, the tautological bundles $\cRR_i$ and
$\cRR_i'$ are related as follows:
\begin{itemize}
\item
Case $+$: if $R(\theta(\rho_1),\ldots,\theta(\rho_r))>0$ then
\begin{equation*}
\cRR_i'=
\begin{cases}
\cRR_i & \text{if $a_i=0$,} \\
\cRR_i\otimes\cO_{\cM_{\theta'}}(D) & \text{if $a_i\neq 0$;}
\end{cases}
\end{equation*}
\item
Case $-$: if $R(\theta(\rho_1),\ldots,\theta(\rho_r))<0$ then
\begin{equation*}
\cRR_i'=
\begin{cases}
\cRR_i & \text{if $a_i=0$,}\\
\cRR_i\otimes\cO_{\cM_{\theta'}}(-D) & \text{if $a_i\neq0$.}
\end{cases}
\end{equation*}
\end{itemize}
\item
$W$ of type I: $\cM_{\theta'}$ is the variety obtained from
$\cM_{\theta}$ by the flop of the curve $\ell$ determining the wall;
the tautological bundles $\cRR_i'$ are the proper transform of $\cRR_i$.
\item
$W$ of type III: $\cM_{\theta'}$ is isomorphic to $\cM_{\theta}$; if
$D$ is the divisor contracted in $\cM_{\theta_0}$, the
tautological bundles $\cRR_i$ and $\cRR_i'$ are related as follows:
\begin{itemize}
\item
Case $+$: if $\{\text{deg}(\cRR_i|_\ell)\}=\{0,1\}$ then
\begin{equation*}
\cRR_i'=
\begin{cases}
\cRR_i & \text{if deg$(\cRR_i|_\ell)=0$,} \\
\cRR_i\otimes\cO_{\cM_{\theta'}}(D) & \text{if deg$(\cRR_i|_\ell)=1$;}
\end{cases}
\end{equation*}
\item
Case $-$: if $\{\text{deg}(\cRR_i|_\ell)\}=\{0,-1\}$ then
\begin{equation*}
\cRR_i'=
\begin{cases}
\cRR_i & \text{if deg$(\cRR_i|_\ell)=0$,}\\
\cRR_i\otimes\cO_{\cM_{\theta'}}(-D) & \text{if deg$(\cRR_i|_\ell)=-1$.}
\end{cases}
\end{equation*}
\end{itemize}
\end{itemize}
Thus, crossing walls of type I induces flops, while walls of type 0
and III induce self-equivalence of the derived category of the
resolved variety. One can start from the chamber of the $G$-Hilb
resolution, follow the change of the tautological bundles crossing
the walls and reconstruct the chamber structure of $\Theta$.

In our example the tautological bundles for the $\ZZ_6$-Hilb are
\begin{eqnarray}
&& \cRR_0=\cO_X,\quad \cRR_1=\cO_X(D_1),\quad \cRR_2=\cO_X(D_2),\quad
\cRR_3=\cO_X(D_3),\cr
&& \cRR_4=\cO_X(-D_1+D_2+D_3-D_4),\quad \cRR_5=\cRR_2\otimes\cRR_3=\cO_X(D_2+D_3).
\end{eqnarray}
We write parameters $\theta$ as $(\theta_0,\ldots,\theta_5)$, where
$\theta_i:=(\theta(\rho_i))$. The inequalities defining the
$\ZZ_6$-Hilb chamber are
\begin{align}
\label{flop1}
&\theta_1>0 &&
\text{wall of type I related to the flop of the curve $C_1$;}\\
&\theta_2+\theta_5>0 &&
\text{wall of type III+ related to the contraction of the divisor $D_5$;}\\
\label{flop2}
&\theta_3+\theta_5>0 &&
\text{wall of type I related to the flop of the curve $C_3$;}\\
\label{flop3}
&\theta_4>0 &&
\text{wall of type I related to the flop of the curve $C_4$;}\\
&\theta_5>0 &&
\text{wall of type 0+ defined by
$\theta(\varphi_C(\cRR^{-1}_{5}|_{D_7}))>0$;}\\
&\theta_2+\theta_3+\theta_4+\theta_5>0 &&
\text{wall of type 0+ defined by
$\theta(\varphi_C(\cRR^{-1}_{5}\otimes\omega_{D_7}))<0$.}
\end{align}
Any other inequality is redundant. As it is proven in section 9 of
\cite{Craw-Ishii}, the flop of any single curve in the $G$-Hilb is
achieved by crossing a wall of the chamber (generally if we are in a
chamber different from the $G$-Hilb's it may be necessary first cross
a type 0 wall to realize a flop).

\subsection*{Resolution \boldmath{$\RRC$}}
Starting from the $G$-Hilb chamber we obtain this resolution by
crossing the wall (\ref{flop1}). The tautological bundles are again
\begin{eqnarray}
&& \cRR_0=\cO_X,\quad \cRR_1=\cO_X(D_1),\quad \cRR_2=\cO_X(D_2),\quad
\cRR_3=\cO_X(D_3),\cr
&& \cRR_4=\cO_X(-D_1+D_2+D_3-D_4),\quad \cRR_5=\cRR_2\otimes\cRR_3=\cO_X(D_2+D_3).
\end{eqnarray}
while the pure D-brane basis is
\begin{eqnarray}
B_0:= \cO_p\ ; \quad\ B_a:=\cO_{C_a} (-T_a)\ ;
\quad\ B_5:=\cO_{D_7}(-T_2-T_3)\ ,
\end{eqnarray}
with $a=1,\ldots,4$. In terms of the $\cRR_i$ and
their duals $\cS_i$, the $B_i$-basis of $K(X)$ and its dual
$\Phi$-basis of $K^c(X)$ are thus:
\begin{align}
& B_0=\cS_0+\cS_1+\cS_2+\cS_3+\cS_4+\cS_5\ ,
&& \Phi_0=\cRR_0 \ ,\cr
& B_1=-\cS_1\ ,
&& \Phi_1=-\cRR_0-\cRR_1+\cRR_3+\cRR_4\ ,\cr
& B_2=\cS_2+\cS_5\ ,
&& \Phi_2=-\cRR_0+\cRR_2\ ,\cr
& B_3=\cS_1+\cS_3+\cS_5\ ,
&& \Phi_3=-\cRR_0+\cRR_3\ ,\cr
& B_4=\cS_1+\cS_4\ ,
&& \Phi_4=-\cRR_0+\cRR_4\ ,\cr
& B_5=\cS_5\ .
&& \Phi_5=\cRR_0-\cRR_2-\cRR_3+\cRR_5\ .
\end{align}
The symplectic form in the selected basis is
\begin{eqnarray}\label{2simplecticform}
\chi(B_i,B_j)=\left(
\begin{array}{cccccc}
0 & 0 & 0 & 0 & 0 & 0 \\
0 & 0 & 0 & 0 & 0 & -1 \\
0 & 0 & 0 & 0 & 0 & 0 \\
0 & 0 & 0 & 0 & 0 & 0 \\
0 & 0 & 0 & 0 & 0 & 0 \\
0 & 1 & 0 & 0 & 0 & 0
\end{array}
\right)\ .
\end{eqnarray}

\subsection*{Resolution \boldmath{$\RSC$}}
Starting from the $G$-Hilb chamber we obtain this resolution by
crossing the wall (\ref{flop2}). The tautological bundles are again
\begin{eqnarray}
&& \cRR_0=\cO_X,\quad \cRR_1=\cO_X(D_1),\quad \cRR_2=\cO_X(D_2),\quad
\cRR_3=\cO_X(D_3),\cr
&& \cRR_4=\cO_X(-D_1+D_2+D_3-D_4),\quad \cRR_5=\cRR_2\otimes\cRR_3=\cO_X(D_2+D_3).
\end{eqnarray}
while the pure D-brane basis is
\begin{eqnarray}
&B_0:= \cO_p\ ; \quad\ B_a:=\cO_{C_a} (-T_a)\ ;\quad\
B_5:=\cO_{D_7}(-T_1-T_2+T_3-T_4)\ ,
\end{eqnarray}
with $a=1,\ldots,4$. In terms of the $\cRR_i$ and
their duals $\cS_i$, the $B_i$-basis of $K(X)$ and its dual
$\Phi$-basis of $K^c(X)$ are thus:
\begin{align}
& B_0=\cS_0+\cS_1+\cS_2+\cS_3+\cS_4+\cS_5\ ,
&& \Phi_0=\cRR_0 \ ,\cr
& B_1=\cS_1+\cS_3+\cS_5\ ,
&& \Phi_1=-\cRR_0+\cRR_1\ ,\cr
& B_2=\cS_2+\cS_5\ ,
&& \Phi_2=-\cRR_0+\cRR_2\ ,\cr
& B_3=-\cS_3-\cS_5\ ,
&& \Phi_3=-\cRR_0+\cRR_1-\cRR_3+\cRR_4\ ,\cr
& B_4=\cS_3+\cS_4+\cS_5 ,
&& \Phi_4=-\cRR_0+\cRR_4\ ,\cr
& B_5=\cS_5\ .
&& \Phi_5=\cRR_0-\cRR_2-\cRR_3+\cRR_5\ .
\end{align}
The symplectic form in the selected basis is
\begin{eqnarray}\label{3simplecticform}
\chi(B_i,B_j)=\left(
\begin{array}{cccccc}
0 & 0 & 0 & 0 & 0 & 0 \\
0 & 0 & 0 & 0 & 0 & 0 \\
0 & 0 & 0 & 0 & 0 & 0 \\
0 & 0 & 0 & 0 & 0 & 1 \\
0 & 0 & 0 & 0 & 0 & -2 \\
0 & 0 & 0 & -1 & 2 & 0
\end{array}
\right)\ .
\end{eqnarray}

\subsection*{Resolution \boldmath{$\RTC$}}
Starting from the $G$-Hilb chamber we obtain this resolution by
crossing the wall (\ref{flop3}). The tautological bundles are again
\begin{eqnarray}
&& \cRR_0=\cO_X,\quad \cRR_1=\cO_X(D_1),\quad \cRR_2=\cO_X(D_2),\quad
\cRR_3=\cO_X(D_3),\cr
&& \cRR_4=\cO_X(-D_1+D_2+D_3-D_4),\quad \cRR_5=\cRR_2\otimes\cRR_3=\cO_X(D_2+D_3).
\end{eqnarray}
while the pure D-brane basis is
\begin{eqnarray}
B_0:= \cO_p\ ; \quad\ B_a:=\cO_{C_a} (-T_a)\ ;\quad\
B_5:=\cO_{D_7}(-T_2-T_3)\ ,
\end{eqnarray}
with $a=1,\ldots,4$. In terms of the $\cRR_i$ and
their duals $\cS_i$, the $B_i$-basis of $K(X)$ and its dual
$\Phi$-basis of $K^c(X)$ are thus:
\begin{align}
& B_0=\cS_0+\cS_1+\cS_2+\cS_3+\cS_4+\cS_5\ ,
&& \Phi_0=\cRR_0 \ ,\cr
& B_1=\cS_1+\cS_4\ ,
&& \Phi_1=-\cRR_0+\cRR_1\ ,\cr
& B_2=\cS_2+\cS_4+\cS_5\ ,
&& \Phi_2=-\cRR_0+\cRR_2\ ,\cr
& B_3=\cS_3+\cS_4+\cS_5\ ,
&& \Phi_3=-\cRR_0+\cRR_3\ ,\cr
& B_4=-\cS_4\ ,
&& \Phi_4=-\cRR_0+\cRR_1-\cRR_4+\cRR_5\ ,\cr
& B_5=\cS_4+\cS_5\ .
&& \Phi_5=\cRR_0-\cRR_2-\cRR_3+\cRR_5\ .
\end{align}
The symplectic form in the selected basis is
\begin{eqnarray}\label{4simplecticform}
\chi(B_i,B_j)=\left(
\begin{array}{cccccc}
0 & 0 & 0 & 0 & 0 & 0 \\
0 & 0 & 0 & 0 & 0 & 0 \\
0 & 0 & 0 & 0 & 0 & -1 \\
0 & 0 & 0 & 0 & 0 & -2 \\
0 & 0 & 0 & 0 & 0 & 1 \\
0 & 0 & 1 & 2 & -1 & 0
\end{array}
\right)\ .
\end{eqnarray}

\subsection*{Resolution \boldmath{$\RUC$}}
Starting from the above chamber of the resolution $\RTC$ we obtain this
resolution by crossing a single wall of type I. The tautological
bundles are again
\begin{eqnarray}
&& \cRR_0=\cO_X,\quad \cRR_1=\cO_X(D_1),\quad \cRR_2=\cO_X(D_2),\quad
\cRR_3=\cO_X(D_3),\cr
&& \cRR_4=\cO_X(-D_1+D_2+D_3-D_4),\quad \cRR_5=\cRR_2\otimes\cRR_3=\cO_X(D_2+D_3).
\end{eqnarray}
while the pure D-brane basis is
\begin{eqnarray}
&B_0:= \cO_p\ ; \quad\ B_a:=\cO_{C_a} (-T_a)\ ;\quad\
B_5:=\cO_{D_7}(T_2-2T_3-T_4)\ ,
\end{eqnarray}
with $a=1,\ldots,4$. In terms of the $\cRR_i$ and
their duals $\cS_i$, the $B_i$-basis of $K(X)$ and its dual
$\Phi$-basis of $K^c(X)$ are thus:\\
\begin{equation}
\begin{aligned}
& B_0=\cS_0+\cS_1+\cS_2+\cS_3+\cS_4+\cS_5\ ,
&& \Phi_0=\cRR_0 \ ,\cr
& B_1=\cS_1+\cS_4\ ,
&& \Phi_1=-\cRR_0+\cRR_1\ ,\cr
& B_2=-\cS_2-\cS_4-\cS_5\ ,
&& \Phi_2=-\cRR_0+\cRR_1-\cRR_2+\cRR_3-\cRR_4+\cRR_5 ,\cr
& B_3=\cS_2+\cS_3+2\cS_4+2\cS_5\ ,
&& \Phi_3=-\cRR_0+\cRR_3\ ,\cr
& B_4=\cS_2+\cS_5\ ,
&& \Phi_4=-\cRR_0+\cRR_1-\cRR_4+\cRR_5\ ,\cr
& B_5=\cS_4+\cS_5\ .
&& \Phi_5=\cRR_0-\cRR_2-\cRR_3+\cRR_5\ .
\end{aligned}
\end{equation}
The symplectic form in the selected basis is
\begin{eqnarray}\label{5simplecticform}
\chi(B_i,B_j)=\left(
\begin{array}{cccccc}
0 & 0 & 0 & 0 & 0 & 0 \\
0 & 0 & 0 & 0 & 0 & 0 \\
0 & 0 & 0 & 0 & 0 & 1 \\
0 & 0 & 0 & 0 & 0 & -3 \\
0 & 0 & 0 & 0 & 0 & 0 \\
0 & 0 & -1 & 3 & 0 & 0
\end{array}
\right)\ .
\end{eqnarray}

\subsection{The cohomological hypergeometric series and $GW$--invariants}
The hypergeometric series are specified by the $\ell$ vectors
corresponding to large K\"ahler parameters and the hypergeometric
coefficients are determined expanding them with respect to the basis
$Q_i={\rm ch}(\Phi_i)$ of $H^*(X,\QQ)$.
\subsection*{Invariants for \boldmath{$\RRC$}}
The vectors $\ell_a$, $a=1,\ldots,4$ are
\begin{eqnarray}
C_1 \quad\ &:& \quad\  \ell_1=(-1,0,0,1,0,1,-1)\ ,\cr
C_2 \quad\ &:& \quad\  \ell_2=(0,1,0,1,-2,0,0)\ ,\cr
C_3 \quad\ &:& \quad\  \ell_3=(1,0,1,0,0,-2,0)\ ,\cr
C_4 \quad\ &:& \quad\  \ell_4=(1,0,0,-2,1,0,0)\ .
\end{eqnarray}
The selected basis of the cohomology is
\begin{eqnarray}
&& Q_0 =1\ , \quad Q_1 =J_1-2C\ , \quad Q_2 =J_2 \ ,\cr
&&Q_3 =J_3 -\frac 12 C\ ,\quad Q_4 =J_4-C\ , \quad Q_5 =-C .
\end{eqnarray}
If we make this change of basis and use the mirror symmetry identification
\begin{eqnarray}
w\left(\vec x, \frac {\vec J}{2\pi i} \right)= Q_0 1 +\sum_{a=1}^4 Q_a t_a +Q_5 g(t_1,\ldots,t_4)\ ,
\end{eqnarray}
then we find
\begin{eqnarray}
&& 2\pi i t_1= \log x_1 -\Psi(x_3) +\Phi(x_2,x_4)-\aleph (\vec x) \ ,\cr
&& 2\pi i t_2= \log x_2 +\Phi(x_2,x_4) -2\Phi(x_4,x_2) \ ,\cr
&& 2\pi i t_3= \log x_3 +2\Psi(x_3) \ ,\cr
&& 2\pi i t_4= \log x_4 -2\Phi(x_2,x_4) +\Phi(x_4,x_2) \ ,\label{2moduli}
\end{eqnarray}
and
\begin{eqnarray}
&& g(\vec t ) =P_2(\vec t)+\frac 1{(2\pi i)^2} \phi(\vec t)\ ,
\end{eqnarray}
where $P_2$ is the degree two polynomial part
\begin{eqnarray}
&&P_2(\vec t )=
-2t_1 -\frac 12 t_3 -t_4 +3t_1^2 +\frac 12 t_3^2+t_4^2\cr
&& \qquad +2t_1 t_2 +3t_1 t_3 +4t_1 t_4 +t_2 t_3 +t_2 t_4 +2t_3 t_4 \ ,
\end{eqnarray}
and
\begin{eqnarray}
&& \phi(\vec t)=
   6\aleph^{(1)}(\vec x)-3\aleph^{(2)}(\vec x)-2\aleph^{(3)}(\vec x)
   -\aleph^{(6)}(\vec x)-\Lambda_4(\vec x)-\Lambda_5(\vec x)\cr
&& \qquad +\Psi_6(x_2,x_4)-2\Psi_4(x_3)+2\Psi_5(x_3)+2\Psi_1(x_2,x_4)
+2\Psi_1(x_4,x_2)\cr
&& \qquad -\Psi_2(x_2,x_4)-\Psi_2(x_4,x_2)-\Psi_3(x_2,x_4)-\Psi_3(x_4,x_2)\cr
&& \qquad -3\aleph^2(\vec x)+\Psi^2(x_3)+\Phi^2(x_2,x_4)+\Phi^2(x_4,x_2)-\Phi(x_2,x_4)\Phi(x_4,x_2),
\end{eqnarray}
with $\vec x$ expressed as a function of $\vec t$ by inverting system (\ref{2moduli}), is the part corresponding to instantonic contributions.
Following Hosono and using (\ref{2simplecticform}) we find
\begin{eqnarray}
(-\partial_{t_1}) F(\vec t) =g(\vec t )\ ,\label{2struttura}
\end{eqnarray}
where $F$ is the prepotential. Setting
\begin{eqnarray}
q_k :=e^{2\pi i t_k}\ ,
\end{eqnarray}
we then find
\begin{eqnarray}
&& F(\vec t)=
t_1^2 +\frac 12 t_1 t_3 -t_1 t_4 -t_1^3 -\frac 12 t_1
t_3^2 -t_1 t_4^2 -t_1^2 t_2 -\frac 32t_1^2 t_3 -2t_1^2 t_4 \cr
&& \qquad -t_1 t_2 t_3-t_1 t_2 t_4 -2t_1 t_3 t_4
+F_{{\rm inst}} (\vec q) +P_{{\rm class}} (t_2, t_3, t_4) 
+Q_{{\rm inst}} (q_2, q_3, q_4) \ . \label{2prepotential}
\end{eqnarray}
$P$ and $Q$ are the undetermined parts.
We list the Gromov-Witten invariants for rational curves up to degree six.
The curves with degree in the integer cone generated by $[0,1,0,0]$, $[0,0,1,0]$, $[0,0,0,1]$
must be excluded, because corresponding to the undetermined part of the prepotential. The only nonvanishing invariants
in the considered range are
\begin{eqnarray}
&& GW_{[1,0,0,0]}=GW_{[1,0,1,0]}=GW_{[1,0,0,1]}=\cr
&& GW_{[1,0,1,1]}=GW_{[1,1,0,1]}=GW_{[1,1,1,1]}=1;\cr
&& GW_{[2,0,1,1]}=GW_{[2,1,1,1]}=GW_{[2,1,1,2]}=-2.
\end{eqnarray}

\subsection*{Invariants for \boldmath{$\RSC$}}
The vectors $\ell_a$, $a=1,\ldots,4$ are
\begin{eqnarray}
C_1 \quad\ &:& \quad\  \ell_1=(1,0,1,0,0,-2,0)\ ,\cr
C_2 \quad\ &:& \quad\  \ell_2=(0,1,0,1,-2,0,0)\ ,\cr
C_3 \quad\ &:& \quad\  \ell_3=(0,0,-1,-1,0,1,1)\ ,\cr
C_4 \quad\ &:& \quad\  \ell_4=(0,0,1,0,1,0,-2)\ .
\end{eqnarray}
The selected basis of the cohomology is
\begin{eqnarray}
&& Q_0 =1\ ,\quad Q_1 =J_1 \ ,\quad Q_2 =J_2 \ ,\quad Q_3 =J_3 \ ,\quad Q_4 =J_4+C \ , \quad Q_5 =C .
\end{eqnarray}
Making use of the mirror symmetry identification
\begin{eqnarray}
w\left(\vec x, \frac {\vec J}{2\pi i} \right)= Q_0 1 +\sum_{a=1}^4 Q_a t_a +Q_5 g(t_1,\ldots,t_4)\ ,
\end{eqnarray}
then we find
\begin{eqnarray}
&& 2\pi i t_1= \log x_1 +2\Psi(x_1)  \ ,\cr
&& 2\pi i t_2= \log x_2 -\Phi(x_2,x_1 x_3^2 x_4) +2\Phi(x_1 x_3^2 x_4,x_2) \ ,\cr
&& 2\pi i t_3= \log x_3 -\Psi(x_1)+\aleph (\vec x) \ ,\cr
&& 2\pi i t_4= \log x_4 -\Phi(x_1 x_3^2 x_4,x_2)-2\aleph (\vec x) \ ,\label{3moduli}
\end{eqnarray}
and
\begin{eqnarray}
&& g(\vec t ) =P_2(\vec t)+\frac 1{(2\pi i)^2} \phi(\vec t)\ ,
\end{eqnarray}
where $P_2$ is the degree two polynomial part
\begin{eqnarray}
P_2(\vec t )=\frac 16 -t_4+t_4^2 +t_2t_4  \ ,
\end{eqnarray}
and
\begin{eqnarray}
&& \phi(\vec t)=
8\aleph^{(1)}(\vec x)-4\aleph^{(2)}(\vec x)-2\aleph^{(3)}(\vec x)
-\aleph^{(4)}(\vec x)+\Lambda_6(\vec x)-2\Lambda_7(\vec x)+\Lambda_8(\vec x)\cr
&& \qquad +\Psi_3(x_2,x_1x_3^2x_4)+\Psi_2(x_1x_3^2x_4,x_2)
+\Psi_3(x_1x_3^2x_4,x_2)-2\Psi_1(x_1x_3^2x_4,x_2)\cr
&& \qquad -4\aleph^2(\vec x)-2\aleph(\vec x)\Phi(x_2,x_1x_3^2x_4)
-\Phi(x_2,x_1x_3^2x_4)\Phi(x_1x_3^2x_4,x_2)\cr
&& \qquad +\Phi^2(x_1x_3^2x_4,x_2),
\end{eqnarray}
with $\vec x$ expressed as a function of $\vec t$ by inverting system (\ref{3moduli}), is the part corresponding to instantonic contributions.
Using (\ref{3simplecticform}) we find
\begin{eqnarray}
(\partial_{t_3}-2\partial_{t_4}) F(\vec t) =g(\vec t )\ ,\label{3struttura}
\end{eqnarray}
where $F$ is the prepotential. Setting
\begin{eqnarray}
q_k :=e^{2\pi i t_k}\ ,
\end{eqnarray}
we then find
\begin{eqnarray}
&& F(\vec t)= \frac 16 t_3 +\frac 14 t_4^2 -\frac 16 t_4^3 -\frac 14 t_2t_4^2 \cr
&& \qquad\ +F_{{\rm inst}} (\vec q)
+P_{{\rm class}} (t_1, t_2, 2t_3+t_4) +Q_{{\rm inst}} (q_1, q_2, q_3^2 q_4) \ . \label{3prepotential}
\end{eqnarray}
$P$ and $Q$ are the undetermined parts.
We list the $GW$--invariants up to degree six.
The curves with degree in the integer cone generated by $[1,0,0,0]$, $[0,1,0,0]$, $[0,0,2,1]$
must be excluded, because corresponding to the undetermined part of the prepotential. The only nonvanishing invariants
in the considered range are
\begin{eqnarray}
&& GW_{[0,0,1,0]}=GW_{[0,0,1,1]}=GW_{[0,1,1,1]}=\cr
&& GW_{[1,0,1,0]}=GW_{[1,0,1,1]}=GW_{[1,1,1,1]}=1;\cr
&& GW_{[0,0,0,1]}=GW_{[0,1,0,1]}=GW_{[1,1,2,2]}=-2;
   \quad GW_{[0,1,1,2]}=GW_{[1,1,1,2]}=3;\cr
&& GW_{[0,1,0,2]}=-4; \quad GW_{[0,1,1,3]}=GW_{[1,1,1,3]}=GW_{[0,2,1,3]}=5;\cr
&& GW_{[0,1,0,3]}=GW_{[0,2,0,3]}=-6; \quad GW_{[0,1,1,4]}=7;
   \quad GW_{[0,1,0,4]}=-8;\cr
&& GW_{[0,1,0,5]}=-10; \quad GW_{[0,2,0,4]}=-32.\label{gwf2}
\end{eqnarray}

\subsection*{Invariants for \boldmath{$\RTC$}}
The vectors $\ell_a$, $a=1,\ldots,4$ are
\begin{eqnarray}
C_1 \quad\ &:& \quad\  \ell_1=(1,0,0,-2,1,0,0)\ ,\cr
C_2 \quad\ &:& \quad\  \ell_2=(0,1,0,0,-1,1,-1)\ ,\cr
C_3 \quad\ &:& \quad\  \ell_3=(0,0,1,0,1,0,-2)\ ,\cr
C_4 \quad\ &:& \quad\  \ell_4=(0,0,0,1,-1,-1,1)\ .
\end{eqnarray}
The selected basis of the cohomology is
\begin{eqnarray}
&& Q_0 =1\ ,\quad Q_1 =J_1\ ,\quad Q_2 =J_2 \ ,\cr
&& Q_3 =J_3 +\frac 12 C\ ,\quad Q_4 =J_4\ , \quad Q_5 =C .
\end{eqnarray}
Via the mirror symmetry identification
\begin{eqnarray}
w\left(\vec x, \frac {\vec J}{2\pi i} \right)= Q_0 1 +\sum_{a=1}^4 Q_a t_a +Q_5 g(t_1,\ldots,t_4)\ ,
\end{eqnarray}
we get
\begin{eqnarray}
&& 2\pi i t_1= \log x_1 +2\Phi(x_1,x_2 x_4) -\Phi(x_2 x_4, x_1)  \ ,\cr
&& 2\pi i t_2= \log x_2 +\Phi(x_2 x_4, x_1) -\Psi(x_1 x_3 x_4^2)-\aleph (\vec x) \ ,\cr
&& 2\pi i t_3= \log x_3 -\Phi(x_2 x_4, x_1)-2\aleph (\vec x) \ ,\cr
&& 2\pi i t_4= \log x_4 +\Psi(x_1 x_3 x_4^2) -\Phi(x_1, x_2 x_4)+\Phi(x_2 x_4,x_1)+\aleph (\vec x) \ ,\label{4moduli}
\end{eqnarray}
and
\begin{eqnarray}
&& g(\vec t ) =P_2(\vec t)+\frac 1{(2\pi i)^2} \phi(\vec t)\ ,
\end{eqnarray}
where $P_2$ is the degree two polynomial part
\begin{eqnarray}
P_2(\vec t )= \frac 16 -\frac 12 t_3 +\frac 12 t_3^2 +t_2 t_3  \ ,
\end{eqnarray}
and
\begin{eqnarray}
&& \phi(\vec t)=
8\aleph^{(1)}(\vec x)-3\aleph^{(2)}(\vec x)-2\aleph^{(3)}(\vec x)
-2\aleph^{(5)}(\vec x)-\aleph^{(7)}(\vec x)\cr
&& \qquad -\Lambda_{9}(\vec x)+\Lambda_{10}(\vec x)-2\Lambda_{11}(\vec x)\cr
&& \qquad -\Psi_1(x_2x_4,x_1) +\Psi_2(x_2x_4,x_1)+\Psi_5(x_1x_3x_4^2)\cr
&& \qquad -4\aleph^2(\vec x)-2\aleph(\vec x) \Psi(x_1 x_3x_4^2)-\aleph(\vec x) \Phi(x_2x_4,x_1)\cr
&& \qquad\ -\Psi(x_1x_3x_4^2) \Phi(x_2x_4,x_1) +\frac 12 \Phi^2(x_2x_4,x_1)
\end{eqnarray}
with $\vec x$ expressed as a function of $\vec t$ by inverting system (\ref{4moduli}), is the part corresponding to instantonic contributions.
Using (\ref{4simplecticform}) we find
\begin{eqnarray}
(\partial_{t_4}-\partial_{t_2}-2\partial_{t_3}) F(\vec t) =g(\vec t )\ ,\label{4struttura}
\end{eqnarray}
where $F$ is the prepotential. Setting
\begin{eqnarray}
q_k :=e^{2\pi i t_k}\ ,
\end{eqnarray}
we then find
\begin{eqnarray}
&& F(\vec t)= \frac 16 t_4 +\frac 18 t_3^2 -\frac 1{12} t_3^3 -\frac 12 t_2^2 t_3+\frac 13 t_2^3  \cr
&& \qquad\ +F_{{\rm inst}} (\vec q) +P_{{\rm class}} (t_1,t_4+t_2,2t_4+ t_3) +Q_{{\rm inst}} (q_1,q_4 q_2, q_4^2 q_3) \ , \label{4prepotential}
\end{eqnarray}
$P$ and $Q$ being the undetermined parts.
We list the $GW$--invariants up to degree six.
The curves with degree in the integer cone generated by $[0,1,0,1]$, $[0,0,1,2]$, $[1,0,0,0]$
must be excluded, because corresponding to the undetermined part of the prepotential. The only nonvanishing invariants
in the considered range are
\begin{eqnarray}
&& GW_{[0,0,0,1]}=GW_{[0,0,1,1]}=GW_{[0,1,0,0]}=\cr
&& GW_{[1,0,0,1]}=GW_{[1,0,1,1]}=GW_{[1,1,1,2]}=1;\cr
&& GW_{[0,0,1,0]}=GW_{[0,1,1,1]}=GW_{[1,1,1,1]}=-2;
   \quad GW_{[0,1,1,0]}=GW_{[1,1,2,2]}=3;\cr
&& GW_{[0,1,2,1]}=GW_{[1,1,2,1]}=-4;
   \quad GW_{[0,1,2,0]}=GW_{[0,2,2,1]}=GW_{[1,2,2,1]}=5;\cr
&& GW_{[0,1,3,1]}=GW_{[0,2,2,0]}=GW_{[1,1,3,1]}=-6;
   \quad GW_{[0,1,3,0]}=7;\cr
&& GW_{[0,1,4,1]}=-8; \quad GW_{[0,1,4,0]}=9;
   \quad GW_{[0,1,5,0]}=11; \quad GW_{[0,3,3,0]}=27;\cr
&& GW_{[0,2,3,0]}=-32; \quad GW_{[0,2,3,1]}=35; 
   \quad GW_{[0,2,4,0]}=-110. \label{gwf1}
\end{eqnarray}

\subsection*{Invariants for \boldmath{$\RUC$}}
The vectors $\ell_a$, $a=1,\ldots,4$ are
\begin{eqnarray}
C_1 \quad\ &:& \quad\  \ell_1=(1,0,0,-2,1,0,0)\ ,\cr
C_2 \quad\ &:& \quad\  \ell_2=(0,-1,0,0,1,-1,1)\ ,\cr
C_3 \quad\ &:& \quad\  \ell_3=(0,1,1,0,0,1,-3)\ ,\cr
C_4 \quad\ &:& \quad\  \ell_4=(0,1,0,1,-2,0,0)\ .
\end{eqnarray}
The selected basis of the cohomology is
\begin{eqnarray}
&& Q_0 =1\ ,\quad Q_1 =J_1\ ,\quad Q_2 =J_2 \ ,\cr
&& Q_3 =J_3 +\frac 12 C\ ,\quad Q_4 =J_4\ , \quad Q_5 =C .
\end{eqnarray}
If we make this change of basis and use the mirror symmetry identification
\begin{eqnarray}
w\left(\vec x, \frac {\vec J}{2\pi i} \right)= Q_0 1 +\sum_{a=1}^4 Q_a t_a +Q_5 g(t_1,\ldots,t_4)\ ,
\end{eqnarray}
we get
\begin{eqnarray}
&& 2\pi i t_1= \log x_1 +2\Phi(x_1,x_4) -\Phi(x_4,x_1) \ ,\cr
&& 2\pi i t_2= \log x_2 +\Psi(x_1 x_2^3 x_3 x_4^2) -\Phi(x_4,x_1)-\aleph (\vec x) \ ,\cr
&& 2\pi i t_3= \log x_3 -\Psi(x_1 x_2^3 x_3 x_4^2)+3\aleph (\vec x) \ ,\cr
&& 2\pi i t_4= \log x_4 -\Phi(x_1,x_4) +2\Phi(x_4,x_1) \ ,\label{5moduli}
\end{eqnarray}
and
\begin{eqnarray}
&& g(\vec t ) =P_2(\vec t)+\frac 1{(2\pi i)^2} \phi(\vec t)\ ,
\end{eqnarray}
where $P_2$ is the degree two polynomial part
\begin{eqnarray}
P_2(\vec t )= \frac 14 -\frac 12 t_3 +\frac 12 t_3^2 \ ,
\end{eqnarray}
and
\begin{eqnarray}
&& \phi(\vec t)=
9\aleph^{(1)}(\vec x)-3\aleph^{(2)}(\vec x)-3\aleph^{(3)}(\vec x)
-3\aleph^{(5)}(\vec x)+\Lambda_{12}(\vec x)+\Lambda_{13}(\vec x)\cr
&& \qquad +\Psi_4(x_1x_2^3x_3x_4^2)+\Psi_5(x_1x_2^3x_3x_4^2)\cr
&& \qquad -\frac 12 \Psi^2(x_1 x_2^3 x_3 x_4^2)-\frac 92 \aleph^2(\vec x)
+3\Psi(x_1 x_2^3 x_3 x_4^2)\aleph (\vec x)
\end{eqnarray}
with $\vec x$ expressed as a function of $\vec t$ by inverting system
(\ref{5moduli}), is the part corresponding to instantonic
contributions.
Using (\ref{5simplecticform}) we find
\begin{eqnarray}
(\partial_{t_1}-3\partial_{t_3}) F(\vec t) =g(\vec t )\ ,\label{5struttura}
\end{eqnarray}
where $F$ is the prepotential. Setting
\begin{eqnarray}
q_k :=e^{2\pi i t_k}\ ,
\end{eqnarray}
we then find
\begin{eqnarray}
&& F(\vec t)=\frac 14 t_1+\frac 1{12} t_3^2-\frac 1{18} t_3^3\cr
&& \qquad \ +F_{{\rm inst}} (\vec q) + P_{{\rm class}}
(3t_1+t_3, t_2,t_4) +Q_{{\rm inst}} (q_1^3 q_3, q_2, q_4) \ .
\label{5prepotential}
\end{eqnarray}
We list the $GW$--invariants up to degree six.
The curves with degree in the integer cone generated by $[0,3,1,0]$, $[1,0,0,0]$, $[0,0,0,1]$
must be excluded, because corresponding to the undetermined part of the prepotential. The only nonvanishing invariants
in the considered range are
\begin{eqnarray}
&& GW_{[0,1,0,0]}=GW_{[0,1,0,1]}=GW_{[1,2,1,1]}
   =GW_{[1,2,1,2]}=GW_{[1,1,0,1]}=1;\cr
&& GW_{[0,1,1,0]}=GW_{[1,1,1,1]}=-2; \quad GW_{[0,0,1,0]}=3;\cr
&& GW_{[0,2,2,1]}=GW_{[1,2,2,1]}=-4;
   \quad GW_{[0,1,2,0]}=GW_{[0,1,2,1]}=GW_{[1,1,2,1]}=5;\cr
&& GW_{[0,0,2,0]}=GW_{[0,1,2,0]}=-6;\quad GW_{[0,2,3,0]}=7;
   \quad GW_{[0,0,3,0]}=27;\cr
&& GW_{[0,1,3,0]}=GW_{[0,1,4,1]}=GW_{[0,1,4,0]}=-32; 
   \quad GW_{[0,2,3,1]}=35;\cr
&& GW_{[0,2,4,0]}=-110; \quad GW_{[0,0,4,0]}=-192;
   \quad GW_{[0,1,4,0]}=GW_{[0,1,4,1]}=286;\cr
&& GW_{[0,0,5,0]}=1695; \quad GW_{[0,1,5,1]}=3038;
   \quad GW_{[0,0,6,0]}=-17064. \label{gwp2}
\end{eqnarray}

\section{Conclusions}\label{sec:conclusion}
As stated in the introduction, this paper is the first one of a short
series devoted to a detailed analysis of some aspects of local
(homological) mirror symmetry in relation to its physical meaning. As
much of such a project results to be quite technical, we preferred to
begin with a preparatory article, where we mainly fix our notations
and the objects of study. For this reason we tried here to be as
elementary as possible and included an introductory section which
obviously does not pretend to be neither exhaustive nor self
contained. However, we also included the first step of our analysis,
that is the application of local mirror symmetry to the construction
of a prepotential accounting for the lower genus Gopakumar-Vafa
invariants (which we simply called the Gromov-Witten invariants or
$GW$--invariants). In particular, for this purpose, we have adopted a
particularly elegant way introduced by Hosono in \cite{Hosono:2004jp}
and which we dubbed the ``Hosono conjecture''. Our results can be thus
interpreted also as a positive (partial) check of the Hosono
conjecture for the case of an orbifold with multiple resolutions.\\
Indeed, we applied the Hosono conjecture to an orbifold model
admitting five distinct crepant resolutions, showing that, for each
resolution, it partially determines a prepotential encoding
information about the Gromov-Witten invariants. As we seen, not all
$GW$--invariant are determined. Indeed it is not even clear how they
could be defined as some ambiguity is introduced by noncompactness of
the varieties considered. However, we can note that for all
resolutions, the only non computable invariants are the one associated
to curves having zero intersection with the compact divisor
$D_7$. Curves having negative intersection with $D_7$ cannot deform
out of $D_7$. When they have non negative intersection with all the
other (non compact) divisors, then we expect for the invariant numbers
to count the number of deformations in $D_7$. However, when the
intersection with some of the noncompact divisors is negative, then
the deformations are constrained on the intersection between the
divisors and we expect for the G--W numbers to vanish.  \\ The
invariants predicted by the prepotential agree with the ones computed
directly by means of the methods described in \cite{Chiang}.  In place
of repeating such computations here, we will simply compare some of
the invariants of our examples with the ones provided in
\cite{Chiang}.  Let us start with resolution five. It contains a
$\mathbb {P}^2$ associated the compact divisor $D_7$ and the curves
$C_{27}$, $C_{37}$ and $C_{67}$, all equivalent. Then, let us fix
$b:=C_3\equiv C_{67}$. It has intersection $-3$ with $D_7$ so that it
is the null section of the normal bundle of $D_7$ in $X$. It also has
intersection numbers $1$ with $D_3$, $D_5$, $D_6$ and $0$ with the
other noncompact divisors. Thus it freely deforms out from all the
noncompact divisors, in the sense discussed above, and then one
expects that the number of its deformations is just the number of
deformations inside $\mathbb {P}^2$.  {F}rom the list (\ref{gwp2}) we
see that, up to degree six, the corresponding numbers are
$GW[d]=GW[0,0,d,0]$ with
\begin{eqnarray}
&& GW[1]=3,\quad GW[2]=-6, \quad GW[3]=27,\cr
&& GW[4]=-192, \quad GW[5]=1695, \quad GW[6]=-17064,
\end{eqnarray}
which indeed coincide with the $GW$--numbers of $\mathcal
{O}(-3)\rightarrow \mathbb {P}^2$, see table $1$ in \cite{Chiang}.\\
Let us move to the fourth resolution. It contains the Hirzebruch
surface $\mathbb {F}_1$ associated to $D_7$ and the curves $C_{27}$,
$C_{37}$, $C_{57}$, $C_{67}$. The independent curves are $b:=C_2\equiv
C_{57}$ and $f:=C_3\equiv C_{67}$ which define the base and the fiber
of the Hirzebruch fibration. Note that $b$ has intersection $-1$ with
$D_5$ so that we expect for its eventual deformations in $\mathbb
{F}_1$ to be constrained. This does not happens for $f$ or for all
combinations $[d_B,d_F]\equiv [0,d_B,d_F,0]$ with $d_F\geq d_B$ which
have negative intersections with $D_7$ only. Thus we again expect for
the $GW$--invariants corresponding to $[d_B,d_F]$ to be the same as in
$F_1$. Indeed from table $10$ in \cite{Chiang} we see that
deformations appears for $K_{\mathbb {F}_1}$ only for $d_F\geq d_B$
(apart from the case $[1,0]$). We can see that our results, as listed
in (\ref{gwf1}), are in perfect agreement
\begin{eqnarray}
&& GW[1,0]=1, \quad GW[0,1]=-2, \quad GW[1,1]=3, \quad GW[1,2]=5,\cr
&& GW[2,2]=-6, \quad GW[1,3]=7, \quad GW[1,4]=9, \quad GW[1,5]=9,\cr
&& GW[3,3]=27, \quad GW[2,3]=-32, \quad GW[2,4]=-110,
\end{eqnarray}
and $GW[i,j]=GW[0,k]=0$ for $j<i$, $i>2$, $i+j\leq 6$ and
$k=2,3,4,5$.\\ A similar comparison can be done for resolution
three. In that case, $D_7$ and the curves $C_{27}$, $C_{37}$, $C_{47}$
and $C_{57}$ define an Hirzebruch surface $\mathbb {F}_2$ with base
$b:=C_2\equiv C_{57}$ and fiber $f:=C_4\equiv C_{47}$. Note that $b$
has intersection $0$ with $D_7$ so that curves $[d_B, 0]$ are not
countable. These correspond to the first column of table $11$ in
\cite{Chiang}. Next curves $[d_B,d_F]=[0,d_B,0,d_F]$ with positive
$d_F$ are computable in $D_7$, but only for $d_F>d_B$ their
intersections with $D_i$ are negative only when they intersect the
compact divisor. We then expect for the curves $[d_B, d_B+1+k]$ to
determine the same numbers as for $K_{\mathbb {F}_2}$, whereas for
$d_F\leq d_B$ they can be constrained by the fact they have negative
intersection with $D_5$ also. However, as follows from table $11$ in
\cite{Chiang} all such numbers vanish (excluding the case
$[d_B,d_F]=[1,1]$) and again our results, collected in (\ref{gwf2}),
agree with the numbers of $K_{\mathbb {F}_2}$
\begin{eqnarray}
&& GW[1,1]=-2, \quad GW[1,2]=-4, \quad GW[1,3]=GW[2,3]=-6,\cr
&& GW[1,4]=-8, \quad GW[1,5]=-10, \quad GW[2,4]=-32,
\end{eqnarray}
and $GW[i,j]=0$ for all the remaining ones up to degree $6$ (and with
$j\neq 0$). \\ All these are only a part of the numbers predicted by
means of the Hosono construction. Indeed, these are the ones
corresponding to curves having negative intersection number with the
compact divisor and thus admitting a representant contained in
it. However, as yet remarked, the constructed potential results to
determine much more numbers and in particular we note that, for all
cases, the only non computable numbers are the ones associated to
curves having null intersection with the compact divisor. Actually the
true meaning of these facts are not completely clear to us and deserve
a deeper analysis, which is left as part of a future work. One way to
proceed in such a direction is to search for an extended GKZ system
whose solutions permit to extend the computation of the invariants to
all curves, as proposed for example in \cite{FoJi1},\cite{FoJi2}. This
also should provide a slight improvement of the Hosono
conjecture. Such analysis are actually under investigation.
Furthermore, Hosono conjecture goes beyond the determination of the
prepotential (or the central charge), involving the monodromy
properties of the hypergeometric components and a concrete
determination of the mirror map at list at the $K$--theoretical level,
and partial information on the homological mirror map Mir. The
multiple resolutions of our example, corresponding to a single mirror
family, are related by flop transformations and must be related by
Fourier--Mukay transforms at the level of derived categories (see
section \ref{K-theory generators flop}). In this contest it could be
helpful to find the solutions of our GKZ system in the full $B$-moduli
space using the approach of \cite{Aganagic:2006wq,Brini:2008rh}. This
is the deeper aspect of the conjecture and will be discussed in the
third paper.

\section*{Acknowledgments}
We would like to thank B. van Geemen and E. Schlesinger for many
useful discussions. We also are grateful to S. Hosono for explaining
us some points about his conjecture, A. Craw and A. Degeratu for
explanations about flop transitions and McKay correspondence and
R. Cavalieri for enlightening us on the meaning of our results.  We
also thank C. Rizzi, M. Penegini and A. Canonaco for interesting
discussions and L. Mauri for his participation to the initial steps of
this work.


\begin{appendix}
\section{The cohomology valued hypergeometric series}\label{appendix}
Here we compute the coefficient hypergeometric series.
\subsection{Computation of the coefficients}
First note that $J_i^3=0$ so that we need the terms up to order two. Also at order zero is survives only the term with $\vec m=\vec 0$,
because for non positive integer argument the $\Gamma$ function diverges. Thus, at order zero $w=1$.
\subsubsection{Some properties of the Gamma function}
The Euler Gamma function has integral representation
\begin{eqnarray}
\Gamma(z)= \int_0^{\infty} e^{-t} t^{z-1} dt \ , \quad\ Re(0)>0\ ,
\end{eqnarray}
and admits analytical continuation to the whole complex plane
excluding the non positive integers.
Indeed it admit the very useful Weierstrass representation
\begin{eqnarray}
\frac 1{\Gamma(z)} =ze^{\gamma z}\prod_{n=1}^\infty \left( 
1+\frac zn\right) e^{-\frac zn}\ , \label{weierstrass}
\end{eqnarray}
where
\begin{eqnarray}
&\gamma=\lim_{N\rightarrow \infty} \left( 1+\frac 12 +\frac 13+\ldots+\frac 1N -\log (N+1)\right) \sim 0.5772156649\ldots
\end{eqnarray}
is the Euler-Mascheroni constant.\\
 {F}rom (\ref{weierstrass}) it follows easily the duplication formula
\begin{eqnarray}
\Gamma(2z) =\frac {2^{2z-1}}{\sqrt \pi}
\Gamma(z) \Gamma(z+1/2)\ .\label{duplication}
\end{eqnarray}
Another useful function is the Psi function
\begin{eqnarray}
\psi(z)=\frac {\Gamma'(z)}{\Gamma(z)}\ .
\end{eqnarray}
 {F}rom (\ref{weierstrass})
\begin{eqnarray}
&\psi(z)=-\gamma -\frac 1z +\sum_{n=1}^{\infty} \frac z{n(n+z)}
\qquad\Rightarrow\qquad
\psi'(z)= \sum_{n=0}^\infty \frac 1{(z+n)^2} =\zeta_z (2)\ .
\end{eqnarray}
Here $\zeta_a (z)=\zeta (a;z)$ is the usual Hurwitz Zeta function.
In particular $\psi'(1)=\pi^2/6$ and, if $N$ is a non negative
integer, using these relations we have
\begin{enumerate}
\item\label{uno}
$$\frac 1{\Gamma(N+1)} =\frac 1{N!}\ , \qquad\ \frac 1{\Gamma(-N)}=0\ ;$$
\item\label{due}
$$\left. \partial_\rho \frac 1{\Gamma(1+N+a\rho)}\right|_{\rho=0} =\frac a{N!}
 \psi(N+1)=\frac a{N!} \left(\gamma-1-\frac 12-\ldots-\frac 1N \right) $$
if $N\neq 0$, and
$$\left. \partial_\rho \frac 1{\Gamma(1+a\rho)}\right|_{\rho=0} =a\gamma\ ;$$
also
$$\left. \partial_\rho \frac 1{\Gamma(-N+\rho)}\right|_{\rho=0} =(-1)^N N!\ .$$
\item\label{tre}
$$ \left. \partial_\rho \partial_\sigma \frac
1{\Gamma(1+N+a\rho+b\sigma)}\right|_{\rho=\sigma=0}=\frac {ab}{N!}
(\psi(N+1)^2-\psi'(N+1))\ ; $$
In particular for $N=0$
$$ \left. \partial_\rho \partial_\sigma \frac1{\Gamma(1+a\rho+b\sigma)}
\right|_{\rho=\sigma=0}=ab\left(\gamma^2-\frac {\pi^2}6\right)\ ; $$
\item\label{quattro}
$$ \left. \partial_\rho \partial_\sigma \frac1{\Gamma(-N+a\rho+b\sigma)}
\right|_{\rho=\sigma=0}=-2ab (-1)^N N! \psi(N+1)\ . $$
\end{enumerate}

\subsubsection{Order one}
To compute the coefficients at order one, we can distinguish three cases:
\begin{itemize}
\item the derivative acts on the numerator. This gives a term $\log x$ only, because the sum contributes only with the term $\vec m=0$;
\item the derivative acts on a factor of the form $1/\Gamma(N+1)$. Then the remaining factors force again $\vec m$ to zero and by (\ref{due})
we see that it contribute with a factor $a\gamma$. There is one such factor for any Gamma factor, and all sum up to zero. This is due
to the fact that for any fixed $\ell$ the sum of its components vanishes;
\item the main contributions come out when the derivative act on a factor $1/\Gamma(-N+1)$. In this case such factor does not contribute
to limiting the allowed values for $\vec m$, which is no more constrained to zero.
\end{itemize}
In conclusion, all the results can be expressed in terms of the
following functions
\begin{eqnarray}
&&\label{psifunc} \Psi(x)=\sum_{n=1}^\infty \frac {(2n-1)!}{(n!)^2} x^n\ ,\\
&&\label{phifunc} \Phi(x,y)=
\sum_{ {\begin{array}{c}(m,k)\in \ZZ_{\geq} \\(m,k)\neq
(0,0)\end{array}}}
\frac {(2k+3m-1)!}{m! (k+2m)! k!}(-x)^m y^{k+2m} \ ,\\
&&\label{aleffunc} \aleph(\vec x)=-\sum_{ {\begin{array}{c} \vec n 
\in\ZZ^4_\geq \\ \vec n \neq 0 \end{array}}}
\frac {(6n_1+4n_2 +2n_3+3n_4-1)!}{n_1!n_2! n_3! n_4! (2n_1+n_2+n_4)!
      (3n_1+2n_2+n_3+n_4)!}\cdot\\
&& \qquad \qquad \qquad \qquad \cdot\ (x_1 x_2^2 x_3^3 x_4^4)^{n_1} 
(x_2 x_3^2 x_4^2)^{n_2} (x_3 x_4)^{n_3} (-x_2 x_3 x_4^2)^{n_4}\nonumber.
\end{eqnarray}
\subsubsection{Order two}
The second order term is obtained applying the second order operator
\begin{eqnarray}
O_2=\frac 12 \partial^2_{\rho_3}+\partial^2_{\rho_4} +\partial^2_{\rho_2\rho_3} +\partial^2_{\rho_2\rho_4} +2\partial^2_{\rho_3\rho_4}
\end{eqnarray}
at $\vec \rho=0$.
In this case there are several contributions\footnote{for simplicity we will call {\it regular} the Gamma factors with positive
argument, and {\it singular} the Gamma factors with negative argument;}:
\begin{itemize}
\item both derivatives acts on the numerator in the terms of the series. This gives rise to terms of the form
$$
(\log x_i)^2\ ,\qquad\ \log x_i \log x_j\ ;
$$
\item one derivative acts on the numerator and the other one acts on the Gamma factors. This gives terms of the form
$$
\log x_i w_j^{(1)}\ ,
$$
where $w_j^{(1)}$ is one of the first order terms computed before;
\item both derivatives acts on two regular Gamma factors. These can be two distinct factors or the same factor. In both the cases
it contributes only the $\vec m=0$ term. For two distinct factors we will find a contribution proportional to $\gamma^2$ and
for the same factor one finds a term proportional to $\gamma^2-\pi^2/6$. A simple argumentation similar to the first order case shows
that the terms in $\gamma^2$ sum up to zero. Thus we expect only a term proportional to $\pi^2$;
\item one derivative acts on a regular term and the other one acts on a singular term. This gives rise to a contribution very similar to
(\ref{psifunc}), (\ref{phifunc}), where the terms of the series are corrected by a multiplicative factor of the form $\psi(N+1)$;
\item both derivatives acts on the same singular term. This gives a contribution very similar to the previous point;
\item the derivatives acts on two distinct singular Gamma terms. These give the more complicated series, because there are minimal constrictions for the range of $\vec m$ in the sums.
\end{itemize}
\subsubsection{Second order functions}
Here we collect the functions which appear in the second order terms
of the cohomological hypergeometric functions.
When the ranges $\vec m\in \mathbb{Z}^4_>$ are intended to be
restricted to the subsets where all factorials and psi functions are
well defined. The results can then be expressed in terms of the
following twenty six functions
\begin{eqnarray}
&& \Psi_1(x,y)=\!\!\!\!\!
\sum_{ {\begin{array}{c}(m,k)\in \ZZ_{\geq} \\(m,k)\neq
(0,0)\end{array}}} \!\!\!\!\!
\frac {(2k+3m-1)!}{m! (k+2m)! k!}[\psi(2k+3m)-\psi(1)] (-x)^m y^{k+2m} \ ;\\
&& \Psi_2(x,y)=\!\!\!\!\!
\sum_{ {\begin{array}{c}(m,k)\in \ZZ_{\geq} \\(m,k)\neq
(0,0)\end{array}}}\!\!\!\!\!
\frac {(2k+3m-1)!}{m! (k+2m)! k!}[\psi(1+k+2m)-\psi(1)] (-x)^m y^{k+2m} \ ;\\
&& \Psi_3(x,y)=\!\!\!\!\!
\sum_{ {\begin{array}{c}(m,k)\in \ZZ_{\geq} \\(m,k)\neq
(0,0)\end{array}}}\!\!\!\!\!
\frac {(2k+3m-1)!}{m! (k+2m)! k!}[\psi(1+k)-\psi(1)] (-x)^m y^{k+2m} \ ;\\
&& \Psi_4 (x) =\sum_{m=1}^\infty \frac {(2m-1)!}{(m!)^2} [\psi(2m)-\psi(1)] x^m \ ;\\
&& \Psi_5 (x) =\sum_{m=1}^\infty \frac {(2m-1)!}{(m!)^2} [\psi(m+1)-\psi(1)] x^m \ ;\\
&& \Psi_6(x,y) =\sum_{ {\begin{array}{c} (m,n)\in\ZZ^2_\geq \\ 2n-m>0\\ 2m-n>0 \end{array}}}
\frac {(2n-m-1)! (2m-n-1)!}{m!n!}(-x)^m(-y)^n\ ;\\
&& \aleph^{(i)}(\vec x)=\sum_{ {\begin{array}{c} \vec n \in\ZZ^4_\geq
\\ \vec n \neq 0 \end{array}}}
\frac {(x_1 x_2^2 x_3^3 x_4^4)^{n_1} (x_2 x_3^2 x_4^2)^{n_2} 
(x_3 x_4)^{n_3} (-x_2 x_3 x_4^2)^{n_4} }{n_1!n_2! n_3! n_4! 
(2n_1+n_2+n_4)! (3n_1+2n_2+n_3+n_4)!} \chi^{(i)}_{\vec n}\ ,\\
&& i=1,\ldots,7;\nonumber
\end{eqnarray}
\begin{eqnarray}
&& \chi^{(1)}_{\vec n} =(6n_1+4n_2 +2n_3+3n_4-1)![\psi(6n_1+4n_2+2n_3+3n_4)-\psi(1)];\\
&& \chi^{(2)}_{\vec n} =(6n_1+4n_2 +2n_3+3n_4-1)![\psi(1+3n_1+2n_2+n_3+n_4)-\psi(1)];\\
&& \chi^{(3)}_{\vec n} =(6n_1+4n_2 +2n_3+3n_4-1)![\psi(1+2n_1+n_2+n_4)-\psi(1)];\\
&& \chi^{(4)}_{\vec n} =(6n_1+4n_2 +2n_3+3n_4-1)![\psi(1+n_2)-\psi(1)];\\
&& \chi^{(5)}_{\vec n} =(6n_1+4n_2 +2n_3+3n_4-1)![\psi(1+n_4)-\psi(1)];\\
&& \chi^{(6)}_{\vec n} =(6n_1+4n_2 +2n_3+3n_4-1)![\psi(1+n_1)-\psi(1)];\\
&& \chi^{(7)}_{\vec n} =(6n_1+4n_2 +2n_3+3n_4-1)![\psi(1+n_3)-\psi(1)];
\end{eqnarray}
\begin{eqnarray}
&&\Lambda_1(\vec x)=
\sum_{ {\begin{array}{c} \vec n \in\ZZ^4_\geq \\ n_1+n_3\neq 0\end{array}}}
\frac {n_1!n_2! n_3! n_4! (n_1+2n_2+n_3+n_4)!}
{(n_1+3n_2+2n_4)!(n_1+2n_3)!x_1^{n_1}}\cdot\\
&& \qquad\qquad\qquad\qquad\quad\cdot\
(-x_1^2 x_2 x_4^2)^{n_2} (x_1 x_3)^{n_3} (x_1 x_4)^{n_4}\ ;\nonumber \\[3mm]
&&\Lambda_2(\vec x)=
\sum_{\vec m\in \mathbb{Z}^4_>}
\frac {(m_1-m_2-m_3+m_4-1)! (m_3+m_4-m_1-1)!}
{m_1! m_2! m_3! (m_4-2m_2)! (m_1-m_3-m_4)!}\cdot\\
&& \qquad\qquad\quad\ \cdot\ 
x_1^{m_1} (-x_2)^{m_2} x_3^{m_3} x_4^{m_4} \ ;\nonumber\\[3mm]
&&\Lambda_3(\vec x)=\sum_{\vec m\in \mathbb{Z}^4_>}
\frac {(m_1+m_3-m_4-1)!(m_3+m_4-m_1-1)!}
{m_1!m_2!m_3!(m_2+m_3-m_1-m_4)!(m_4-2m_2)!}\cdot\\
&& \qquad\qquad\quad\ \cdot\ 
x_1^{m_1} x_2^{m_2} x_3^{m_3} x_4^{m_4};\nonumber\\[3mm]
&&\Lambda_4 (\vec x)=
\sum_{\vec m\in \mathbb{Z}^4_>}
\frac {(m_1-m_3-m_4-1)!(m_1-1)!}
{m_2!m_3!(m_1+m_2-2m_4)!(m_4-2m_2)!(m_1-2m_3)!}\cdot\\
&& \qquad\qquad\quad\ \cdot\ 
x_1^{m_1} x_2^{m_2} (-x_3)^{m_3} (-x_4)^{m_4};\nonumber\\[3mm]
&&\Lambda_5 (\vec x)=
\sum_{\vec m\in \mathbb{Z}^3_>}
\frac {(2m_3-m_1-1)!(2m_2-1)!}
{m_1!m_2!(m_2+m_3)!(m_3-2m_1)!}\cdot\\
&& \qquad\qquad\quad\ \cdot\ 
(-x_2)^{m_1} x_3^{m_2} (-x_4)^{m_3};\nonumber\\[3mm]
&&\Lambda_6 (\vec x)=
\sum_{\vec m\in \mathbb{Z}^4_>}
\frac {(m_3-m_4-m_1-1)!(m_3-m_2-1)!}
{m_1!m_2!(m_4-2m_2)!(m_3-2m_2)!(m_3-2m_4)!}\cdot\\
&& \qquad\qquad\quad\ \cdot\ 
(-x_1)^{m_1} (-x_2)^{m_2} x_3^{m_3} (-x_4)^{m_4};\nonumber
\end{eqnarray}
\begin{eqnarray}
&&\Lambda_7 (\vec x)=
\sum_{\vec m\in \mathbb{Z}^4_>}
\frac {(m_3-m_2-1)!(2m_4-m_3-1)!}
{m_1!m_2!(m_1-m_3+m_4)!(m_4-2m_2)!(m_3-2m_2)!}\cdot\\
&& \qquad\qquad\quad\ \cdot\ 
x_1^{m_1} (-x_2)^{m_2} x_3^{m_3} x_4^{m_4};\nonumber\\[3mm]
&&\Lambda_{8} (\vec x)=
\sum_{\vec m\in \mathbb{Z}^4_>}
\frac {(m_3-m_2-1)!(2m_2-m_4-1)!}
{m_1!m_2!(m_1-m_3+m_4)!(m_3-2m_2)!(m_3-2m_4)!}\cdot\\ 
&& \qquad\qquad\quad\ \cdot\ 
x_1^{m_1} (-x_2)^{m_2} (-x_3)^{m_3} (-x_4)^{m_4};\nonumber\\[3mm]
&&\Lambda_{9} (\vec x)=
\sum_{\vec m\in \mathbb{Z}^4_>}
\frac {(m_4-m_3+m_2-m_1-1)!(2m_3+m_2-m_4-1)!}
{m_1!m_2!m_3!(m_4-2m_1)!(m_2-m_4)!}\cdot\\
&& \qquad\qquad\quad\ \cdot\ 
(-x_1)^{m_1} x_2^{m_2} (-x_3)^{m_3} x_4^{m_4};\nonumber\\[3mm]
&&\Lambda_{10} (\vec x)=
\sum_{\vec m\in \mathbb{Z}^4_>}
\frac {(m_4-m_3+m_2-m_1-1)!(m_4-m_2-1)!}
{m_1!m_2!m_3!(m_4-2m_1)!(m_4-2m_3-m_2)!}\cdot\\
&& \qquad\qquad\quad\ \cdot\ 
(-x_1)^{m_1} x_2^{m_2} (-x_3)^{m_3} x_4^{m_4};\nonumber\\[3mm]
&&\Lambda_{11} (\vec x)=
\sum_{\vec m\in \mathbb{Z}^4_>}
\frac {(m_4-m_2-1)!(2m_3+m_2-m_4-1)!}
{m_1!m_2!m_3!(m_4-2m_1)!(m_1+m_3-m_2-m_4)!}\cdot\\
&& \qquad\qquad\quad\ \cdot\ 
x_1^{m_1} x_2^{m_2} x_3^{m_3} x_4^{m_4};\nonumber\\[3mm]
&&\Lambda_{12} (\vec x)=
\sum_{\vec m\in \mathbb{Z}^4_>}
\frac {(m_2-m_3-m_4-1)!(m_2-m_3-1)!}
{m_1!m_3!(m_4-2m_1)!(m_1+m_2-2m_4)!(m_2-3m_3)!}\cdot\\
&& \qquad\qquad\quad\ \cdot\ 
x_1^{m_1} x_2^{m_2} x_3^{m_3} (-x_4)^{m_4};\nonumber\\[3mm]
&&\Lambda_{13} (\vec x)=
\sum_{\vec m\in \mathbb{Z}^4_>}
\frac {(m_2-m_3-1)!(3m_3-m_2-1)!}
{m_1!m_3!(m_3+m_4-m_2)!(m_4-2m_1)!(m_1+m_2-2m_4)!}\cdot\\
&& \qquad\qquad\quad\ \cdot\ 
x_1^{m_1} x_2^{m_2} x_3^{m_3} x_4^{m_4}.\nonumber
\end{eqnarray}

\end{appendix}

\newpage

\bibliographystyle{my-h-elsevier}

\end{document}